\documentclass{jfm}
\usepackage{latexsym}
\usepackage{amsmath}
\usepackage{amssymb}
\usepackage{float}
\usepackage{graphics}
\usepackage{graphicx}
\usepackage{mathrsfs}
\usepackage[toc,page]{appendix}
\usepackage{subfigure}
\usepackage{caption}
\usepackage{hyperref}
\hypersetup{
	colorlinks = true,
	urlcolor   = blue,
	citecolor  = blue,
}

\newcommand{\bea}{\begin{eqnarray}}
	\newcommand{\eea}{\end{eqnarray}}

\newcommand{\bc}{\begin{center}}
	\newcommand{\ec}{\end{center}}

\newcommand{\be}{\begin{enumerate}}
	\newcommand{\ee}{\end{enumerate}}
\newcommand{\bt}{\begin{tabbing}}
	\newcommand{\et}{\end{tabbing}}
\newcommand{\btb}{\begin{tabular}}

\title{Diffusiophoresis of a non-polar fluid droplet laden with soluble ionic surfactants}

\author{Subrata Majhi\aff{1,2}, \and Somnath Bhattacharyya\aff{1}%
	\corresp{\email{somnath@maths.iitkgp.ac.in}}}

\affiliation{
	\aff{1}\parbox[t]{0.95\textwidth}{Department of Mathematics, Indian Institute of Technology Kharagpur, Kharagpur 721302, India}
	
	\aff{2}\parbox[t]{0.95\textwidth}{Multicomponent Fluids Group, Center for Complex Flows and Soft Matter Research, Department of Mechanics and Aerospace Engineering, Southern University of Science and Technology, Shenzhen 518055, Guangdong, P.R. China}
}
\begin{document}
	\maketitle
\begin{abstract}
We investigate the diffusiophoresis of a non-polarizable droplet laden with soluble ionic surfactant, for which the surface charge arises from adsorption of surfactant at the fluid–fluid interface. Unlike previous studies that assume either a fixed surface charge or instantaneous equilibrium between the interface and the adjacent electrolyte, we formulate the interfacial transport based on the mass-balance framework incorporating Langmuir adsorption–desorption kinetics and finite surface diffusivity. The coupled electrokinetic problem is solved using a perturbation approach with imposed ionic concentration as a perturbation parameter. Analytical expressions for the droplet mobility and interfacial velocity are derived based on the Debye--H\"uckel approximation as well as thin layer consideration for the insoluble surfactant, and are found to agree with the numerical results in the corresponding limits. We demonstrate that assuming uniform, immobile surface charge leads to unphysical predictions, including negative chemiphoresis and singular mobility, whereas allowing the surface charge to evolve through interfacial surfactant redistribution yields continuous and physically consistent droplet diffusiophoresis. Interfacial kinetic exchange is found to play a central role. Increasing the desorption rate enhances surfactant redistribution and Marangoni stress, weakens the negative mobility, reverses the direction of motion through competition between electrophoretic and chemiphoretic contributions, and subsequently leads to a strong enhancement of positive mobility before eventual saturation in the transport-limited regime. The dependence of mobility on viscosity ratio and electrolyte composition of different salts further reveals how mixed electrolytes provides a robust means of tuning droplet motion. This study highlights the critical role of finite-rate surfactant dynamics and interfacial transport in determining the diffusiophoresis of fluid particles, with implications for manipulating droplets in microfluidic and varying-salinity environments.
\end{abstract}
	
\section{\textbf{Introduction}}
The stability of dispersed fluid particles, such as nanobubbles and oil droplets, is governed largely by ion enrichment and charge distribution at the fluid–liquid interface. For nanobubbles, the adsorbed charge generates a size-dependent restoring force against thermodynamic perturbations \citep{zhang2020surface}. A high $\zeta$-potential typically creates repulsive forces that prevent coalescence, contributing to stabilization \citep{nirmalkar2018interpreting}. However, interpreting these potentials is complex; for instance, the large negative $\zeta$-potentials observed at gas–liquid interfaces are often attributed to charged surface-active impurities \citep{uematsu2020nanomolar}. Similarly, the charging mechanisms of non-polar oil droplets with low dielectric constants are dominated by the specific adsorption of ionic surfactants. Recent experiments \citep{wang2025movement} demonstrate that oil droplets in aqueous solutions under salinity gradients are stabilized primarily by these surface-active components. Crucially, $\zeta$-potentials at fluid–fluid interfaces are often calculated using the Henry or Smoluchowski models derived for rigid solid–liquid interfaces. These models fail to account for fluid interface dynamics, meaning measured $\zeta$-potentials may not accurately reflect the true surface charge state \citep{hill2020electrokinetic,hill2020dynamic}. This shows that characterization of surfactant adsorption at the fluid-liquid interface at both equilibrium and dynamic state is important in analyzing the electrokinetics of fluid droplets in an aqueous medium. 

\par 
The diffuisophoresis \citep{derjaguin1993diffusiophoresis} is an electrokinetic mechanisms to create a directed transport of a charged colloid mediated by an ionic concentration gradient. The osmotic pressure gradient created by the solute concentration gradient polarizes the electric double layer (EDL) around the charged particle. In this process, the charged particle experiences a momentum by the chemiphoresis mechanisms. Another part of the diffuisophoresis mechanisms, the electrophoresis, arises due to the difference in diffusivity of ionic species. Under a thin Debye layer consideration, \cite{prieve1984motion} determined an expression for the velocity of a rigid charged particle in diffusiophoresis. Since then, various aspects of diffusiophoresis have been investigated through theoretical, numerical, and experimental approaches \citep{ebel1988diffusiophoresis,velegol2016origins,shim2022diffusiophoresis,ault2025physicochemical}.
\par 
The diffusiophoresis of nanodroplets has relevance in several practical context such as, water purification, soil and petroleum industry \citep{yang2018diffusiophoresis,park2021microfluidic,chuang2025diffusiophoresis}. In addition, the directed transport of nano- or micro-droplets has several importance in droplet microfluidics such as, controllable drug delivery, and food products dispersion \citep{yang2010manipulation,simon2012microfluidic,sohrabi2020retracted,majhi2023diffusiophoresis}. Over the years several studies on diffuisophoresis of a charged droplet is reported in the literature by considering the surfactant-free droplet surface \citep{yang2018diffusiophoresis,wu2021diffusiophoresis,fan2022diffusiophoresis,samanta2023diffusiophoresis}.  In those studies demonstrate that the Maxwell traction created by the macroscopic electric field at the interface plays a major role in determining the diffusiophoretic mobility of the droplet. For a negatively charged droplet in a concentrtaion gradient of an electrolyte with lower cationic diffusivity the electrophoresis part creates a positive mobility, which may cooperates with the chemiphoresis part. However, for a low viscous droplet the chemiphoresis part may drag the droplet in the direction opposite to the imposed concentration gradient. Such phenomena is termed as the double layer polarization of second kind ( DLP-II), in which the strong surface charge repels the coins and prevents the diffusion across the double layer. This leads to repulsive force to the droplet, creating a force counetracting with the electrophoresis part. In this case the droplet can have a mobility opposite to the direction of the concentration gradient (negative mobility).
\par 
When the charging mechanisms of the non-polar droplet is considered to be due to the specific adsorption of ionic surfactant then the phoretic transport, either electrophoresis ( governed by an externally imposed electric field) or diffusiophoresis, becomes different from the case of droplet surface with constant charge density. In such cases the dynamics become complicated owing to the Marangoni stress due to the non-uniform surfactant distribution around the droplet. Theoretical study on the impact of ionic surfactant on droplet diffusiophoresis was addressed by \citet{baygents1988migration}. In this case the dynamics becomes complicated owing to the Marangoni stress due to the non-uniform surfactant distribution. The adsorption/ desorption of surfactant at the interface creates a non-homogeneous surface charge, which generates a non-uniform EDL around the particle. This non-uniform EDL augments the DLP effect, creating a stronger chemiphoresis. \citet{baygents1988migration} has shown that the Marangoni stress due to the formation of the nonuniform ionic surfactant may enhance the positive mobility of the droplet. However, their formulation assumed zero surface diffusivity $(D_s=0)$ and instantaneous equilibrium ($k_a\to\infty$, $k_d\to\infty$) between the interface and the immediately adjacent electrolyte. These assumptions correspond to vanishing interfacial-charge mobility and infinitely fast kinetic exchange, which are difficult to justify physically because adsorption and desorption rates are thermodynamically coupled to the equilibrium charge density and $\zeta$-potential. Their model uses a Linear (Henry-type) adsorption relation, which is valid only in the dilute limit. Real surfactants exhibit nonlinear adsorption due to finite site availability and lateral interactions, well described by the Langmuir or Frumkin isotherms. The linear assumption neglects saturation effects, leading to inaccurate prediction of the surface tension gradient, and thus of Marangoni stresses and interfacial slip at high surface coverage. Moreover, \citet{baygents1988migration} effectively treated ordinary electrolyte ions (e.g. K$^{+}$) as the adsorbing species, overlooking the role of true surface-active ions such as DS$^{-}$ that control interfacial tension and compositional Marangoni stresses. Consequently, their theory captures the electrocapillary limit of electrophoresis but not the finite-rate, surfactant-driven dynamics relevant to real fluid interfaces. More recently, \citet{hill2025roles} developed a comprehensive framework for the electrophoresis of fluid droplets that includes interfacial kinetic exchange and charge mobility of ionic surfactants. Unlike earlier models such as the uniform-charge formulation of \citet{booth1951cataphoresis}, the mercury-drop model of \citet{ohshima1984electrokinetic}, and the equilibrium-interface theory of \citet{baygents1988migration}, Hill’s analysis incorporates finite adsorption–desorption kinetics (although using the Henri model), lateral charge transport, and Marangoni stresses within the weak-field approximation. His study demonstrated that finite interfacial mobility regularizes the singular electrophoretic mobilities reported for highly charged drops and bubbles and that adsorption kinetics crucially determine whether an interface behaves as mobile or effectively rigid. Despite these advances, a corresponding theoretical framework for diffusiophoresis that accounts for finite interfacial kinetics and charge mobility remains lacking.
\par 
In this study, we investigate the diffusiophoresis of a non-polarizable droplet laden with soluble ionic surfactants. The surface charge of the droplet arises from the adsorption of ionic surfactant molecules at the fluid–fluid interface. The interfacial transport of surfactant is formulated through a mass-balance framework incorporating Langmuir adsorption kinetics, where both adsorption and desorption fluxes are explicitly modeled. The electrokinetic problem couples Stokes flow inside and outside the droplet with ion transport governed by the Nernst–Planck equation and electrostatics described by the Poisson equation. This coupled system is solved through a first-order perturbation approach, limited to a linear order in the externally imposed ionic concentration gradient. Analytical expressions for the diffusiophoretic mobility and interfacial velocity are obtained under the Debye–Hückel and thin double-layer approximations in the regime corresponding to negligible desorption. Our results show that the double-layer polarization of type-II, reported in the literature \citep{wu2021diffusiophoresis,fan2022diffusiophoresis,samanta2023diffusiophoresis} in the context of diffusiophoresis of droplets with constant surface charge density, drives the droplet rapidly in the direction opposite to the imposed ionic concentration gradient and the discontinuity in the mobility at a high surface charge density is an artifact of the consideration of immobile surface charge and neglect of the surface tension gradient. We find that the presence of Marangoni stress and mobility of surface ions create a smooth dependence of the diffusiophoretic mobility on the surface charge density. At a higher desorption kinetics of soluble surfactant, the droplet velocity increases significantly in the positive direction, independent of the diffusion potential. This behavior highlights the critical role of surfactant dynamics in controlling droplet transport, offering potential strategies for optimizing electrokinetic manipulation in applications such as drug delivery and enhanced oil recovery.

\section{\bf Mathematical Formulation}\label{Math_model}
\begin{figure}
 \centering
 \includegraphics[width=5.0in]{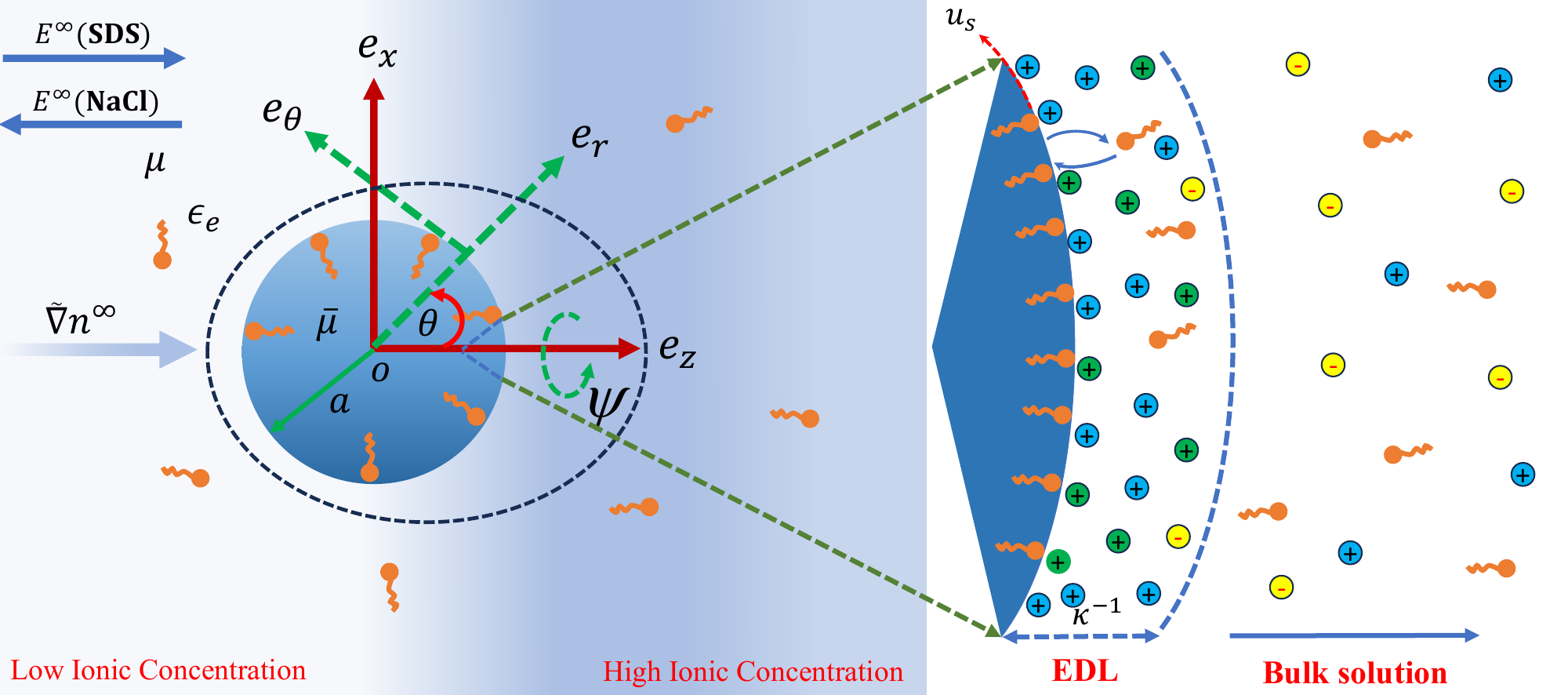}
 \caption{Problem description and spherical coordinate system for the diffusiophoresis of a surfactant-laden liquid droplet in an electrolyte solution subjected to an ionic concentration gradient.}
 \label{fig1}
\end{figure}

We consider the diffusiophoresis of a non-conducting, immiscible, and neutrally buoyant spherical droplet of a viscous dielectric liquid with radius $a$ and viscosity $\overline{\mu}$. The droplet is suspended in an electrolyte solution of viscosity $\mu$ and permittivity $\varepsilon_e$. The surrounding electrolyte consists of a simple supporting salt (e.g. NaCl, KCl, or HCl) and a soluble ionic surfactant, such as sodium dodecyl sulfate (SDS) or sodium dodecylbenzene sulfonate (SDBS). The surfactant dissociates into an anionic surface-active species and its counterion. We assume that only the anionic species adsorbs at the droplet interface and exchanges with the adjacent electrolytes. The droplet is translating with diffusiophoretic velocity $\tilde{U}_{D}$, unknown a priori, in response to an external ionic concentration gradient $\tilde{\nabla}{n^{\infty}}$. Variables with a tilde are dimensional. The spherical polar coordinate ($r,\theta,\psi$) is adopted with its origin fixed at the center of the droplet and $z$-axis ($\theta$ = 0) is along the imposed concentration gradient $\tilde{\nabla}{n^{\infty}}$ (Fig.\ref{fig1}). We assume that the problem is axially symmetric with $z$-axis as the axis of symmetry. The uniform concentration gradient at $r=\infty$ is given by $\tilde{\nabla}{n^{\infty}}=|\tilde{\nabla}{n^{\infty}}|\boldsymbol{e_z}$, where $\boldsymbol{e_z}$ is the unit vector along the $z$-axis and $|\tilde{\nabla}{n^{\infty}}|$ is the magnitude of the imposed concentration gradient.
The motion of ionized incompressible Newtonian electrolyte outside the droplet can be expressed in non-dimensional form as,
\begin{subequations}\label{eq:eq1}
	\begin{gather}
		\label{eq:eq1:a}
		-\nabla p+\nabla^{2}\textbf{u}-(1/2)(\kappa a)^{2} \rho_e \nabla \psi =0,\\
		\label{eq:eq1:b}
		\nabla \cdot \textbf{u}=0 
	\end{gather}
\end{subequations}
where $\textbf{u}=v \boldsymbol{e}_r+ u \boldsymbol{e}_\theta$ is the velocity vector, with $v$ and $u$ being the radial and polar velocity components, respectively; $t$ is the time and $p$ is the pressure. The governing equations are nondimensionalised using the droplet radius $a$ as the characteristic length scale, the velocity scale $U_{0}=\varepsilon_{e}\psi_{0}^{2}/(a\mu)$, the time scale $\tau=a/U_{0}$, and the pressure scale $\varepsilon_{e}\psi_{0}^{2}/a^{2}$. Note that $\mu$ is the viscosity of the electrolyte medium and $\psi_{0}=k_{B}T/e$ is the thermal potential. The electric potential $\psi$ is scaled by $\psi_{0}$. The space charge density $\rho_e$, defined as $\rho_e=\textstyle\sum_{i}{z_{i}n_{i}}$, scaled by $eI^{\infty}$, where $e$ is the elementary charge and $I^{\infty}$ is the bulk ionic strength. $I^{\infty}$ is defined by $I^{\infty}=(1/2)\sum_{i=1}^{N}{z^{2}_{i}n^{\infty}_{i}}$ with $n_{i}^{\infty}$ denoting the dimensional bulk concentration of the $i$th ionic species. The variable $n_{i}$ is the dimensionless ionic concentration of i$^{\text{th}}$ ionic species, scaled by $I^{\infty}$. Here, $\kappa^{-1}$ is the Debye length defined as $\kappa^{-1}=\sqrt{\epsilon_e\psi_{0}/2I^{\infty}e}$. Further, the nondimensional equation for the flow inside the droplet is,
\begin{subequations}\label{eq:eq2}
	\begin{gather}
		\label{eq:eq2:a}
		\mu_{r} \nabla^{2}\overline{\textbf{u}}-\nabla \overline{p}=0,\\
		\label{eq:eq2:b}
		\nabla \cdot \overline{\textbf{u}}=0
	\end{gather}
\end{subequations}
where $\overline{\textbf{u}}=\overline{v} \boldsymbol{e}_r
+\overline{ u} \boldsymbol{e}_\theta$ is the velocity vector and $\overline{p}$ is the pressure inside the droplet. $\mu_{r}~(=\overline{\mu}/\mu)$ is the viscosity ratio between the viscosity of the drop and the viscosity of the ionized fluid. The droplet is assumed to contain no free ions and thus, the electric body-force term does not appear in eq.(\ref{eq:eq2:a}). In eqs.(\ref{eq:eq1:a}) and~(\ref{eq:eq2:a}), the inertial terms are neglected since the Reynolds numbers $Re$ and $\overline{Re}$ are $O(10^{-4})$ and therefore inertia has no influence on the fluid motion or on the droplet dynamics. The boundary conditions for the velocity field at the droplet surface ($r=1$) are the tangential velocity continuity and fluid impermeability conditions i.e.,
\begin{equation}
	\textbf{u}=\overline{\textbf{u}},~~~~~\textbf{u}\cdot\textbf{n}=0.
	\label{eq6}
\end{equation}
Here, $\textbf{n}$ is the outward unit normal vector at the surface. The stress balance condition is crucial in droplet electrophoresis and it's accurate derivation is needed in presence of ionic surfactants at the interface. The tangential stress balance condition is given by \citep{slattery2007interfacial,manikantan2020surfactant} 
\begin{equation}\label{SBC:1}
	\textbf{n}\cdot(S_{H}-\overline{S}_{H})\cdot\textbf{t}=-Ma\nabla_{s}\gamma\cdot \textbf{t}
\end{equation}
Here, $S_{H}$ denotes the hydrodynamic stress tensor, defined as $S_{H}=-p\boldsymbol{I}+[\nabla \textbf{u} + (\nabla \textbf{u})^T]$, scaled by $\mu U_{0}/a$ and $\mathbf{t}$ is the unit tangential vector at the interface. The variable $\gamma$ represents the dimensionless surface tension and $\nabla_{s}=(\boldsymbol{I}-\textbf{n}\textbf{n})\cdot\nabla$ is the gradient along the surface. The surface tension is scaled with $\Gamma^{\infty} k_{B}T$, where $\Gamma^{\infty}$ is the maximum surface packing concentration of the surfactant. $Ma$ is the Marangoni number defined as $Ma=ea\Gamma^{\infty}/\epsilon_e\psi_{0}$ or equivalently, $Ma=k_{B}T\Gamma^{\infty}/{\mu U_{0}}$, which measures the ratio between the characteristic surface-pressure stress generated by surfactant accumulation, to the viscous stress scale. In the present study, the deformation of the droplet is not considered as it has negligible impact on the droplet dynamics. This assumption is justified by the fact that the capillary number remains very small as the applied concentration gradient $\alpha\sim O(10^{-3})$, as discussed later. The surface tension and surfactant concentration are related through the Gibbs adsorption isotherm \citep{kralchevsky1997chemical,manikantan2020surfactant},
\begin{equation}
	-\nabla_{s}\gamma=\nabla_{s}\Pi=\sum_{i}{\Gamma_{i}\nabla_{s}\mu_{s,i}},
\end{equation}
where $\Pi$ is the surface pressure arising from the adsorption of surfactant components $i$ at the interface with interfacial concentrations $\Gamma_{i}$ and $\mu_{s,i}$ denotes the interfacial electrochemical potential of species $i$. In the present study, we restrict ourselves to a single adsorbed species characterized by the surface concentration $\Gamma$, electrochemical potential $\mu_{s}$, and valency $z_{s}$. The interfacial electrochemical potential of the surfactant can be obtained as 
\begin{equation}\label{eq:secp}
	\mu_{s}=\mu_{s0}+\ln[1/(1-\Gamma)]+z_{s}\psi_{s},
\end{equation}
in which the second term in the right hand side (RHS) can be interpreted as the contribution of the change in surface tension due to uncharged surfactant adsorbing to the interface and it modeled through the Langmuir isotherm model, whereas the last term as the contribution due to charging of the adsorbed molecules. Substituting the stresses in the eq.(\ref{SBC:1}) and performing algebraic simplification, we get
\begin{equation}\label{eq7}
	\left(\partial_{r}u-u/r\right)|_{r=1^+}-\mu_r \left(\partial_{r} \overline{u}-\overline{u}/r\right)|_{r=1^-} ={\left[\sigma\partial_{\theta} \psi\right]}_{r=1}+Ma(1-\Gamma)^{-1}\partial_{\theta} \Gamma
\end{equation}
The adsorbed surfactant creates a surface charge density $\sigma=z_{s}Ma\Gamma$ and thus, creates the Maxwell stress as well as Marangoni stress. Dimensionless surface charge density $\sigma$ is given by $\sigma=z_{s}Ma\Gamma$ in which the surface surfactant concentration $\Gamma^{*}$ is scaled by $\Gamma^{\infty}$. In eq.(\ref{eq7}), the left hand side provides the difference in the hydrodynamic shear stresses at the interface. The first term of RHS presents accounts the tangential Maxwell stress created by the charged surfactant and the second term corresponds to the Marangoni stress. An interplay between theses stresses govern the droplet motion. The normal stress balance condition at the surface is not required, as the droplet is considered undeformed. Instead, the impermeability condition given by eq.(\ref{eq6}) is employed as the second boundary condition for fluid flow outside the droplet. 
\par 
The surfactant concentration distribution at the interface is governed by mass balance principle as
\begin{equation}\label{eq:sef}
	\nabla_{s}\cdot\left[Pe_{s}\Gamma u_{s}-\nabla_{s}\Gamma/(1-\Gamma)-z_{s}\Gamma\nabla_{s}\psi_{s}\right]=C_{a}{n_{N,s}}(1-\Gamma)-C_{d}\Gamma
\end{equation}
where, $\Gamma=\Gamma^*/\Gamma^\infty$ and $Pe_s=U_0 a/D_s$, surface Peclet number. The RHS accounts for the interfacial mass exchange arising from adsorption–desorption kinetics, i.e., $J_{\textbf{n}}=C_{a}{n_{N,s}}(1-\Gamma)-C_{d}\Gamma$. The coefficients $C_{a}$ and $C_{d}$ are defined as $C_{a}=k_{a}I^{\infty}a^{2}/D_{s}$ and $C_{d}=k_{d}a^{2}/D_{s}$, in which $k_{a}$ and $k_{d}$ are the the kinetic rate coefficients for adsorption and desorption of surfactant. Note that the RHS of the eq.(\ref{eq:sef}) arises from the finite adsorption and desorption rate during the flow. Here, $n_{N}$ denotes the concentration of the surface-active anionic surfactant species in the surrounding electrolyte, and $n_{N,s}~(=n_N|_{r=1})$ represents its value evaluated at the interface. The quantity $n_{N,s}$ is distinct from the surface concentration of specifically adsorbed surfactant molecules, $\Gamma$. The two are interfacial variables of different dimensionality and are coupled through adsorption kinetics. In the limiting case where the droplet is coated with insoluble, compressible ionic surfactant, that is, when the surfactant are irreversibly adsorbed at the interface ($k_{d}=0$) and do not exchange with the surrounding electrolytic medium during diffusiophoresis, the governing equation for the interfacial surfactant concentration simplifies to
\begin{equation}\label{eq:sef_insoluble}
	\nabla_{s}\cdot\left[Pe_{s}\Gamma u_{s}-\nabla_{s}\Gamma/(1-\Gamma)-z_{s}\Gamma\nabla_{s}\psi_{s}\right]=0.
\end{equation}
\par 
The governing equations for the concentration of the \(i^{\text{th}}\) ionic species $n_{i}$, derived from the conservation of ionic flux, is given by  
\begin{equation}
	\nabla\cdot[Pe_{i}n_{i}\textbf{u}-n_{i}\nabla \mu_{i}]=0,
	\label{eq5}
\end{equation}
in which $\mu_{i}$ denotes the dimensionless electrochemical potential, defined as $\mu_{i}=\mu_{i0}+z_{i}\psi+\ln{n_{i}}$. Here, $Pe_i=\varepsilon_e \psi_0^{2}/\mu D_i$ is the P\'eclet number of the $i^{\text{th}}$ ionic species, characterizing the ratio of advective to diffusion transport of ions in which $D_i$ is the diffusion coefficient of the  $i^{\text{th}}$ ionic species. The boundary condition at the interface for the ionic species are 
\begin{equation}
	\bigl(Pe_{i}n_{i}\mathbf{u}-n_{i}\nabla \mu_{i}\bigr)\cdot \mathbf{n} 
	= \delta_{iN}\,\bigl(I_{d}\Gamma - I_{a}n_{N,s}(1-\Gamma)\bigr),~~~i=1,\dots,N.
\end{equation}
where $ \delta_{iN}$ is the Kronecker delta and $i=N$ refers to the anionic surfactant species that adsorbs at the droplet interface and undergoes adsorption-desorption exchange with the adjacent electrolyte. Here, the coefficients $I_{a}$ and $I_{d}$, which arise due to the kinetics exchange between the interface and the adjacent electrolyte, are defined as $I_{a}=k_{a}\Gamma^{\infty}a/D_{N}$ and $I_{d}=k_{d}\Gamma^{\infty}a/(D_{N}I^{\infty})$.
\par 
The governing equation for the electric potential ($\psi$) is the Poisson equation described as
\begin{equation}\label{MPE:1}
	\nabla^{2}\psi=-(1/2)(\kappa a)^{2}\textstyle\sum_{i}{z_{i}n_{i}},
\end{equation}
derived from the Gauss's Law. The boundary condition at the surface for electric potential is  
\begin{equation}\label{MPE:5}
	\textbf{n}\cdot\nabla\psi=-\sigma,
\end{equation} 
appropriate when the electric permittivity is much larger in the electrolyte than in the droplet. 
\par 
An externally imposed ionic concentration gradient $|\tilde{\nabla} n^{\infty}|\boldsymbol{e}_z$ is imposed in the far field, at distances much larger than the droplet radius $(r=R \gg 1)$. Accordingly, the boundary conditions for ionic concentration $n_{i}$ and electric potential $\phi$ are obtained as 
\begin{equation}
	n_i=({n^{\infty}_{i}}/I^{\infty})(1+\alpha R\cos\theta)~~~~~~\text{and}~~~~~\psi=-\alpha\beta R\cos\theta.
	\label{eq11_a}
\end{equation}
The parameter ${\alpha={|\tilde{\nabla}{n^{\infty}}|a}/{n^{\infty}}}$ is the scaled externally imposed concentration gradient. Due to the zero net-current condition under no externally imposed electric field, an induced electric field, the diffusion field, develops when the diffusivity of the ionic species are different i.e., $\beta={\sum_{i}{z_{i}D_{i}n^{\infty}_{i}}}/{\sum_{i}{z^{2}_{i}D_{i}n^{\infty}_{i}}}$ is non-zero and the scaled electric potential is $-\alpha\beta R\cos\theta$.
\par 
The problem is formulated in a reference frame translating with the droplet at the diffusiophoretic velocity $U_D$. In this case, the particle is fixed and the fluid far from the particle is moving with a uniform velocity $-U_D \boldsymbol{e}_z$. Thus, the boundary condition far the particle $(r=R \gg 1)$ for the fluid flow is 
\begin{equation}
	\textbf{u}=-\mu_{D}\alpha \boldsymbol{e_z}.
	\label{eq11_b}
\end{equation}
The diffusiophoretic mobility, denoted by $\mu_{D}$, quantifies the velocity of the droplet per unit concentration gradient \citep{ganguly2024unified} and is defined as $\mu_{D}=U_{D}/{\alpha}$. Here, $U_{D}$ is the dimensionless diffusiophoretic velocity, scaled by a characteristic velocity scale $U_{0}$, which is not known a priori.  Under steady-state conditions, the diffusiophoretic velocity $U_{D}$ is determined by imposing the force-free condition, requiring that the net hydrodynamic and electrical forces on an arbitrary control volume enclosing the droplet vanish.
\section{Perturbation Analysis}\label{Perturbed}
If the imposed ionic concentrations at the two ends of the domain of length $2\tilde{R}$ are denoted by $n^{\infty,L}$ and $n^{\infty,R}$, then associated scaled concentration gradient is obtained by the parameter as $\alpha=(n^{\infty,R}-n^{\infty,L})a/(0.5(n^{\infty,R}+n^{\infty,L})2\tilde{R})<a/\tilde{R}$, as $|\tilde{\nabla}{n^{\infty}}|=(n^{\infty,R}-n^{\infty,L})/2\tilde{R}$ and $n^{\infty}=0.5(n^{\infty,R}+n^{\infty,L})$. Here, $\tilde{R}$ is the radius of the computational domain, which must be chosen sufficiently large compared to the droplet radius $a$ (i.e., $a/\tilde{R} \ll 1$) for any arbitrary values of $\kappa a$. Consequently, we have $\alpha<a/\tilde{R}\ll1$. This range of $\alpha$ is consistent with several experimental studies on diffusiophoresis, where the imposed dimensionless concentration gradient of electrolytes is typically varied in the range $\alpha=0.001~\text{to}~0.0065$ \citep{ebel1988diffusiophoresis,shin2016size}. Under such a small $\alpha$, the variables can be expressed as a linear perturbation from its equilibrium i.e., $\phi=\phi^{0}(r)+\delta\phi(r,\theta)$, $n_{i}=n^{0}_{i}(r)+\delta n_{i}(r,\theta)$, and $\mu_{i}=\mu^{0}_{i}+\delta\mu_{i}(r,\theta)$, where the quantities with the superscript $0$ refer to those at equilibrium i.e., at $\alpha=0$ and the perturbed variables with prefix $\delta$ are of $O(\alpha)$. The interfacial surfactant concentration $\Gamma$ can be expressed as $\Gamma=\Gamma^{0}+\delta\Gamma$ and consequently, the surface charge density is given by $\sigma=\sigma^{0}+\delta\sigma=\sigma^{0}+z_{s}Ma\delta\Gamma$, where $z_{s}$ is the valence of the ionic surfactant species and $Ma$ is the Marangoni number based on $\Gamma^{0}$. In the following subsection, we derive the governing equations for both the equilibrium state and the perturbed regime.
\subsection{Equilibrium state}\label{equilibrium}
The equilibrium state refers to the situation before applying any external ionic concentration gradient i.e., $\alpha=0$. At this situation, the ion velocities as well the fluid velocity vanishes, which implies that the electrochemical potential gradient must be zero. This gives 
\begin{equation}\label{eq:sm2A}
	z_{i}n^{0}_{i}(r)\psi^{0}_{r}(r)+n^{0}_{i,r}(r)=0.
\end{equation}
Here and throughout the paper, the subscripts $r$ and $rr$ denote the first and second derivatives with respect to the radial coordinate $r$, respectively. Upon integration to the far field, the above eq.(\ref{eq:sm2A}) gives 
\begin{equation}\label{eq:sm2}
	n^{0}_{i}(r)=(n^{\infty}_{i}/I^{\infty})\exp\left[-z_{i}\psi^{0}(r)\right].
\end{equation}
Eq.(\ref{eq:sm2}) is obtained with the condition that the equilibrium potential vanishes at the far field. Applying eq.({\ref{eq:sm2}}) into the Poisson equation ({\ref{MPE:1}}), the equilibrium potential outside the droplet is governed by 
\begin{equation}\label{eq:sm3}
\psi^{0}_{rr}(r) + (2/r)\psi^{0}_{r}(r)= -(1/2)(\kappa a)^2 \textstyle\sum_{i} z_{i} n^{0}_{i}(r)
\end{equation}
with the boundary conditions
\begin{equation}\label{eq:sm4}
\psi^{0}_{r}(1)=-z_{s}Ma\Gamma^{0}, \quad \psi^{0}(r\to \infty)\to 0
\end{equation}
Note that at equilibrium the adsorption and desorption rates are equal, yielding $C_{a}n^{0}_{N,s}(1-\Gamma^{0})=C_{d}\Gamma^{0}$. Here, $n^{0}_{N,s}$ can be found from eq.(\ref{eq:sm2}) as $n^{0}_{N,s}=(n^{\infty}_{N}/I^{\infty})e^{-z_{N}\zeta}$, in which $\zeta=\psi^{0}_{s}=\psi^{0}(r=1)$ is the equilibrium interfacial potential. Thus, the adsorption rate coefficient $k_{a}$ can be determined for a prescribed desorption rate coefficient $k_{d}$ as
\begin{equation}\label{eq_k_a}
k_{a}=k_{d}\Gamma^{0}/[n^{\infty}_{N}(1-\Gamma^{0})e^{-z_{N}\zeta}].
\end{equation} 
\subsection{Linear order analysis}
Upon substituting the linear perturbations described above into the governing equations presented in Section~\ref{Math_model} and separating the perturbed components, we obtain the following set of equations governing the perturbations.
\begin{subequations}\label{eq:pe}
	\begin{gather}
		\label{eq:pe2}
		\nabla^{2}\delta\psi(r,\theta) = -(1/2)(\kappa a)^2\textstyle\sum_{i}z_{i}\delta{n_{i}}(r,\theta),\\
		\label{eq:pe3}
		n^{0}_{i}(r)\nabla^{2}\delta\mu_{i}(r,\theta) 
			+\nabla n^{0}_{i}(r)\cdot \nabla\delta\mu_{i}(r,\theta)= Pe_{i}\delta\textbf{u}(r,\theta)\cdot\nabla n^{0}_{i}(r),\\
		\label{eq:pe4}
		\nabla \times \left[\nabla \times\nabla \times\delta\overline{\textbf{u}}(r,\theta)\right]=0,\\
		\label{eq:pe5}
		\nabla\times \{\nabla\times\nabla\times\delta\textbf{u}(r,\theta)\}=(1/2)(\kappa a)^2 \textstyle\sum_{i}{\nabla\delta\mu_{i}(r,\theta)\times\nabla n^{0}_{i}(r)}.
	\end{gather}
\end{subequations}
Exploiting the axial symmetry imposed by the externally applied concentration gradient, the first-order perturbation variables can be expressed as separable functions of the radial coordinate multiplied by prescribed angular variations \citep{ohshima1983approximate,prieve1987diffusiophoresis}. Accordingly, we write
\begin{subequations}\label{eq:sm5}
	\begin{gather}
		\label{eq:sm5a}
		\delta\psi(r,\theta)=-\Psi(r)\alpha\cos\theta,\\
		\label{eq:sm5c}
		\delta\mu_{i}(r,\theta)=-z_{i}\Phi_{i}(r)\alpha\cos\theta,\\
		\label{eq:sm5d}
		\delta\overline{\textbf{u}}(r,\theta)=\left[-(2/r)\overline{h}(r)\alpha \cos\theta \right] \boldsymbol{e}_r+\left[(\overline{h}_{r}(r)+(\overline{h}(r)/r))\alpha\sin\theta \right]\boldsymbol{e}_\theta,\\
		\label{eq:sm5e}
		\delta\textbf{u}(r,\theta)=\left[-(2/r)h(r)\alpha \cos\theta\right]\boldsymbol{e}_r+\left[(h_{r}(r)+(h(r)/r))\alpha \sin\theta\right]\boldsymbol{e}_\theta.
	\end{gather}
\end{subequations}
\par 
The corresponding perturbation of the interfacial surfactant concentration is written as
\begin{equation}\label{eq:pse}
	\delta\Gamma(\theta)=\Delta_s\alpha\cos\theta.
\end{equation}
Substituting eq.(\ref{eq:sm5a}) in eq.(\ref{eq:pe2}), the $O(\alpha)$ equation of perturbed electric potential outside the droplet is obtained as
\begin{equation}\label{eq:sm6}
\mathcal{L}\Psi(r)=(1/2)(\kappa a)^2\textstyle\sum_{i}z^{2}_{i}n^{0}_{i}\left[\Psi(r)-\Phi_{i}(r)\right].
\end{equation}
The operator $\mathcal{L}$ is defined as $\mathcal{L}=h_{rr}+(2/r)h_{r}-2/r^{2}$, where the subscript indicates a derivative with respect to the radial coordinate. Note that $\delta n_{i}(r,\theta)$ present on the RHS of eq.(\ref{eq:pe2}) is obtained using eqs.(\ref{eq:sm5a},\ref{eq:sm5c}) from the expression $\delta n_{i}(r,\theta)=n^{0}_{i}(r)[\delta\mu_{i}(r,\theta)-z_{i}\delta\psi(r,\theta)]$. The boundary conditions corresponding to the second order ODE (\ref{eq:sm6}) are 
\begin{equation}\label{eq:sm7}
\Psi_{r}|_{r=1}=z_{s}Ma\Delta_{s},~~~~~~~~~\Psi(r)=\beta r ~~~as~~~~ r \to \infty.	
\end{equation}
 Applying perturbation approximation on the eq.(\ref{eq:sef}), we get 
\begin{equation}
	Pe_{s}\Gamma^{0}\partial_{\theta}u_{s}-\partial_\theta\left[({1-\Gamma^{0}})^{-1}\partial_{\theta}\delta\Gamma\right]-z_{s}\Gamma^{0}\partial^{2}_\theta\delta\psi_{s}=C_{a}(1-\Gamma^{0})\delta n_{N,s}-(C_{a}n^{0}_{N,s}+C_{d})\delta\Gamma
\end{equation}
Substituting the expression of $u_{s}$ and $\delta\psi_{s}$ from equation (\ref{eq:sm5}) and $\delta\Gamma$ from eq.(\ref{eq:pse}) and doing algebraic simplification, $\Delta_{s}$ can be obtained as 
\begin{equation}\label{eq:pscd}
	\Delta_{s}=\Gamma_{C}\Gamma^{0}\left[z_{s}\Psi(1)-Pe_{s}h_{r}|_ {r=1}\right]+\Gamma_{C}\left[C_{a}(1-\Gamma^{0})z_{N}n^{0}_{N,s}(\Psi(1)-\Phi_{N}(1))\right].
\end{equation} 
where $\Gamma_{C}$ is defined as $\Gamma_{C}=(1-\Gamma^{0})/[1+(C_{a}n^{0}_{N,s}+C_{d})(1-\Gamma^{0})]$. Consequently, the surfactant concentration at the interface of the droplet can be expressed as 
\begin{equation}\label{eq:SEM1}
\begin{split}
\Gamma=&\Gamma^{0}+\big\{\Gamma_{C}\Gamma^{0}\left[z_{s}\Psi(1)-Pe_{s}h_{r}|_ {r=1}\right]+\\
& \Gamma_{C}\left[C_{a}(1-\Gamma^{0})z_{N}n^{0}_{N,s}(\Psi(1)-\Phi_{N}(1))\right]\big\}\alpha\cos\theta.
\end{split}
\end{equation} 
Upon inserting the perturbation expansions (\ref{eq:sm5}) into the linearized Nernst–Planck equation (\ref{eq:pe3}), which governs the $O(\alpha)$ response, one obtains a second-order boundary-value problem given by
\begin{equation}\label{eq:sm6c}
	\mathcal{L}\Phi_{i}+(n^{0}_{i,r}/n^{0}_{i})\Phi_{i,r}=(2Pe_{i}/z_{i})(n^{0}_{i,r}/n^{0}_{i})(h/r)
\end{equation}
corresponds to the boundary conditions 
\begin{subequations}\label{eq:sm7c}
	\begin{gather}
		\label{eq:sm7c:a}
		\Phi_{i,r}|_{r=1} 
		= \delta_{iN}\left[(I_{d}+I_{a}n^{0}_{N})(z_{N}n^{0}_{N,s})^{-1}\,\Delta_{s}(1)
		- I_{a}(1-\Gamma^{0})(\Psi(1)-\Phi_{N}(1))\right],\\
		\label{eq:sm7c:b}
		\Phi_{i}(r \to \infty) = \left[-(1/z_{i})+\beta\right] r
	\end{gather}
\end{subequations}
Substituting eqs.({\ref{eq:sm5}) into the eqs.(\ref{eq:pe4},\ref{eq:pe5}), we get 
	\begin{subequations}\label{eq:sm_h}
		\begin{gather}
			\label{eq:sm_h1}
			\mathcal{L}(\mathcal{L}\overline{h})=0,\\
			\label{eq:sm_h2}
			\mathcal{L}(\mathcal{L}h)=(1/2r)(\kappa a)^2\textstyle\sum_{i}z_{i}n^{0}_{i,r}\Phi_{i}
		\end{gather}
	\end{subequations}
	subject to the boundary conditions 
	\begin{subequations}\label{eq:sm_hb}
		\begin{gather}
			\label{eq:sm_hba}
			\overline{h}(1^{-})=0,~~~~~~~~~\overline{h}_{r}(1^{-})=h_{r}(1^{+}),\\
			\label{eq:sm_hbb}
			h(1^{+})=0, ~~~~~~~~~h_{rr}(r=1^{+})-\mu_{r}\overline{h}_{rr}(r=1^{-})=z_{s}Ma\Gamma^{0} \Psi(1)-Ma(1-\Gamma^{0})^{-1}\Delta_{s},\\
			h(r)\to (\mu_{D}/2)r+O(r^{-1})  ~~~as ~~~~r \to \infty.
		\end{gather}
	\end{subequations}
	Solution of eq.(\ref{eq:sm_h1}) satisfying eqs.(\ref{eq:sm_hba}) can be obtained as \citep{ohshima1984electrokinetic} 
	\begin{equation}\label{eq:sm82}
		\overline{h}(r)=-(1/2)h_{r}(1^{+})[r-r^3].
	\end{equation}
	Using eqs.(\ref{eq:pscd},\ref{eq:sm82}), we can construct the second boundary condition of eq.(\ref{eq:sm_hbb}) as
	\begin{equation}\label{eq:sm9}
		h_{rr}(r=1^{+})-3\mu_{r}h_{r}(r=1^{+})=z_{s}Ma\Gamma^{0} \Psi(1)-Ma(1-\Gamma^{0})^{-1}\Delta_{s}.
	\end{equation}
	We reduce the forth order ODE (\ref{eq:sm_h2}) into two second-order ODE as 
	\begin{subequations}\label{eq:sm10}
		\begin{gather}
			\label{eq:sm10a}
			\mathcal{L}(h)=h_{1},\\
			\label{eq:sm10b}
			\mathcal{L}(h_{1})=(1/2r)(\kappa a)^2\textstyle\sum_{i}z_{i}n^{0}_{i,r}\Phi_{i}.
		\end{gather}
	\end{subequations}
	and the corresponding boundary conditions become 
	\begin{subequations}\label{eq:sm11}
		\begin{gather}
			\label{eq:sm11a}
			h(1)=0, ~~~~~~~~~~h_{rr}({r \to \infty})=0,\\
			\label{eq:sm11b}
			h_{1}(1)=(3\mu_{r}+2)h_{r}(r=1^{+})+z_{s}Ma\Gamma^{0} \Psi(1)-Ma(1-\Gamma^{0})^{-1}\Delta_{s},~~~~~h_{1}({r \to \infty})=0.
		\end{gather}
	\end{subequations}
Using the equilibrium electric potential, $\psi^{0}(r)$ and ionic concentrations, $n^{0}_{i}(r)$ obtained in previous subsection-\ref{equilibrium}, the coupled perturbation equations (\ref{eq:sm6},\ref{eq:sm6c}), and (\ref{eq:sm10}) are solved simultaneously together with their associated boundary conditions using a finite-difference discretization. A detailed description of the numerical methodology is provided in \citet{majhi2025electrophoresis}. Based on these solutions, the diffusiophoretic mobility is obtained as \\
	\begin{equation}\label{eq:sm12}
		\mu_{D}=\lim_{r\to\infty}2h(r)/{r}.
	\end{equation} 
To elucidate the physical mechanisms governing the droplet motion, it is instructive to evaluate the interfacial stresses. The tangential Maxwell stress $\Sigma_{E}$ and the Marangoni stress $\Sigma_{M}$ acting at the droplet interface are given by
\begin{subequations}\label{eq:Stress}
	\begin{gather}
		\label{eq:Stress_1}
		\Sigma_{E}=-\sigma{\partial_{\theta}}\psi=-z_{s}Ma\Gamma^{0}\Psi(1)\alpha\sin\theta,\\
		\label{eq:Stress_2}
		\Sigma_{M}=-Ma(1-\Gamma)^{-1}{\partial_\theta}\Gamma=Ma(1-\Gamma^{0})^{-1}\Delta_{s}\alpha\sin\theta.
	\end{gather}
\end{subequations}
\section{Analytical solution for the case of insoluble surfactant}\label{analytic}
In this section, we derive the analytical solution under the assumption of an insoluble surfactant, where desorption is negligible during the flow, i.e., $k_{d}=0$. Consequently, we get $I_{a},I_{d}, C_{a}, C_{d}=0$. Under this assumption ($k_{d}=0$), the solution of the eq.({\ref{eq:sm_h2}}) can be obtained as 
\begin{equation}\label{eq:sd2}
	\begin{split}
		h(r)=& \frac{1}{9}\left[\left(1+s^{-1}\right)r -\frac{3}{2}+{(1-2s^{-1})}\frac{1}{2r^2}\right]\times{\int_{1}^{\infty}x^{3}G(x)\,dx}\\
		& -\left[\frac{r^3}{30}-\left(1-2s^{-1}\right)\frac{r}{18}+\left(1-5s^{-1}\right)\frac{1}{45r^{2}}\right]{\int_{1}^{\infty}G(x)\,dx}\\
		& -{\int_{1}^{r}\left[\frac{-r^3}{30}+\frac{rx^2}{6}-\frac{x^3}{6}+\frac{x^5}{30r^2}\right]G(x)\,dx}
	\end{split}
\end{equation}
where 
\begin{equation}\label{eq:sd3}
	G(r)=(1/2r)(\kappa a)^2\textstyle\sum_{i}z_{i}n^{0}_{i,r}\Phi_{i},
\end{equation}
Here, $n^{0}_{i}(r)=(n^{\infty}_{i}/I^{\infty})\exp\left[-z_{i}\psi^{0}(r)\right]$ in which $\psi^{0}$ can be obtained by solving the following nonlinear equation (\ref{eq:sm3}) and $\Phi_{i}$ can be obtained by solving the equation (\ref{eq:sm6c}) with the boundary conditions (\ref{eq:sm7c}) as 
\begin{equation}\label{eq:ddd3}
	\begin{split}
		\Phi_{i}(r)=& (-\frac{1}{z_{i}}+\beta)(r+\frac{1}{2r^2})-\frac{1}{3}\left(r+\frac{1}{2r^2}\right) {\int_{1}^{\infty} \frac{d\phi^{0}}{dx}\left(z_{i}\frac{d\Phi_{i}}{dx}-2 Pe_{i}\frac{h}{x}\right) \,dx }\\
		& +{\frac{1}{3}} {\int_{1}^{r}\left(r-\frac{x^3}{r^2} \right)\frac{d\phi^{0}}{dx}\left(z_{i}\frac{d\Phi_{i}}{dx}-2 Pe_{i}\frac{h}{x}\right) \,dx}
	\end{split}           
\end{equation}
In eq.(\ref{eq:sd2}), the parameter $s$ is defined as $s=3\mu_{r}+Ma\Gamma^{0}Pe_{s}+2$. Further, substituting $h(r)$ into the eq.(\ref{eq:sm12}), the elctrophoretic mobility can be expressed as 
\begin{equation}\label{eq:iem}
	\mu_{D}= \frac{1}{9}{\int_{1}^{\infty}\left[1-2s^{-1}-3r^{2}+2\left(1+s^{-1}\right)r^{3} \right]G(r)\,dr}
\end{equation}
The above eq.(\ref{eq:iem}) is the integral form for the mobility $\mu_{D}$, valid under arbitrary $\kappa a$ and equilibrium surface potential. This equation is further used to derive the analytical solutions under the Debye-H{\"u}ckel (D-H) approximation as well as thin layer consideration.
\subsection{Mobility under Debye-H{\"u}ckel approximation}
Debye-H{\"u}ckel approximation refers to the situation where the interface electric potential is less than the thermal potential, which implies $\psi<1$. In such a situation, higher order of $\psi$ can be neglected. In this subsection, we derive the explicit mobility expression of a liquid droplet immersed in monovalent electrolyte solution under D-H approximation without any restriction on the bulk ionic concentration. 
Linearizing the potential equation (\ref{eq:sm3}) under the D-H limit, we get
\begin{equation}\label{c6eq:AE2}
	\psi^{0}_{rr}(r)+(2/r)\psi^{0}_{r}(r)+(\kappa a)^{2}\psi^{0}=0.
\end{equation}
The solution of the above equation subject to the boundary condition for electrostatic potential discussed earlier is given by 
\begin{equation}\label{c6eq:AE4}
	\psi^{0}(r)=\frac{z_{s}Ma\Gamma^{0}}{\kappa a+1}\frac{e^{-\kappa a(r-1)}}{r}.
\end{equation}
At the interface of the droplet $\psi^{0}$ is obtained as $z_{s}Ma\Gamma^{0}(\kappa a+1)^{-1}$, which is termed the equilibrium surface potential or $\zeta$-potential i.e., $\zeta=z_{s}Ma\Gamma^{0}(\kappa a+1)^{-1}$.\\
Under the D-H limit, $\Phi_{\pm}(r)$ can be obtained from the eq.(\ref{eq:ddd3}) in the following explicit form 
\begin{equation}\label{c6eq:AE5}
	\begin{split}
		\Phi_{\pm}(r)=& (\mp 1+\beta)\left[\left(r+\frac{1}{2r^{2}}\right)\bigg\{1 \mp \frac{1}{3}\int_{1}^{\infty}{\frac{d\psi^{0}}{dx}(1-\frac{1}{x^{3}})dx}\bigg\}\right. \\
		& \left.\pm \frac{1}{3}\int_{1}^{r}{\frac{d\psi^{0}}{dx}(1-\frac{1}{x^{3}})(r-\frac{x^{3}}{r^{2}})dx}\right].
	\end{split}
\end{equation}
In order to obtain the explicit expression for diffusiophoretic mobility of the droplet, we first need to find the term $G(r)$ given in eq.(\ref{eq:sd3}). Substituting eq.(\ref{c6eq:AE5}) in the eq.({\ref{eq:sd3}}) and performing the appropriate approximation under the D-H limit, $G(r)$ can be obtained as
\begin{equation}
	\begin{split}
		G(r)=& -(\kappa a)^{2}\frac{d\psi^{0}}{dr}\left[(1+\frac{1}{2r^{3}})\bigg \{\beta+\left(\psi^{0}+\frac{1}{3}\int_{1}^{\infty}{\frac{d\psi^{0}}{dr}(1-\frac{1}{r^{3}})dr}\right)\bigg\}\right.\\
		& \left. -\frac{1}{3}\int_{1}^{r}{\frac{d\psi^{0}}{dx}(1-\frac{x^{3}}{r^{3}})(1-\frac{1}{x^{3}})dx}\right]
	\end{split}
\end{equation}
which further reduces to 
\begin{equation}
	\begin{split}
		G(r)=& -(\kappa a)^{2}\frac{d\psi^{0}}{dr}\left[(1+\frac{1}{2r^{3}})\beta+\bigg\{(1+\frac{1}{2r^{3}})\psi^{0}-\frac{z_{s}Ma\Gamma^{0}}{\kappa a+1}\right.\\
		& \left.\left(\frac{1}{\kappa a r^{3}}+\frac{1}{(\kappa a)^{2}r^{3}}-e^{-\kappa a(r-1)}\left(\frac{1}{\kappa a r^{3}}+\frac{1}{(\kappa a)^{2}r^{3}}\right)\right.\right.\\
		& \left. \left. +\frac{1}{2r^{3}}e^{\kappa a}E_{5}(\kappa a)+\frac{1}{r^{4}}e^{\kappa a}E_{5}(\kappa a r)\right)\bigg\}\right]
	\end{split}
\end{equation}
Here, $E_{n}(x)$ is the exponential integral of order $n$ and is defined as $E_{n}(x)=x^{n-1}\int_{x}^{\infty}(e^{-t}/t^{n})dt$. Substituting the expression of $G(r)$ in eq.(\ref{eq:iem}) and performing the algebraic calculations, one can find explicit expression for the mobility $\mu_{D}$ as 
\begin{equation}\label{eq:mobex_1}
		\mu_{D}=\underbrace{\beta\frac{z_{s}Ma\Gamma^{0}}{\kappa a+1}\Theta_{1}(\kappa a)}_{Electrophoresis}+\underbrace{\frac{z^{2}_{s}Ma^{2}{\Gamma^{0}}^{2}}{8(\kappa a+1)^{2}}\Theta_{2}(\kappa a)}_{Chemiphoresis}
\end{equation}
where the functions $\Theta_{1}(\kappa a)$ and $\Theta_{2}(\kappa a)$ are respectively given by
\begin{subequations}\label{eq:mobex_2}
	\begin{align}
		\label{eq:mobex_2_1}
		\Theta_{1}(\kappa a)=&1-s^{-1}+s^{-1}\kappa a+2e^{\kappa a}E_{5}(\kappa a)-5(1-2s^{-1})e^{\kappa a}E_{7}(\kappa a),\\
		\label{eq:mobex_2_2}
		\begin{split}
			\Theta_{2}(\kappa a)=&1-2s^{-1}+2s^{-1}\kappa a-(8/3)e^{\kappa a}E_{3}(\kappa a)+8e^{\kappa a}E_{4}(\kappa a)\\
			& +(8/3)(1-2s^{-1})e^{\kappa a}E_{5}(\kappa a)-(40/3)(1-2s^{-1})e^{\kappa a}E_{6}(\kappa a)\\
			& -8\Theta_{1}(\kappa a)e^{\kappa a}E_{5}(\kappa a)+(10/3)e^{2\kappa a}E_{6}(2\kappa a)+(7/3)(1-2s^{-1})e^{2\kappa a}E_{8}(2\kappa a).
		\end{split}
	\end{align}
\end{subequations}
Expression (\ref{eq:mobex_1}) is one of the key findings of the present study. It is useful for experimentalists for the correct evaluation of intrinsic hydrodynamic and electrostatic properties of surfactant-laden droplets in the context of diffusiophoresis. We can separate the electrophoresis and chemiphoresis parts in the mobility expression (\ref{eq:mobex_1}). The terms multiplied by $\beta$ as indicated in eq.(\ref{eq:mobex_1}) correspond to the electrophoretic contribution, denoted by $\mu_{E}$ and the remaining terms independent $\beta$ corresponds to the chemiphoretic contribution, which is denoted by $\mu_{C}$ i.e., $\mu_{D}=\mu_{E}+\mu_{C}$.
\par 
Under the H{\"u}ckel limit situation i.e., when $\kappa a \to 0$, the mobility expression (\ref{eq:mobex_1}) reduces to 
\begin{equation}\label{eq:dhr_1}
\mu_{D}=\frac{2}{3}z_{s}Ma\Gamma^{0}(1-s^{-1})\beta
\end{equation}
The above expression clearly shows that at lower range of $\kappa a$, $\mu_{D}$ is driven by the electrophoresis part and chemiphoresis has no impact on $\mu_{D}$.\\
If we consider the situation where the Debye layer is extremely thin i.e., $\kappa a \to \infty$, the expression for $\mu_{D}$ reduces the following simplified form 
\begin{equation}\label{eq:somul}
	\mu_{D}=\beta\zeta(1-2s^{-1}+s^{-1}\kappa a)+{{\zeta}^{2}}/{8}(1-2s^{-1}+2s^{-1}\kappa a)
\end{equation}
where $s=3\mu_{r}+Ma\Gamma^{0}Pe_{s}+2$. This expression cloud be valuable for mapping the measured droplet mobility with the corresponding $\zeta$-potential. For a rigid particle, $\mu_{r}\to\infty$, thus $s^{-1}\to 0$ and consequently, the mobility expression (\ref{eq:somul}) reduces to 
\begin{equation}\label{eq:somul_2}
	\mu_{D}=\beta\zeta+{{\zeta}^{2}}/{8},
\end{equation}
which is the same mobility expression for rigid particle obtained from the analytical expression of \citet{prieve1984motion} for low $\zeta$-potential. Note that $s$ characterizes the effective viscosity ratio and by virtue of eq.(\ref{eq:somul}), mobility is inversely proportional to $s$.
\par 
The interface velocity $u_{s}(\theta)$ can be determined by $u_{s}(\theta)=h_{r}(r)|_{r=1}\alpha\sin\theta$, which can be expressed as 
\begin{equation}\label{c6eq:slip_v2}
	u_{s}(\theta)=s^{-1}\int_{1}^{\infty}\frac{r^{3}-1}{3}G(r)\alpha\sin\theta\,dr 
\end{equation} 
Note that $u_{s}(\theta)$ becomes zero when $\mu_{r}\to \infty$, which corresponds to the rigid particle. Moreover, $s^{-1}$ is maximum when $\mu_{r}\to0$ for the other parameters fixed, which implies maximum interfacial velocity for baubles ($\mu_{r}\to0$). The interface velocity for any arbitrary surface charge density and bulk molar concentration can be obtained by numerical integration of eq.(\ref{c6eq:slip_v2}). Based on the D-H approximation an analytical expression for $u_{s}(\theta)$ can be obtained as 
\begin{equation}\label{c6eq:slip_v3}
	u_{s}(\theta)=s^{-1}\left[\frac{z_{s}Ma\Gamma^{0}}{\kappa a+1}\beta S_{1}(\kappa a)+\frac{z^{2}_{s}Ma^{2}{\Gamma^{0}}^{2}}{8(\kappa a+1)^{2}}S_{2}(\kappa a)\right]\alpha\sin\theta
\end{equation}
where $S_{1}(\kappa a)$ and $S_{2}(\kappa a)$ are given by 
\begin{subequations}\label{c6eq:slip_v4}
	\begin{align}
		\label{c6eq:slip_v4_1}
		S_{1}(\kappa a)=& 1+\kappa a+(1/2){(\kappa a)^{2}}e^{\kappa a}E_{5}(\kappa a),\\
		\label{c6eq:slip_v4_2}
		\begin{split}
			S_{2}(\kappa a)=&5+3\kappa a-16e^{\kappa a}E_{5}(\kappa a)-16 \kappa ae^{\kappa a}E_{5}(\kappa a)\\
			& -4(\kappa a)^{2}e^{2\kappa a}E^{2}_{5}(\kappa a)-7e^{2\kappa a}E_{8}(2\kappa a).
		\end{split}
	\end{align}
\end{subequations}
\subsubsection{Mobility under an uniform surfactant distribution}
When the surfactant distribution is uniform along the droplet interface, the surface charge density remains constant and no Marangoni stress arises, i.e., $\Delta_{s}=0$. Under this condition, the stress balance equation (\ref{eq:sm9}) simplifies to 
\begin{equation}\label{eq:sm9_uniform}
	h_{rr}(r=1^{+})-3\mu_{r}h_{r}(r=1^{+})=z_{s}Ma\Gamma^{0} \Psi(1).
\end{equation}
With this boundary condition, the mobility expression (\ref{eq:iem}) reduces to the form 
 \begin{equation}\label{eq:iem_uniform}
 	\mu_{D}= \frac{1}{9}{\int_{1}^{\infty}\left[1-2s^{-1}-3r^{2}+2\left(1+s^{-1}\right)r^{3} \right]G(r)\,dr}-\frac{2}{3}z_{s}Ma\Gamma^{0}s^{-1}\Psi(1),
 \end{equation}
 and $s$ reduces to $s=3\mu_{r}+2$. Under the Debye-H{\"u}ckel approximation, solving eq.(\ref{eq:sm6}) for the case of $\Delta_{s}=0$ yields the interfacial perturbed potential $\Psi(1)$ as 
 \begin{equation}\label{eq: per_elec_sol}
 \Psi(1)=\frac{3}{2}\beta+\frac{z_{s}Ma\Gamma^{0}}{K_{1}(\kappa a+1)^{2}}\left[\frac{21}{4}-\frac{3}{4}\kappa a+\frac{3}{4}(\kappa a)^{2}-3e^{\kappa a}E_{5}(\kappa a)+15e^{\kappa a}E_{6}(\kappa a)-45 e^{\kappa a}E_{7}(\kappa a)\right],
 \end{equation} 
 where $K_{1}=(\kappa a+1)+(\kappa a+1)^{-1}$. Substituting in eq.(\ref{eq:iem_uniform}), the solution (\ref{eq:mobex_1}) for the mobility $\mu_{D}$ can be reduces to 
 \begin{equation}\label{eq:mobex_1_uniform}
 \begin{split}
 \mu_{D}=&\beta\frac{z_{s}Ma\Gamma^{0}}{\kappa a+1}\Theta^{U}_{1}(\kappa a)+\frac{z^{2}_{s}Ma^{2}{\Gamma^{0}}^{2}}{8(\kappa a+1)^{2}}\Theta^{U}_{2}(\kappa a)-\frac{2z_{s}Ma\Gamma^{0}}{3(3\mu_{r}+2)}\left[\frac{3}{2}\beta+\frac{z_{s}Ma\Gamma^{0}}{K_{1}(\kappa a+1)^{2}}\right.\\
 & \left.\bigg\{\frac{21}{4}-\frac{3}{4}\kappa a+\frac{3}{4}(\kappa a)^{2}-3e^{\kappa a}E_{5}(\kappa a)+15e^{\kappa a}E_{6}(\kappa a)-45 e^{\kappa a}E_{7}(\kappa a)\bigg\}\right]		
 \end{split}
 \end{equation}
 where $\Theta^{U}_{1}(\kappa a)$ and $\Theta^{U}_{2}(\kappa a)$ are obtained from eq.(\ref{eq:mobex_2}) by substituting the reduced form of $s=3\mu_{r}+2$. Substituting $z_sMa\Gamma^{0}=\sigma$~(as $\delta\sigma=0$), the expression (\ref{eq:mobex_1_uniform}) can be recast as
 \begin{equation}\label{eq:mobex_1_uniform}
 	\begin{split}
 		\mu_{D}=&\frac{\beta\sigma}{\kappa a+1}\Theta^{U}_{1}(\kappa a)+\frac{\sigma^{2}}{8(\kappa a+1)^{2}}\Theta^{U}_{2}(\kappa a)-\frac{2\sigma}{3(3\mu_{r}+2)}\left[\frac{3}{2}\beta+\frac{\sigma}{K_{1}(\kappa a+1)^{2}}\right.\\
 		& \left.\bigg\{\frac{21}{4}-\frac{3}{4}\kappa a+\frac{3}{4}(\kappa a)^{2}-3e^{\kappa a}E_{5}(\kappa a)+15e^{\kappa a}E_{6}(\kappa a)-45 e^{\kappa a}E_{7}(\kappa a)\bigg\}\right]		
 	\end{split}
 \end{equation}
 which is identical with the expression derived by \citet{samanta2023diffusiophoresis} for a non-polarizable droplet with constant surface charge density and, hence no Marangoni stress under the Debye-H{\"u}ckel approximation.
\subsection{Mobility under large $\kappa a$ consideration}
When the electric double layer is thin compared to the droplet radius, i.e. $\kappa a \gg 1$, the general mobility expression (\ref{eq:iem}) admits a systematic expansion in powers of $(\kappa a)^{-1}$ This expression yields an asymptotic expression for the diffusiophoretic mobility of a charged irreversibly bound surfactant-laden droplet that remains valid for arbitrary equilibrium surface potential. At a larger $\kappa a \gg 1$, the particle interface can be treated as planar surface, in that case eq.(\ref{eq:sm3}) reduces to 
\begin{equation}\label{eq:tla3}
	\psi^{0}_{rr}(r)=-(1/2I^{\infty})(\kappa a)^{2}\textstyle\sum_{i}{z_{i}n^{\infty}_{i}\exp\left[-z_{i}\psi^{0}(r)\right]}
\end{equation}
Integrating the equation over $r$ to $\infty$ after multiplying by $d\psi^{(0)}/dr$, we obtain 
\begin{equation}\label{eq:tla3}
\psi^{0}_{r}(r)=-sgn(\zeta)({\kappa a}/\sqrt{I^{\infty}})\sqrt{\textstyle\sum_{i}{n^{\infty}_{i}\{\exp\left[-z_{i}\psi^{0}(r)\right]-1\}}}
\end{equation}
where $sgn(\zeta)=1$ for $\zeta>0$ and $-1$ for $\zeta<0$, with $\zeta=\psi^{0}(r)|_{r=1}$ denoting the equilibrium surface potential.
\par 
In the limit $\kappa a \gg 1$, electrostatic and concentration variations are confined to a narrow interfacial layer of thickness $O(\kappa^{-1})$. As a result, the dominant contribution to the mobility arises from this near-surface region, where $r-1 \sim O((\kappa a)^{-1})$. Within this layer, the perturbation fields vary weakly in the radial direction, and the solution of (\ref{eq:ddd3}) may be approximated as $\Phi_i(r) = \Phi_i(1) + O((\kappa a)^{-1})$, where $\Phi_i(1)$ can be evaluated as \citet{ohshima1983approximate}
\begin{equation}\label{eq:tla6}
	\Phi_{i}(1)=\frac{3}{2}\left(-\frac{1}{z_{i}}+\beta \right)+{\int_{1}^{\infty}\left(1-\exp[-z_{i}\psi^{0}(r)]\right) \left(\Phi_{i}(1)-\frac{Pe_{i}}{z_{i}}\frac{d}{dr}\left[\frac{h(r)}{r}\right]\right) \,dr}+\mathcal{O}({\kappa a}^{-1})
\end{equation}
To obtain an asymptotic expression for $\frac{d}{dr}\left[\frac{h(r)}{r}\right]$ in the thin–double-layer limit, we expand the integrand of (\ref{eq:sd2}) about $r=1$ and treat $r-1\sim \mathcal{O}({\kappa a}^{-1})$, which yields
\begin{equation}\label{eq:tla7}
	\begin{split}
		\frac{d}{dr}\left[\frac{h(r)}{r}\right]=&s^{-1}{\int_{1}^{\infty} (x-1)G(x)\,dx }+s^{-1}{\int_{1}^{\infty} (x-1)^{2}G(x)\,dx }\\
		&+(1-4s^{-1})(r-1){\int_{1}^{\infty} (x-1)G(x)\,dx }-\frac{1}{2}(r-1)^{2}{\int_{r}^{\infty}G(x)\,dx}\\
		& +\frac{1}{2}{\int_{1}^{r} (x-1)^{2}G(x)\,dx }-(r-1){\int_{1}^{r} (x-1)G(x)\,dx }
	\end{split}
\end{equation}
where,
\begin{equation}\label{eq:mis1}
	G(r)=(\kappa a)^{2}\frac{1}{1+(r-1)}F(r)=(\kappa a)^{2}[1-(r-1)+(r-1)^{2} \mp \cdots]F(r)
\end{equation}
with $F(r)$ is obtained as  
\begin{equation}\label{eq:F(r)}
	F(r)=-(1/2)\textstyle\sum_{i}{z^{2}_{i}n^{\infty}_{i}\exp[-z_{i}\psi^{0}(r)]\psi^{0}_{r}(r)\Phi_{i}(1)}.
\end{equation}
Following the same procedure, the integral expression (\ref{eq:iem}) for the diffusiophoretic mobility is expanded about $r=1$, yielding
\begin{equation}\label{eq:tla1}
	\mu_{D}= \frac{1}{3}{\int_{1}^{\infty}\left[(r-1)^{2}+2s^{-1}(r-1)\right](\kappa a)^{2}F(r)\,dr}+\mathcal{O}({\kappa a}^{-1}),
\end{equation}
Upon substituting the expression for $F(r)$ from (\ref{eq:F(r)}) into the asymptotic mobility expression (\ref{eq:tla1}), we obtain the diffusiophoretic mobility of a surfactant-laden dielectric droplet in a general electrolyte as
\begin{equation}\label{eq:tla8}
\mu_{D}=-\frac{1}{3}\sum_{j}\left(z_{j}n^{\infty}_{j}/I^{\infty}\right)\Phi_{j}(1)\left[\mathcal{I}_{2}(\zeta;z_{j})+{sgn(\zeta) s^{-1}\kappa a}\mathcal{I}_{1}(\zeta;z_{j})\right]
\end{equation}
with $\Phi_i(1)$ obtained from eq.(\ref{eq:tla6},\ref{eq:tla7}) as
\begin{equation}\label{eq:tla9}
	\begin{split}
		\Phi_{i}(1) = & -\frac{3}{2} \left(\frac{1}{z_{i}} - \beta\right) - \text{sgn}(\zeta) \frac{\Phi_{i}(1)}{\kappa a} \mathcal{I}_{1}(\zeta;z_{j}) \\
		& + \frac{Pe_{i}}{2z_{i}} \sum_{j} z_{j} \left({n^{\infty}_{j}}/I^{\infty}\right) \Phi_{j}(1) \left[\frac{\text{sgn}(\zeta)}{\kappa a} \mathcal{I}_{3}(\zeta;z_{i},z_{j}) - s^{-1} \mathcal{I}_{1}(\zeta;z_{i}) \mathcal{I}_{1}(\zeta;z_{j}) \right]
	\end{split}
\end{equation}
where the integrals $\mathcal{I}_{i}$ appear in eqs. (\ref{eq:tla8}) and (\ref{eq:tla9}) are given as:
\begin{subequations} \label{eq:I_all}
	\begin{align}
		\mathcal{I}_1(\zeta^0; z_i) &= \int_0^{\zeta^0} 
		{\mathcal{J}(\psi^{0};z_{i})}{\chi(\psi_0)}^{-1} \, d\psi^0, \label{eq:I1} \\
		\mathcal{I}_2(\zeta;z_{i}) &= {\int_{0}^{\zeta}\int_{\psi^{0}}^{\zeta} {\mathcal{J}(\psi^{0};z_{i})}{\chi(\psi^0)^{-1}\chi({\psi^0}')^{-1}}\,d{\psi^{0}}'\,d\psi^{0} }, \label{eq:I2}\\
		\mathcal{I}_3(\zeta; z_i, z_j) &= 
		\int_{0}^{\zeta}\int_{\zeta}^{\psi^{0}}\int_{0}^{{\psi^{0}}''}\frac{\mathcal{J}(\psi^{0};z_{i})\mathcal{J}({\psi^{0}}'';z_{j})}{\chi(\psi^{0})\chi({\psi^{0}}')\chi({\psi^{0}}'')}\,d{\psi^{0}}''\,d{\psi^{0}}'\,d\psi^{0}, \label{eq:I3}
	\end{align}
\end{subequations}
with $\mathcal{J}(\psi^{0};z_{i})=[\exp(-z_i \psi^0) - 1]$ and $\chi(\psi^{0})=\sqrt{\sum_{j}(n^{\infty}_{j}/I^{\infty})[\exp(-z_{j}\psi^{0})-1]}$. Eq.(\ref{eq:tla8}) thus provides the expression of the diffusiophoretic mobility for a dielectric droplet bearing adsorbed ionic surfactant in the thin–double-layer limit and it is applicable over the full range of equilibrium surface potentials.
\par 
Although the above result is valid for general electrolyte compositions, including mixtures of ions with arbitrary valence, we now specialize to the case of a symmetric monovalent electrolyte, representative of common systems such as NaCl, KCl, and HCl. For this class of electrolytes, the ionic valences satisfy $z_+=1$ and $z_-=-1$. Without loss of generality, we first consider the case $\zeta>0$, for which the general mobility expression simplifies to
\begin{equation}\label{eq:tla8_1}
	\mu_{D}=-\frac{1}{3}\left[\Phi_+(1)	\mathcal{I}_2(\zeta;1)-\Phi_-(1)	\mathcal{I}_2(\zeta;-1)\right]-\frac{s^{-1}\kappa a}{3}\left[\Phi_+(1)	\mathcal{I}_1(\zeta;1)-\Phi_-(1)\mathcal{I}_1(\zeta;-1)\right]
\end{equation}
Now, the integrals defined in eq.(\ref{eq:I_all}) are obtained for $\zeta>0$ as 
\begin{subequations} \label{eq:I_evaluated}
	\begin{align}
		\mathcal{I}_1(\zeta; \pm 1) &= 2 \left(e^{\mp \zeta/2} - 1\right), \label{eq:I1_evaluated} \\
		\mathcal{I}_2(\zeta; \pm 1) &= 4\ln\left(\frac{1+e^{{\mp}{\zeta/2}}}{2}\right), \label{eq:I2_evaluated} \\
		\mathcal{I}_3(\zeta; z_i, z_i)\big|_{z_i = \pm 1} &= 
		8 \left[ \left(1 - e^{\mp \zeta/2}\right) + 2 \ln \left( \frac{1+e^{\mp \zeta/2}}{2}\right)\right], \label{eq:I2_self} \\
		\mathcal{I}_3(\zeta; z_i, -z_i)\big|_{z_i = \pm 1} &= 
		16 \ln \left(\frac{e^{\zeta/4} + e^{-\zeta/4}}{2}\right). \label{eq:I2_cross}
	\end{align}
\end{subequations}
Several studies \cite{ohshima1983approximate,ohshima2024fundamentals} assume that $e^{|\zeta|}$ grows to the order of $\kappa a$ and consequently we can approximate:
\begin{equation}
	\frac{e^{|\zeta|/2} - 1}{\kappa a} \sim O((\kappa a)^0), \quad 
	\frac{e^{-|\zeta|/2} - 1}{\kappa a} \sim O((\kappa a)^{-1}). \label{eq:large_kappa_a_combined}
\end{equation}
 The evaluation of eq.(\ref{eq:tla9}) leads to:
\begin{align}
	\Phi_+(1) &= -\frac{3}{2}(1-\beta) - \frac{Pe_+}{2}
	\frac{\mathcal{I}_1(\zeta; 1) \left[\Phi_+ \mathcal{I}_1(\zeta; 1) - \Phi_- \mathcal{I}_1(\zeta; -1)\right]}{(\eta+ 2)}
	+ O((\kappa a)^{-1}), \label{eq:Phi_plus} \\
	\Phi_-(1) &= \frac{3}{2}(1+\beta) - D_g \Phi_- + \frac{{Pe}_-}{2}
	\frac{\mathcal{I}_1(\zeta; -1) \left[\Phi_+ \mathcal{I}_1(\zeta; 1) - \Phi_- \mathcal{I}_1(\zeta; -1)\right]}{(\eta+ 2)}
	+ O((\kappa a)^{-1}), \label{eq:Phi_minus}
\end{align}
where
\begin{equation}
	D_g = \frac{\mathcal{I}_1(\zeta; -1) - (1/2){Pe}_- \mathcal{I}_3(\zeta; -1, -1)}{\kappa a}
	= 2(1 + 2{Pe}_-)
	\frac{e^{\zeta/2} - 1}{\kappa a} + O((\kappa a)^{-1}). \label{eq:F1}
\end{equation}
Solving for $\Phi_\pm(1)$ yields:
\begin{align}
	\Phi_+(1) &= -\frac{3(1-\beta)s}{2(s + 3D_l)}
	+ \frac{3 \mathcal{I}_1(\zeta; -1) \left[(1+\beta){Pe}_+ \mathcal{I}_1(\zeta; 1) - (1-\beta)\text{Pe}_- \mathcal{I}_1(\zeta; -1)\right]}{4(1 + D_g)(s + 3D_l)}, \label{eq:Phi_plus_solution} \\
	\Phi_-(1) &= \frac{3(1+\beta)s}{2(1 + D_g)(s + 3D_l)}
	+ \frac{3 \mathcal{I}_1(\zeta; 1) \left[(1+\beta)Pe_+ \mathcal{I}_1(\zeta; 1)-(1-\beta) \text{Pe}_- \mathcal{I}_1(\zeta; -1)\right]}{4(1 + D_g)(s+ 3D_l)}, \label{eq:Phi_minus_solution}
\end{align}
where $D_l$ is obtained as
\begin{equation}
	D_l = \frac{2}{3} \left(
	{Pe}_+ \left[{\mathcal{I}_1(\zeta; 1)}/{2}\right]^2
	+ \frac{{Pe}_-}{1 + D_g} \left[{\mathcal{I}_1(\zeta; -1)}/{2}\right]^2
	\right), \label{eq:F2}
\end{equation}
Substituting the expressions for $\Phi_{j}(1)$, $\mathcal{I}_{1}(\zeta;z_{j})$ and $\mathcal{I}_{2}(\zeta;z_{j})$ into the above expression (\ref{eq:tla8_1}), the explicit mobility expression can be obtained as for $\zeta>0$
 \begin{equation}\label{eq:tla10_1}
	\begin{split}
		\mu_{D}=&\frac{\kappa a}{s+3D_l}\left[(1-\beta)\left(e^{-\frac{\zeta}{2}}-1\right)+\frac{1+\beta}{1+D_g}\left(e^{\frac{\zeta}{2}}-1\right)\right]\\
		&+\frac{4(1-\beta)Pe_{-}}{(1+D_{g})(s+3D_{l})}\left(e^{-\frac{\zeta}{2}}-1\right)^{2}\ln\left(\frac{1+e^{-\frac{\zeta}{2}}}{2}\right)\\
		& +\frac{2s}{s+3D_l}\left[(1-\beta)\ln\left(\frac{1+e^{-\frac{\zeta}{2}}}{2}\right)+\frac{1+\beta}{1+D_{g}}\ln\left(\frac{1+e^{\frac{\zeta}{2}}}{2}\right)\right]
	\end{split}
\end{equation}
Similarly, one can obtain the mobility expression for $\zeta<0$ as 
 \begin{equation}\label{eq:tla10_2}
	\begin{split}
		\mu_{D}=&-\frac{\kappa a}{s+3D_l}\left[\frac{1-\beta}{1+D_g}\left(1-e^{-\frac{\zeta}{2}}\right)+(1+\beta)\left(1-e^{\frac{\zeta}{2}}\right)\right]\\
		&+\frac{4(1+\beta)Pe_{+}}{(1+D_{g})(s+3D_{l})}\left(1-e^{-\frac{\zeta}{2}}\right)^{2}\ln\left(\frac{1+e^{\frac{\zeta}{2}}}{2}\right)\\
		& +\frac{2s}{s+3D_l}\left[\frac{1+\beta}{1+D_{g}}\ln\left(\frac{1+e^{-\frac{\zeta}{2}}}{2}\right)+({1-\beta})\ln\left(\frac{1+e^{\frac{\zeta}{2}}}{2}\right)\right]
	\end{split}
\end{equation} 
where $D_{g}$ and $D_{l}$ are given as 
	\begin{subequations}\label{eq:sup_coeff}
	\begin{gather}
		\label{eq:sup_coeff_1}
		D_{g}=2(1+2Pe_{+})\frac{e^{-\zeta/2}-1}{\kappa a},\\
		\label{eq:sup_coeff_2}
		D_{l}=\frac{2}{3}\left[Pe_{-}(e^{\zeta/2}-1)^{2}+\frac{Pe_{+}}{1+D_{g}}(e^{-\zeta/2}-1)^{2}\right]
	\end{gather}
\end{subequations}
In order to relate the $\zeta$-potential with the equilibrium surfactant concentration, eq.(\ref{eq:tla3}) provide at the droplet interface as 
\begin{equation}\label{eq:zsl1}
\psi^{0}_{r}(1)=-sgn(\zeta)\kappa a\sqrt{\exp(-\zeta)+\exp(\zeta)-2}=-2\kappa a \sinh(\zeta^0/2)
\end{equation}
Thus the $\zeta$-potential in the thin-double-layer limit can be obtained after substituting  $\psi^{0}_{r}(1)$ from the eq.(\ref{eq:sm4}) as
\begin{equation}\label{eq:zsl2}
	\zeta=2\sinh^{-1}\left[{z_{s}Ma\Gamma^{0}}/{(2\kappa a)}\right]
\end{equation}
Eqs.(\ref{eq:tla10_1}) and (\ref{eq:tla10_2}) provide explicit asymptotic expressions for the mobility of surfactant-laden droplets in the thin–double-layer limit. These expressions constitute one of the key findings of the present study, as they enable direct evaluation of the droplet mobility as a function of $\zeta$-potential, without imposing any restriction on its magnitude, and of the electrolyte properties. As such, they offer a convenient theoretical framework for interpreting and guiding experiments on the diffusiophoresis of charged liquid droplets.Comparisons between these analytical solutions and the numerical simulations performed in the present work is provided in subsection~\ref{sub:5.1}.
\section{Results and Discussion}
To obtain the numerical results, the thermal potential is taken as $\psi_{0}=25.8~\rm mV$, the viscosity of the suspension as $\mu = 0.89 \times 10^{-3}~\rm Pa\cdot s$, and the dielectric permittivity as $\epsilon_{e} = 7.083 \times 10^{-10}~\rm C/V\cdot m$, which corresponds to water at room temperature. The droplet radius is set to $a = 50~\rm nm$, and the background electrolyte is chosen as either NaCl, KCl, HCl, SDS, or their mixtures. The physicochemical parameters corresponding to these electrolytes are summarized in Table~\ref{table:electr_para}. The viscosity ratio of the droplet to the surrounding electrolyte, $\mu_{r}$, is varied from 0.01 to $10^{3}$, thereby encompassing the limiting cases from gas bubbles to nearly rigid solid particles. 
\begin{table}
	\caption{Ionic valences, diffusivities, Péclet numbers, and diffusion potential parameter $\beta$ for the electrolytes considered at $25^{\circ}\mathrm{C}$.}
	\label{c3table:2}
	\centering
	\small
	\begin{tabular}{c c c c c c c c c c}
		\hline
		Electrolyte & $z_+$ & $z_-$ & $D_{+}~(\rm m^2s^{-1})$ & $D_{-}~(\rm m^2s^{-1})$ & $Pe_{+}$ & $Pe_{-}$ & $\beta$ \\
		\hline\hline
		NaCl & 1 & -1  & $1.33 \times 10^{-9}$ & $2.03 \times 10^{-9}$ & 0.4 & 0.26 & -0.208 \\
		KCl & 1 & -1  & $1.96 \times 10^{-9}$ & $2.03 \times 10^{-9}$ & 0.23 & 0.26 & -0.01 \\
		HCl & 1 & -1  & $9.31 \times 10^{-9}$ & $2.03 \times 10^{-9}$ & 0.05 & 0.26 & 0.64 \\
		SDS & 1 & -1 & $1.33 \times 10^{-9}$ & $3.94 \times 10^{-10}$ & 0.4 & 1.34 & 0.543 \\
		\hline\hline
	\end{tabular}
	\label{table:electr_para}
\end{table}
\par
The rest of this section is organized as follows. In the first subsection, we compare the present numerical results with existing studies and with the analytical solutions as derived in the Section \ref{analytic}. In subsection~\ref{sub:5.2}, we examine droplet diffusiophoresis in the presence of irreversibly bound surfactant ($k_{d}=0$). The role of interfacial kinetic exchange of soluble ionic surfactant is then investigated in subsection~\ref{sub:5.3}. Subsequently, the impact of the kinetic exchange rate of anionic surfactant species in mixed electrolytes is discussed in subsection~\ref{sub:5.4}.
\subsection{Comparison with existing studies and present analytical solutions}\label{sub:5.1}
\begin{figure}
	\center
	\includegraphics[width=1.72in]{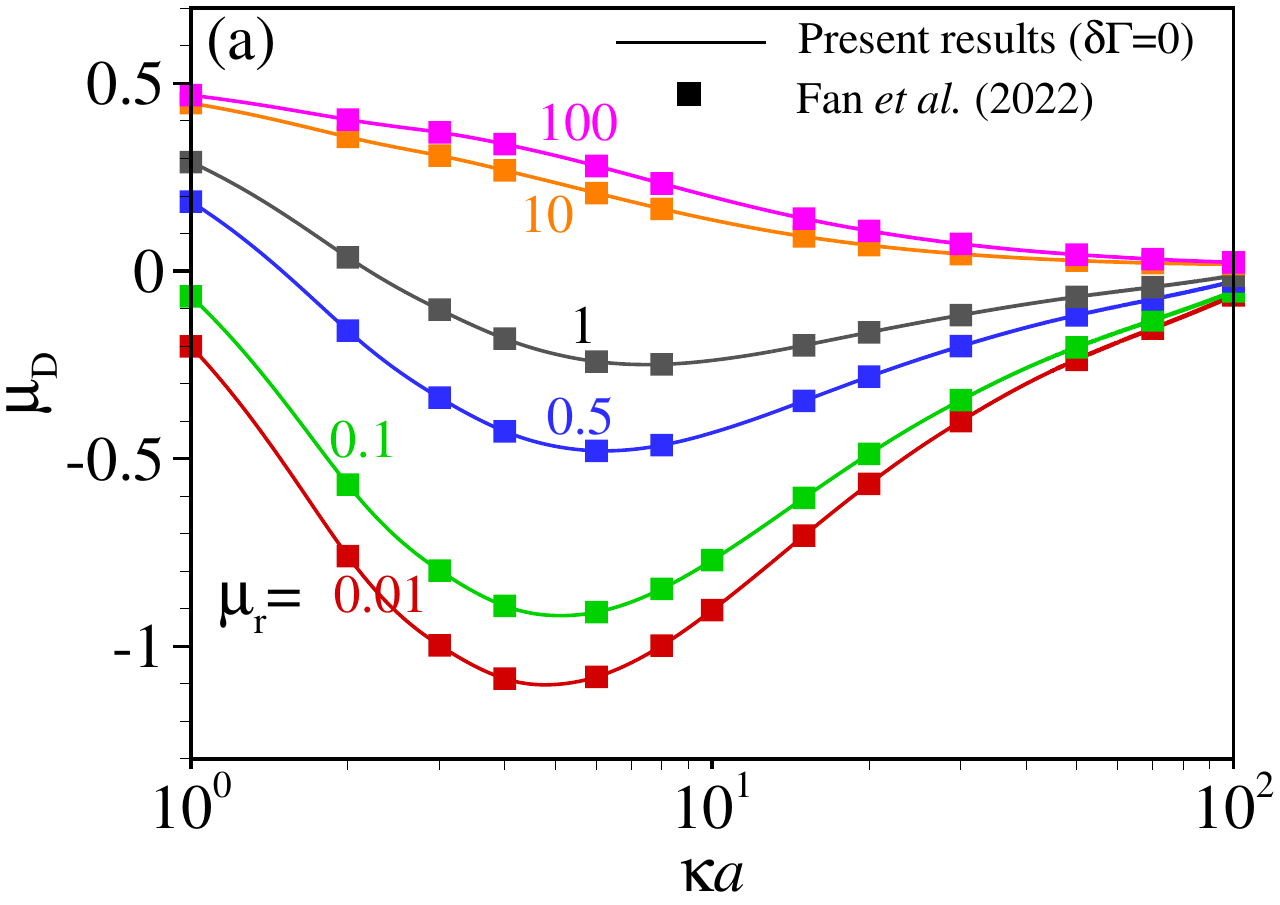}
	\includegraphics[width=1.72in]{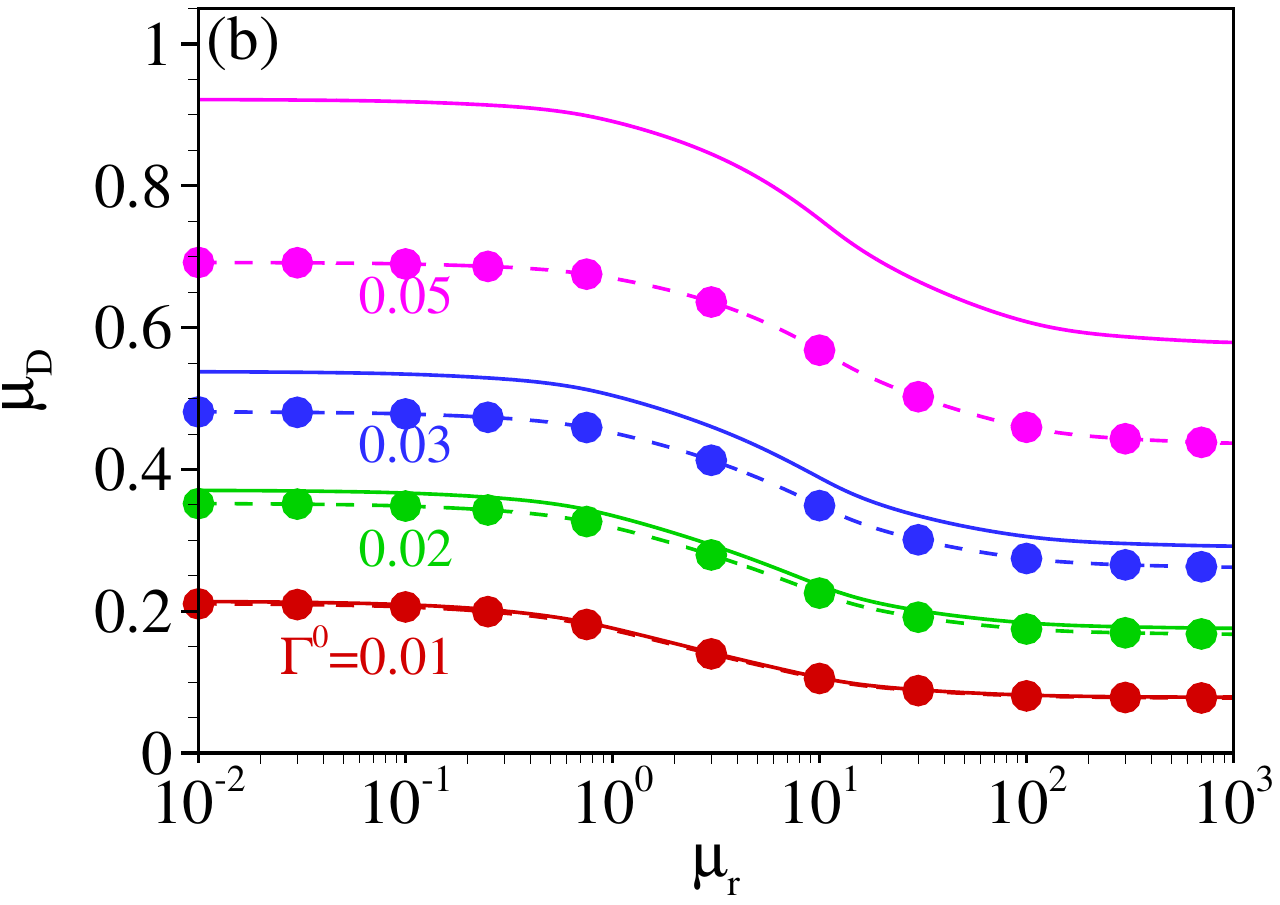}
	\includegraphics[width=1.72in]{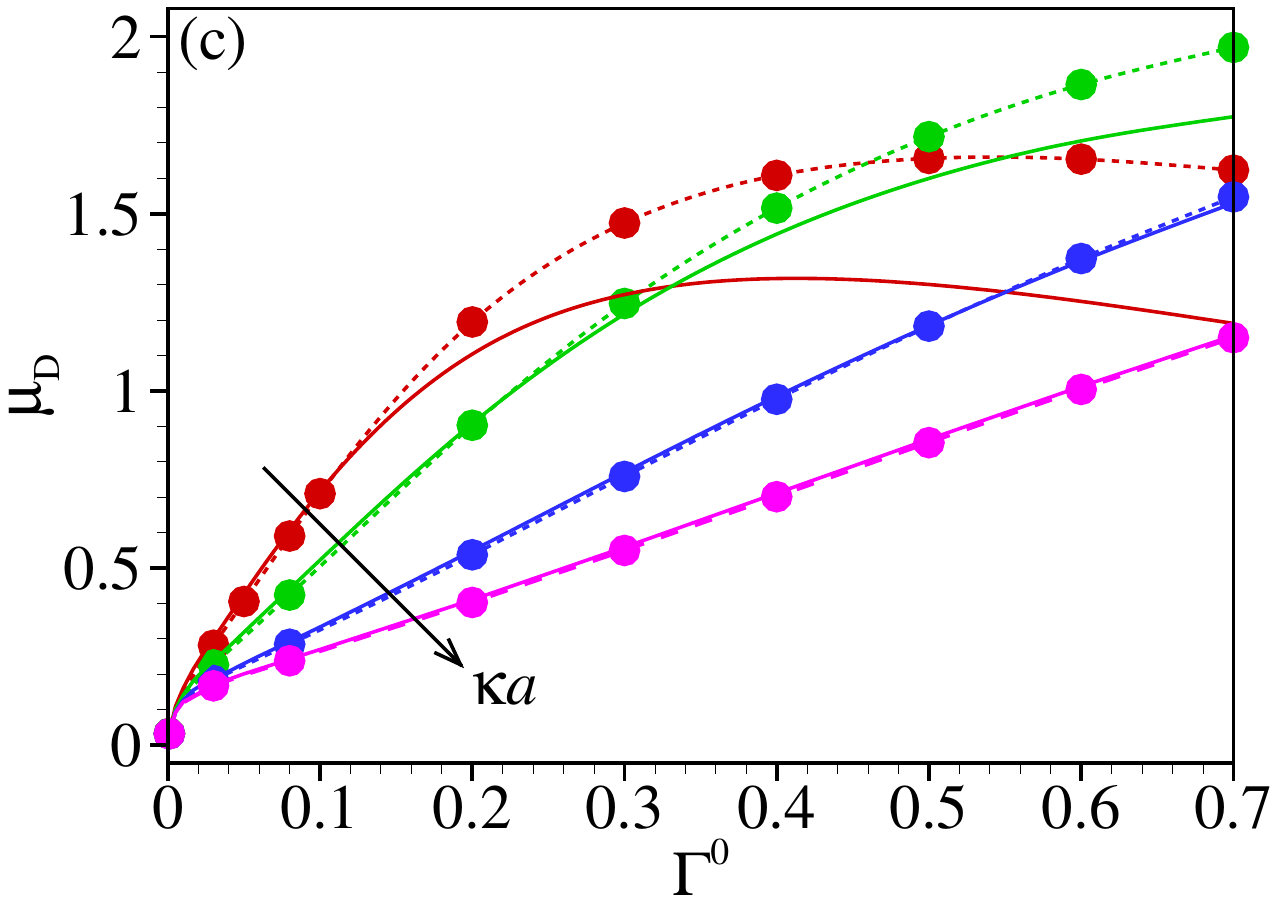}
	\caption{Comparison of diffusiophoretic mobility $\mu_{D}$ with (a) \citet{fan2022diffusiophoresis} for a non-polarizable droplet with uniform surfactant distribution at the interface for different viscosity ratio $\mu_{r}=0.01,0.1,0.5,1,10,100$ with $\Gamma^{0}=0.024~(\sigma^{0}=-10.53)$; (b) the analytical solution (\ref{eq:mobex_1}) for different $\Gamma^{0}=0.01,0.02,0.03,0.05$ at $\kappa a=10$ and (c) with the analytical solution (\ref{eq:tla10_2}) for different $\kappa a=30,50,100,150$ and $\mu_r=0.1$. Here, NaCl is considered as the electrolyte, $\Gamma^{\infty}=1~\rm nm^{-2}~(Ma=438.3)$ and the diffusion coefficient $D_{s}=3.94\times10^{-10}~\rm m^{2}/s$. In (a), square symbols, \citet{fan2022diffusiophoresis} and solid lines, present results with uniform surfactant concentration. In (b,c), solid lines, analytical solution and dashed lines, numerical simulations.}
	\label{fig_AA}
\end{figure}
We begin by validating our numerical results through comparisons with existing studies and with analytical solutions obtained in various limiting situations. Fig.\ref{fig_AA}a compares our numerical results with those of \citet{fan2022diffusiophoresis} for the case in which the surfactant distribution is uniform along the interface, which corresponds to a constant surface charge density. In this comparison, Marangoni stresses are neglected and interfacial kinetic exchange is absent ($k_d = 0$). We find that our results are in exact agreement with those reported by \citet{fan2022diffusiophoresis}  for all values of $\kappa a$. This agreement is expected, as both studies employ a perturbation approach to derive the governing coupled ordinary differential equations. However, \citet{fan2022diffusiophoresis} restricted their analysis to constant surface charge density ($\delta\sigma=0$) and did not account for effects arising from non-uniform surfactant distributions, such as mobile surface charge, Marangoni stress, and interfacial kinetic exchange.
\par 
In Fig.\ref{fig_AA}b, we compare our numerical results for a wide range of viscosity ratios $\mu_{r}$ with the analytical solution (\ref{eq:mobex_1}) derived in the previous section under the Debye--H{\"u}ckel approximation. We find that the numerical results agree exactly with the analytical solution for the lower values of $\Gamma^{0}$. However, as $\Gamma^{0}$ increases, we observe deviations, which become significant for larger values of $\Gamma^{0}$. This behaviour is expected because the analytical solution (\ref{eq:mobex_1}) is derived under the Debye-H{\"u}ckel approximation where the interfacial potential is less than the thermal potential i.e., when $z_{s}Ma\Gamma^{0}(\kappa a+1)^{-1}<1$. For $\kappa a = 10$, this condition requires $\Gamma^{0} < 0.025$. Consequently, deviations from the analytical solution is justified when $\Gamma^{0}$ exceeds this threshold and the consistent increment in difference with further increases in $\Gamma^{0}$.
\par 
Fig.\ref{fig_AA}c presents the diffusiophoretic mobility $\mu_{E}$ as a function of $\Gamma^{0}$ for several values of $\kappa a$ when $\mu_{r}=0.1$. The solid lines represent the analytical solution~(\ref{eq:tla10_2}), while the dashed lines with symbols correspond to the numerical results. For smaller values of $\kappa a$, we find a quantitative discrepancy between the numerical and analytical solutions. These discrepancies diminish as $\kappa a$ increases, and for sufficiently large $\kappa a$ the two solutions are in agreement. This trend is consistent with the assumptions underlying the analytical solution~(\ref{eq:tla10_2}), which is derived under the thin electric double-layer limit and therefore valid for larger values of $\kappa a$.
\subsection{Impact of irreversibly bound surfactants}\label{sub:5.2}
\begin{figure}
	\center
	\includegraphics[width=1.72in]{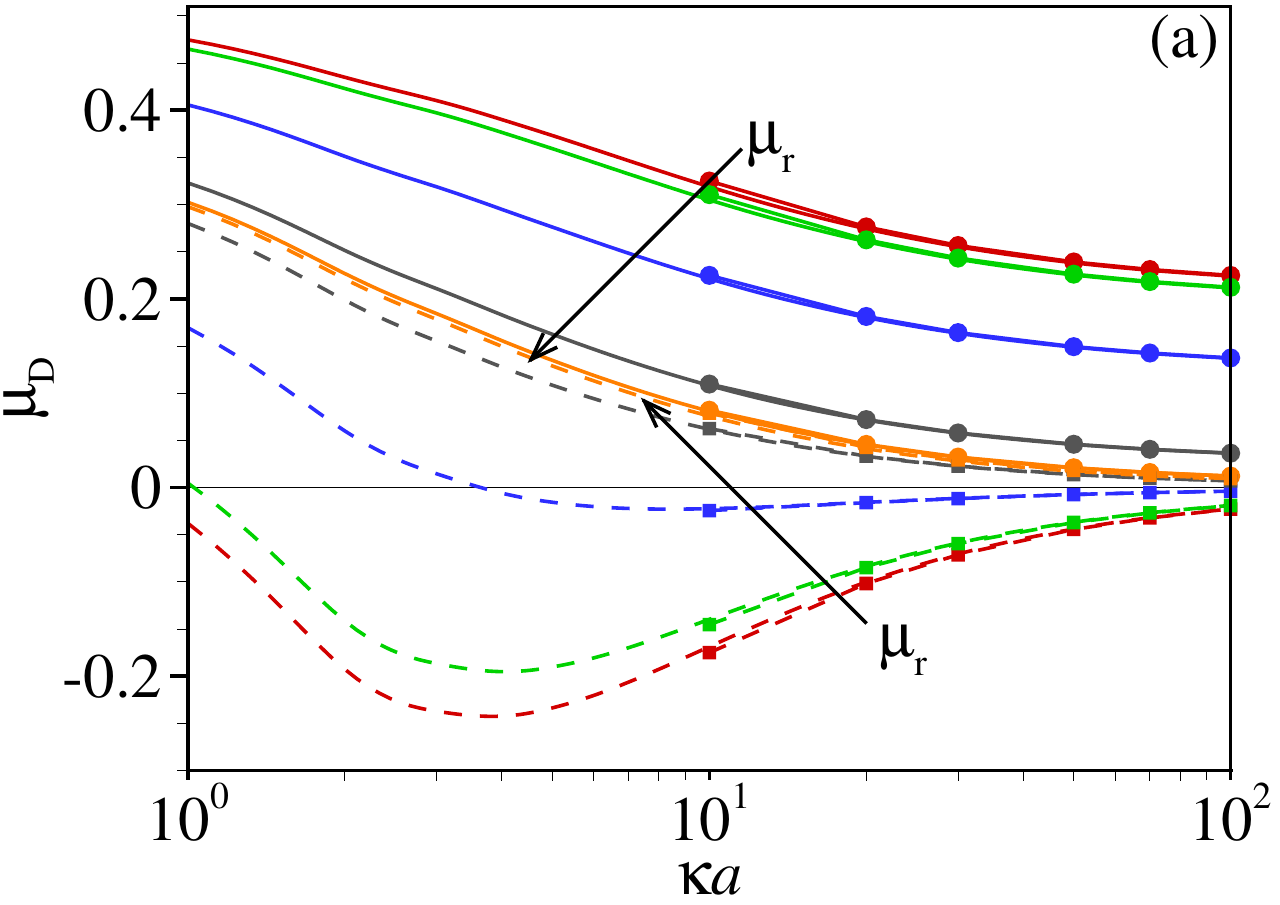}
	\includegraphics[width=1.72in]{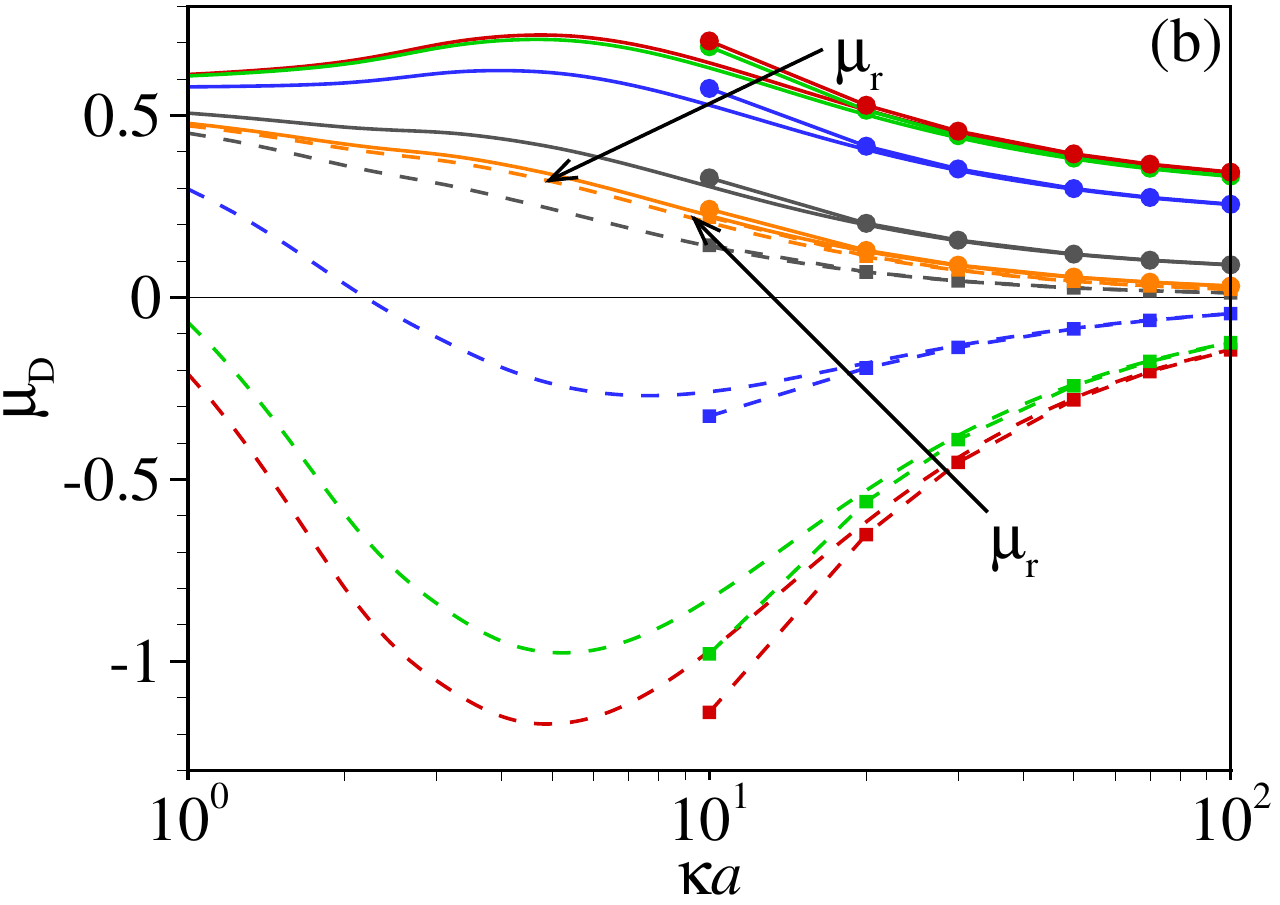}
	\includegraphics[width=1.72in]{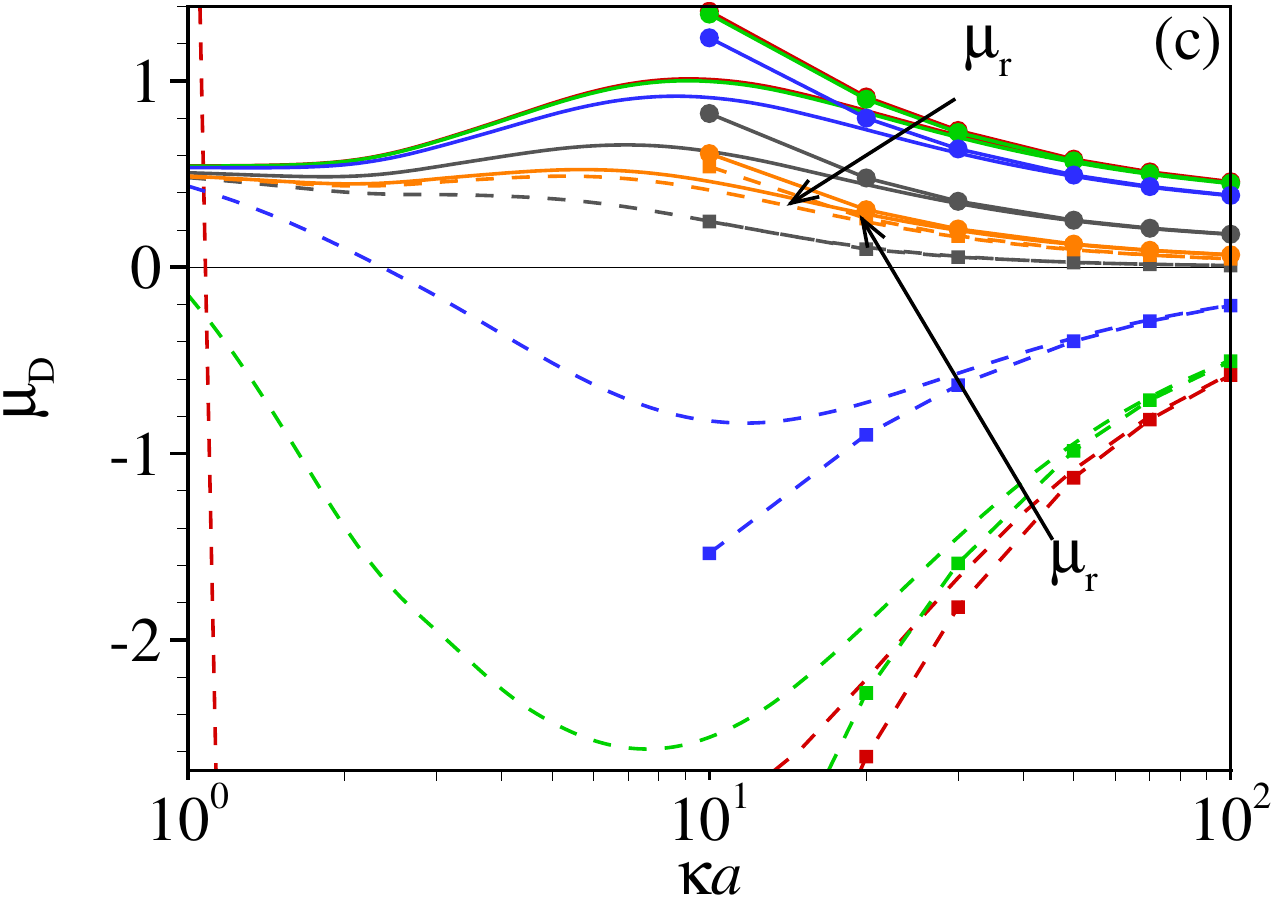}
	\caption{Variation of the diffusiophoretic mobility $\mu_{D}$ as a function of $\kappa a$ at (a) $\Gamma^{0}=0.01~(\sigma^{0}=-4.383)$, (b) $\Gamma^{0}=0.025~(\sigma^{0}=-10.958)$ and (c) $\Gamma^{0}=0.05~(\sigma^{0}=-21.915)$ for different values of $\mu_{r}=0.01,0.1,1,10,100$ when $\Gamma^{\infty}=1\rm ~nm^{-2}~(Ma=438.3)$ and $k_d=0$ in NaCl electrolyte solution. Here, dashed lines, uniform surfactant ($\delta\Gamma=0$); solid lines, non-uniform surfactant ($\delta\Gamma\neq0$). Dashed line with symbols, D-H based analytic solution (\ref{eq:mobex_1_uniform}) for uniform surfactant; solid lines with symbols, D-H based analytic solution (\ref{eq:mobex_1}) for non-uniform surfactant. Arrows indicate increasing values of $\mu_r$ and $D_{s}=10^{-9}~\rm m^{2}/s$.}
	\label{fig_BB}
\end{figure}
Fig.\ref{fig_BB}a-c show the variation of the diffusiophoretic mobility with the Debye-layer thickness for different viscosity ratios $\mu_r$. In order to assess both the qualitative and quantitative differences between the two cases and to elucidate the role of surface-charge non-uniformity in droplet dynamics, we present results for irreversibly adsorbed surfactant alongside those obtained assuming a uniform interfacial surfactant concentration ($\delta\Gamma=0$). In diffusiophoresis, the direction of electro-migration is governed by the sign of the product $\beta\sigma$. For the case of uniform surfactant concentration, in agreement with previous studies~\citep{tsai2022analytical,fan2022diffusiophoresis,majhi2023diffusiophoresis}, we find that for sufficiently small viscosity ratios $\mu_r$ the initially positive mobility decreases with increasing $\kappa a$ and eventually becomes negative. The emergence of negative mobility is attributed to the type-II double-layer polarization (DLP-II) mechanism \citep{wu2021diffusiophoresis,fan2022diffusiophoresis}. This effect arises when surface charge impedes the diffusion of co-ions across the Debye layer, leading to their accumulation near the outer edge of the electric double layer. The resulting excess co-ion concentration generates a repulsive force on the droplet, thereby producing a negative mobility. In contrast, this behaviour is suppressed when the charge density is considered to develop due to the adsorption of ionic surfactant. In this case, the combined effect of Maxwell traction and Marangoni stress due to the non-uniform surfactant distribution at the interface lowers the interfacial speed so as to attenuates the Debye layer polarization, which diminishes the DLP-II effect. Consequently, the chemiphoretic velocity remains positive i.e., along the direction of the imposed concentration gradient. Here, the electrophoretic contribution is positive ($\beta\sigma>0$) and acts cooperatively with the chemiphoretic component, yielding an overall positive mobility. Thus, we find that mobility is positive for the negatively charged droplet with negative diffusion field ($\beta=-0.208$). We find that at a higher $\Gamma^{0}$, a local maxima in $\mu_{D}$ occurs when varied with $\kappa a$, whereas $\mu_{D}$ monotonically declines with $\kappa a$ at a lower $\Gamma^{0}$. These trends are corroborated with the analytical solutions~(\ref{eq:mobex_1_uniform}) for uniform surfactant distribution and~(\ref{eq:mobex_1}) for non-uniform distribution in the regime of sufficiently large $\kappa a$ where the Debye-H\"uckel condition $z_s Ma \Gamma^{0}(\kappa a+1)^{-1}<1$ is satisfied. On the basis of these results, we conclude that the DLP-II effect observed in diffusiophoresis of uniformly charged droplets with immobile surface ions is largely an artefact of the assumption of constant surface charge density.
\par
\begin{figure}
	\center
	\includegraphics[width=1.72in]{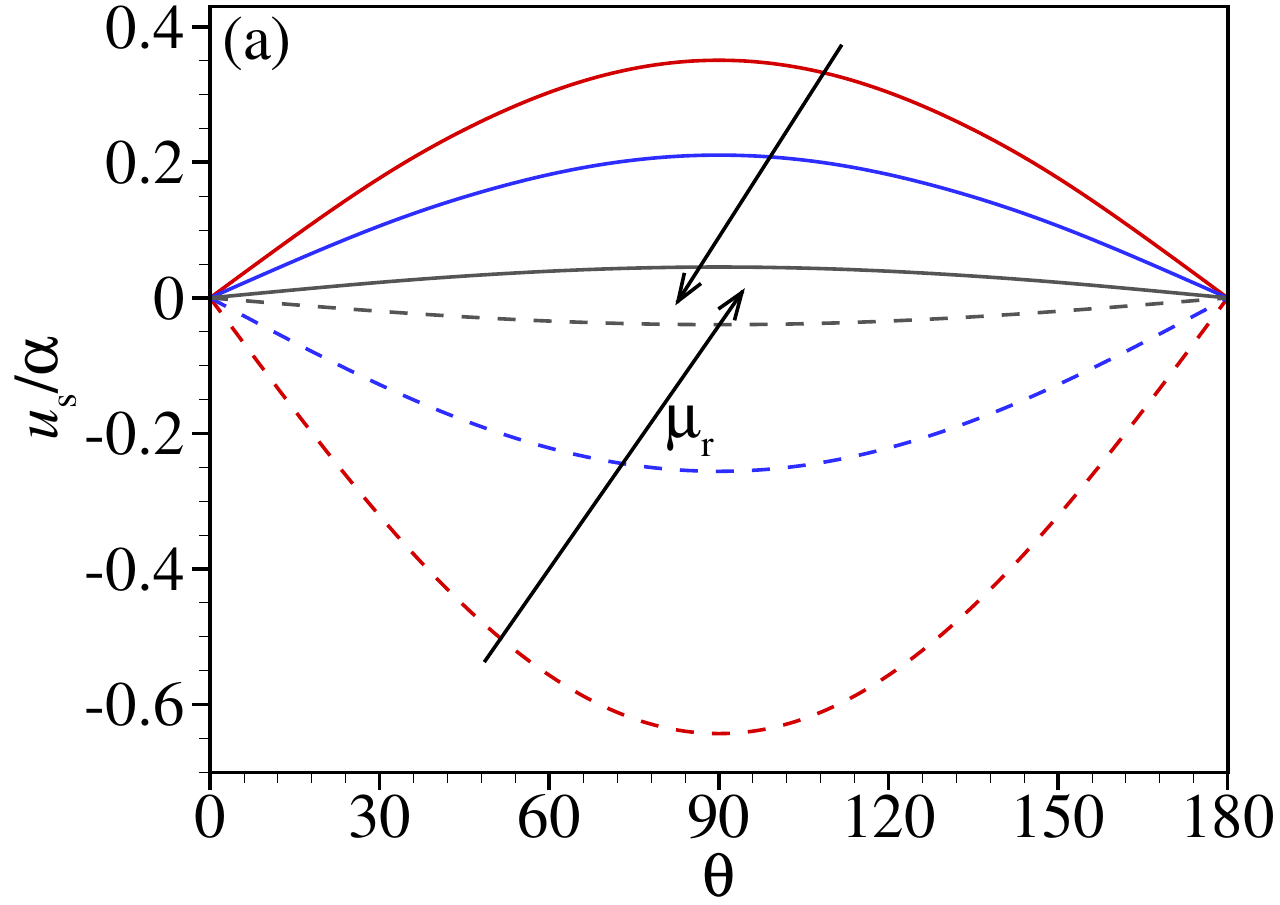}
	\includegraphics[width=1.72in]{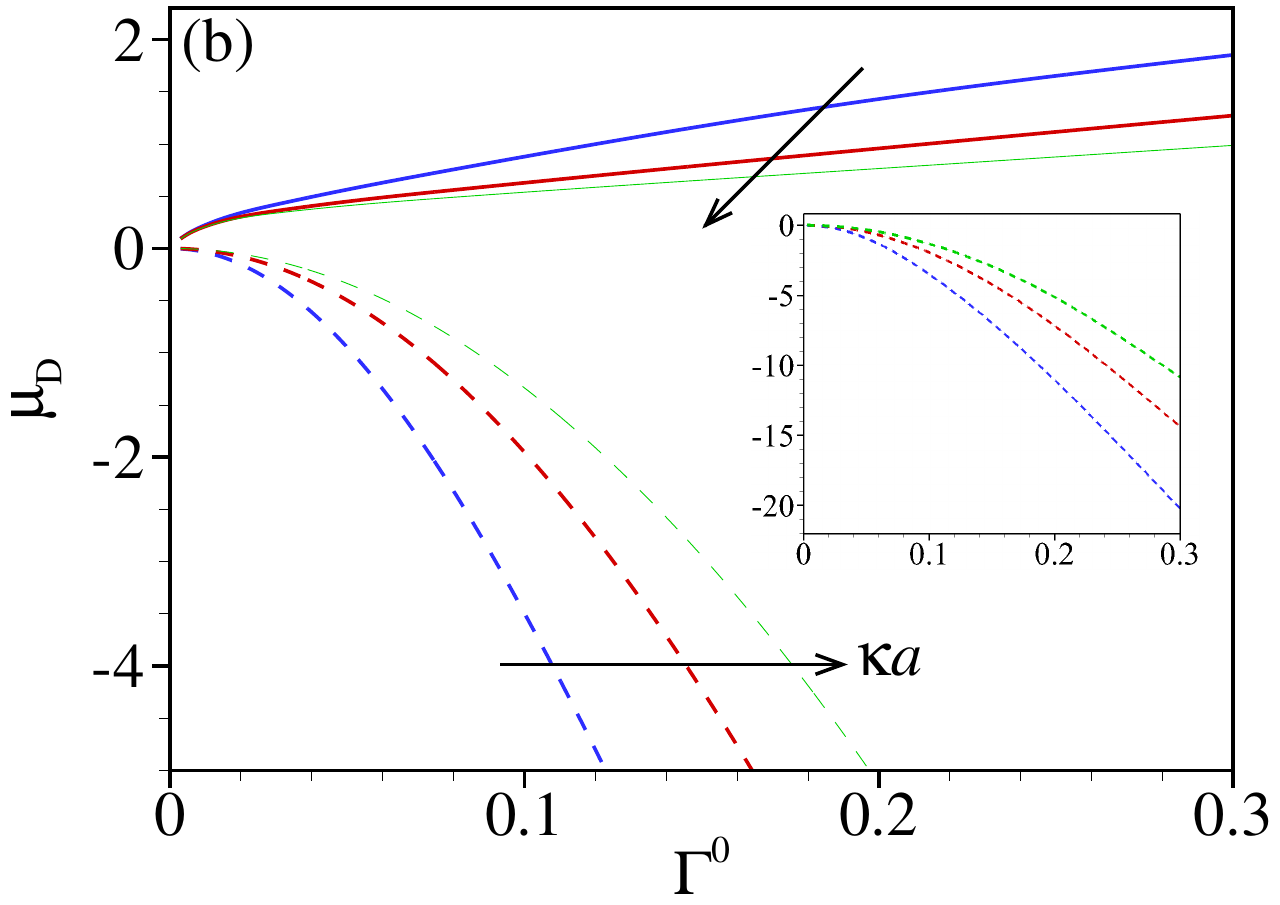}
	\includegraphics[width=1.72in]{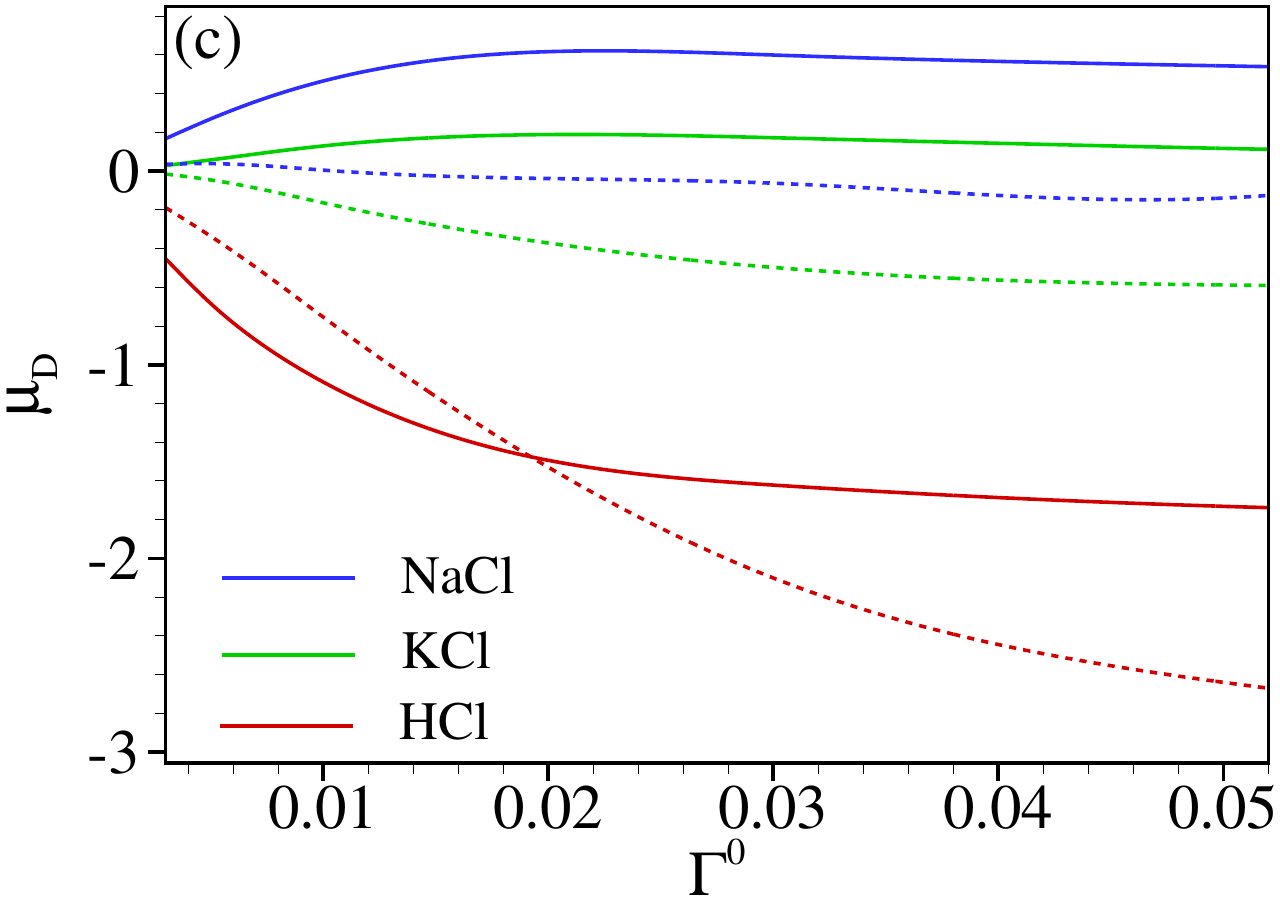}
	\caption{Variation of (a) the interfacial velocity at $\kappa a=3$ and $\Gamma^{0}=0.01$ for different $\mu_{r}=0.01,1,10$; (b) $\mu_{D}$ as a function of $\Gamma^{0}$ for different $\kappa a=50,100,150$ at $\mu_{r}=0.1$ (inset figure shows the complete range for uniform surfactant) and (c) $\mu_{D}$ as a function of $\Gamma^{0}$ for different salt (NaCl, KCl and HCl) at $\kappa a=1$ and $\mu_{r}=0.1$. Here, $\Gamma^{\infty}=1\rm ~nm^{-2}~(Ma=438.3)$, $k_d=0$, $D_{s}=10^{-9}~\rm m^{2}/s$, and dashed lines, uniform surfactant ($\delta\Gamma=0$); solid lines, non-uniform surfactant ($\delta\Gamma\neq0$). In (a,b), the electrolyte is NaCl.}
	\label{fig_CC}
\end{figure}
Fig.\ref{fig_CC}a shows the interfacial velocity distributions for different viscosity ratios $\mu_r$ at $\kappa a = 3$ and $\Gamma^{0} = 0.01$. For a droplet with constant surface charge density ($\delta\Gamma=0$), the interfacial velocity $u_s$ is negative, corresponding to a clockwise rotating flow inside the droplet. This reversal is interpreted as a consequence of the type-II double-layer polarization (DLP-II) mechanism. When the droplet is considered laden with compressible surfactant, it creates a non-uniform and mobile surface charge density, the interfacial velocity becomes positive, corresponding to a counter-clockwise circulation inside the droplet. In this case, surface-charge mobility regularizes the tangential Maxwell stresses at the interface, thereby suppressing the DLP-II mechanism and leading to a positive interfacial slip velocity. As a consequence, the diffusiophoretic mobility becomes positive. In this case, the induced fluid motion in the surrounding electrolyte and the interfacial flow are aligned in the same direction. Consequently, the diffusiophoretic mobility increases as the viscosity ratio decreases, implying that the droplet translates faster than the corresponding rigid particle, as also observed in Fig.\ref{fig_BB}.
\par 
Impact of the low to moderate equilibrium surfactant concentration $\Gamma^{0}$ at different $\kappa a=50,100,150$ on the mobility of a droplet with viscosity ratio $\mu_{r}=0.1$ is presented in Fig.\ref{fig_CC}b. We find that the mobility $\mu_{D}$ enhances rapidly in the negative direction even through $\beta\sigma>0$ for the case of uniform surfactant distribution. However, in the present case of compressible surfactant-laden droplets ($\delta\Gamma\neq 0$), $\mu_{D}$ is positive and it increases at a relatively lower rate with $\Gamma^{0}$. We also find that $\mu_{D}$ decreases as $\kappa a$ increases at higher values of $\kappa a$.
\par 
Fig.\ref{fig_CC}c and Fig.\ref{fig_DD}a illustrate the salt-specific dependence on the diffusiophoretic mobility of the droplet as function of $\Gamma^{0}$ for lower range of $\kappa a=1$ (Fig.\ref{fig_CC}c) and for higher range of $\kappa a=80$ (Fig.\ref{fig_DD}a). The dashed curves correspond to the case of uniform surfactant ($\delta\Gamma=0$ and hence uniform surface charge) distribution, while the solid curves represent the present model with non-uniform surfactant adsorption and hence, mobile surface charge. Three electrolytes are considered namely, NaCl ($\beta=-0.208$), KCl ($\beta=0$), and HCl ($\beta=0.64$). We find a huge difference in mobility of the present case ($\delta\Gamma\neq 0$) from the uniformly charged ($\delta\Gamma=0$) droplets. For uniformly charged droplets at $\kappa a = 1$, the mobility $\mu_D$ is negative in HCl and KCl solutions and positive in NaCl, reflecting a pronounced salt-specific response. However, in a thinner double-layer case $\kappa a = 80$, the mobility for the uniformly charged droplet becomes negative for all three electrolytes (i.e., independent of $\beta$) and the differences in magnitude among the salts become negligible. The underlying reason for $\mu_{D}$ to become independent of $\beta$ is the dominance of the type-II double-layer polarization (DLP-II) mechanism. As both $\kappa a$ and $\Gamma^{0}$ increase, the DLP-II effect strengthens, producing increasingly strong negative chemiphoretic contribution, which overwhelms the electrophoretic component. In contrast, for the present case of non-uniform and mobile interfacial charge density, the DLP-II mechanism is suppressed and the interfacial stress components remain comparable to each other in magnitude. Consequently, the chemiphoretic contribution does not alone governs the diffusiophoresis and the electrophoretic contribution becomes comparable. As a result, the mobility retains the salt dependency and obeys the relation, $\mu_D>0$ for $\beta\sigma>0$ and $\mu_D<0$ for $\beta\sigma<0$, as is evident from Fig.\ref{fig_CC}c and~\ref{fig_DD}a.
\par
\begin{figure}
	\center
	\includegraphics[width=1.72in]{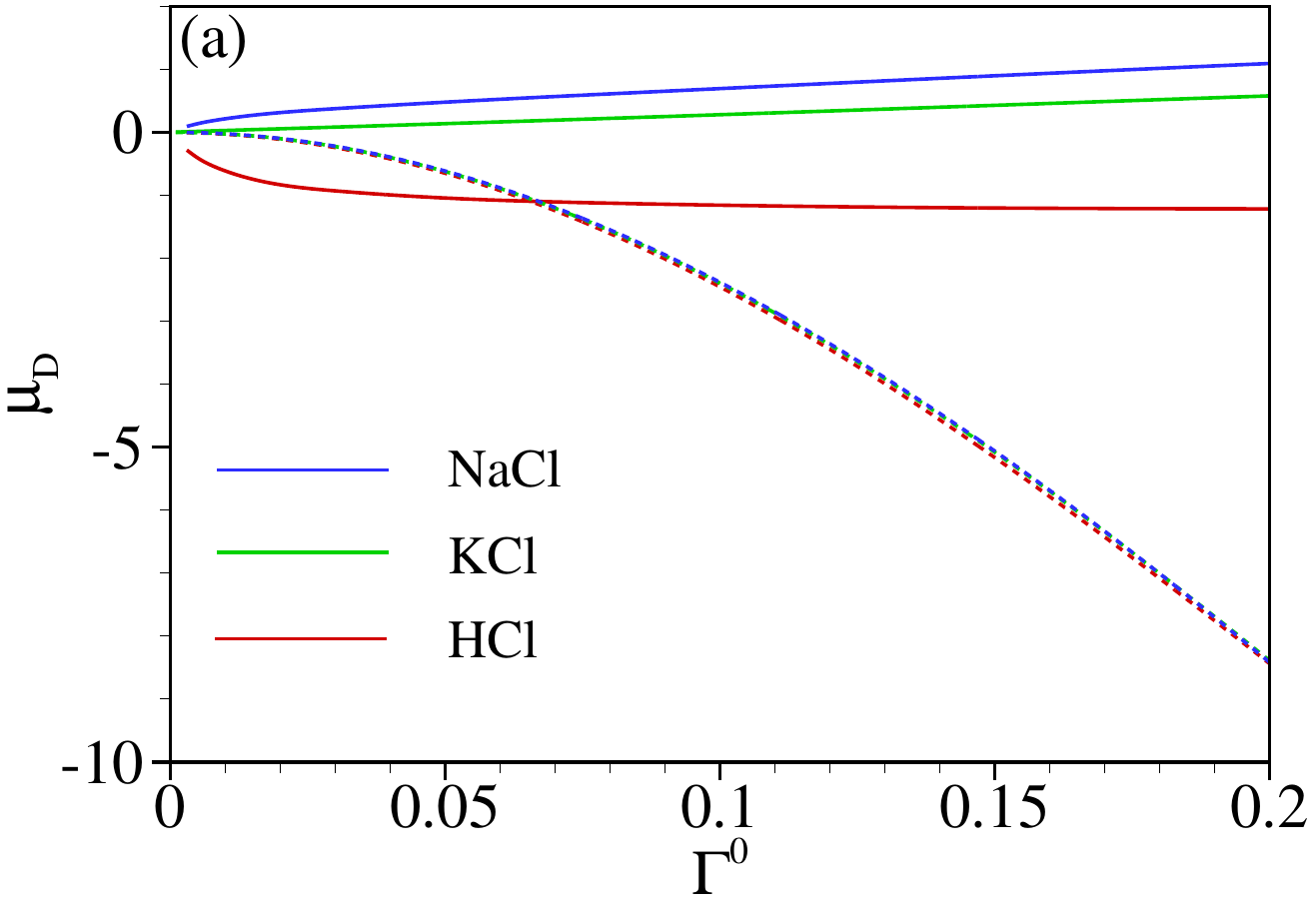}
	\includegraphics[width=1.72in]{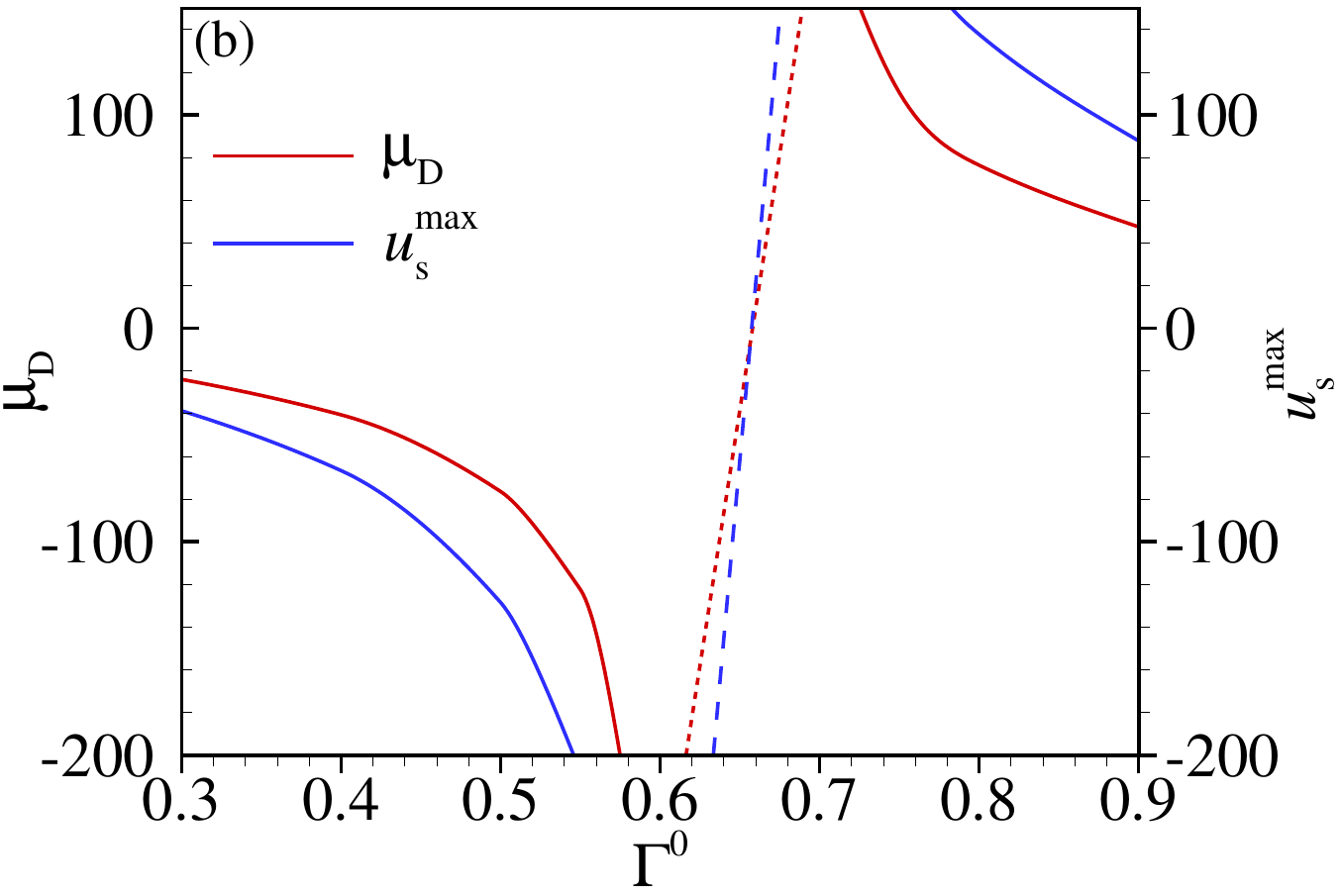}
	\includegraphics[width=1.72in]{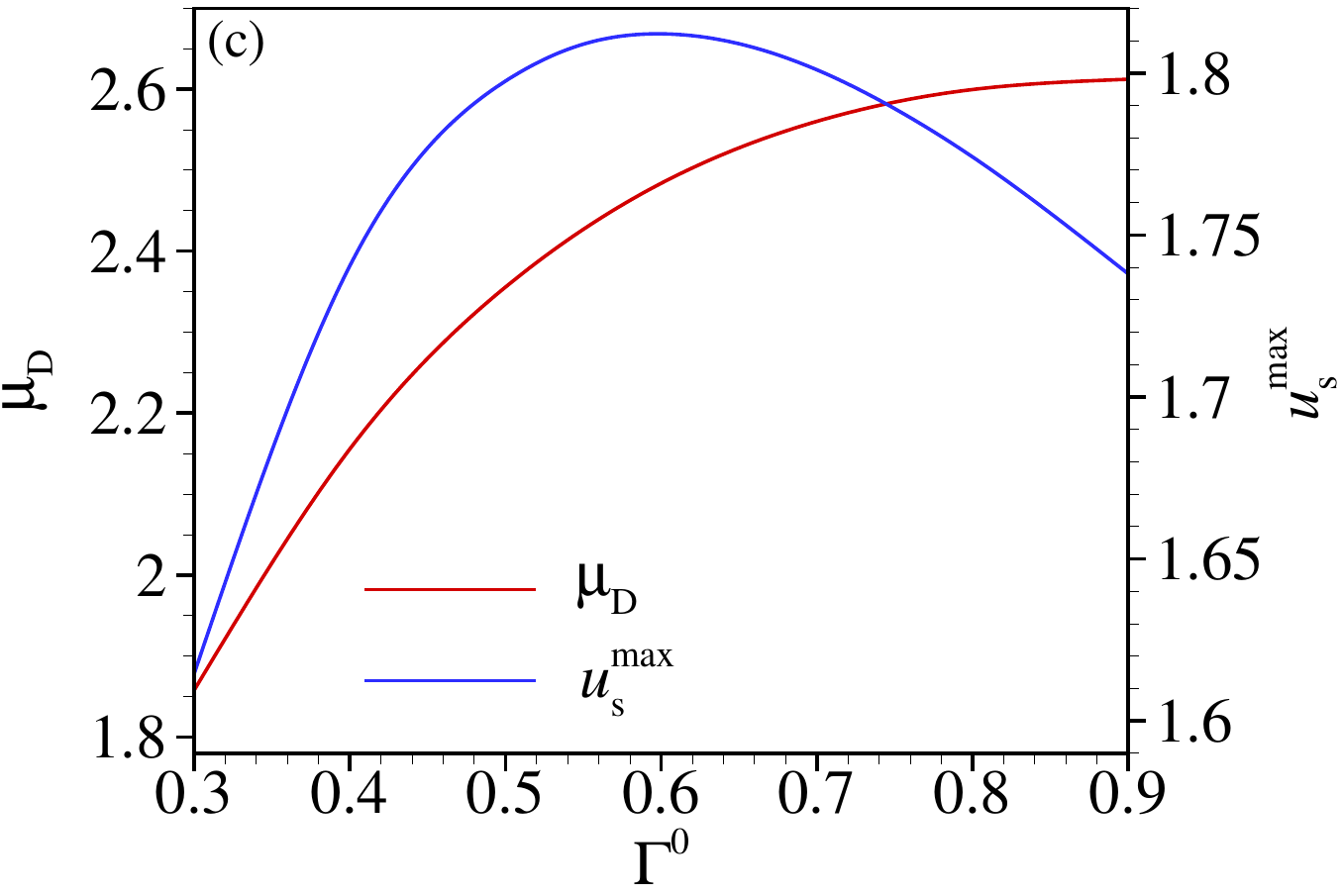}
	\caption{(a) Variation of mobility $\mu_{D}$ as a function of $\Gamma^{0}$ for different salts NaCl, KCl and HCl at $\kappa a=80$ when $\mu_{r}=0.1$. Variation of mobility $\mu_{D}$ and interfacial velocity $u_{s}$ as a function of higher values of $\Gamma^{0}$ at $\kappa a=50$ when $\mu_{r}=0.01$ in NaCl electrolyte solution for (b) uniform surfactant ($\delta\Gamma=0$) and (c)  non-uniform surfactant ($\delta\Gamma\neq0$). Here, $\Gamma^{\infty}=1\rm ~nm^{-2}~(Ma=438.3)$, $k_d=0$, and $D_{s}=10^{-9}~\rm m^{2}/s$. In (a), dashed lines, uniform surfactant; solid lines, non-uniform surfactant. In (b,c), blue lines, interfacial velocity ($u_{s}$); red lines, mobility ($\mu_{D}$).}
	\label{fig_DD}
\end{figure}
Fig.\ref{fig_DD}(b,c) present the variation of the diffusiophoretic mobility and the maximum interfacial velocity ($u_{s}^{max}$) as a function of high equilibrium surfactant concentration ($\Gamma^{0}$) corresponding to higher surface charge density, for a droplet suspended in a NaCl electrolyte solution. Results are presented for both a uniformly charged interface with immobile surface ions (Fig.\ref{fig_DD}b) and a non-uniformly charged interface arising from surfactant adsorption (Fig.\ref{fig_DD}c). For the droplet with constant surface charge density, we find that mobility $\mu_{D}$ is abruptly high for higher $\Gamma^{0}$ and, it attains a unphysical sky-high jump for $\Gamma^{0}$ between $0.6$ to $0.7$ for the considered values of $\kappa a$. This abrupt increase in $\mu_{D}$ raises serious concern regarding the physical validity of the constant-charge model in the high surface-potential regime. The origin of this behaviour can be traced to the interfacial velocity, which is driven by the combined action of hydrodynamic, Maxwell, and Marangoni stresses. Under the assumption of uniform surface charge, the tangential Maxwell stress becomes abruptly high for higher $\Gamma^{0}$, which influences the other stresses as they are coupled and in combination, it makes the interfacial velocity to encounter a step-jump. For this, the mobility $\mu_{D}$ encounters a singularity when varied with the surface charge density ($\Gamma^{0}$) when $\delta\Gamma=0$ i.e., a constant surface charge density is considered. In contrast, when the surface charge density is allowed to vary self-consistently through surfactant adsorption/ desorption, the tangential mobility of the surface ions occurs and the nonunifromity in surfactant is associated with the development of the Marangoni stress. These effects alter the interfacial stress balance condition qualitatively. The Maxwell stress and other interfacial stresses varies smoothly with $\Gamma^{0}$. As a result, the interfacial velocity remains finite and varies smoothly, as shown in Fig.\ref{fig_DD}b. Consequently, the diffusiophoretic mobility remains bounded and exhibits a physically consistent dependence on $\Gamma^{0}$. These results demonstrate that incorporating surface-charge mobility and interfacial Marangoni effects due to the non-uniformity of surfactant are essential for obtaining physically meaningful modeling of droplet mobility at high surfactant concentration.

\subsection{Interfacial kinetic exchange of soluble ionic surfactants}\label{sub:5.3}
\begin{figure}
	\center
     \includegraphics[width=1.72in]{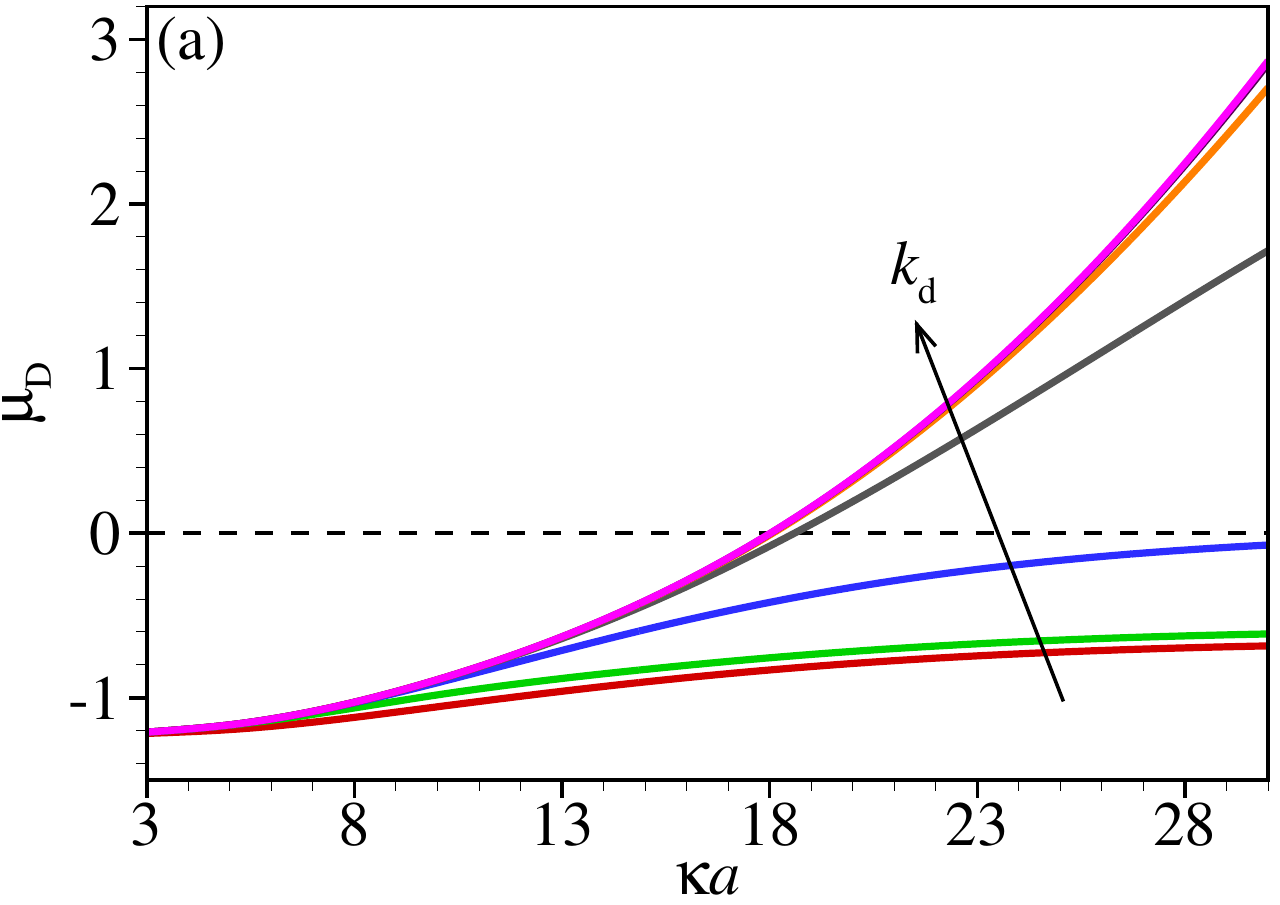}
	\includegraphics[width=1.72in]{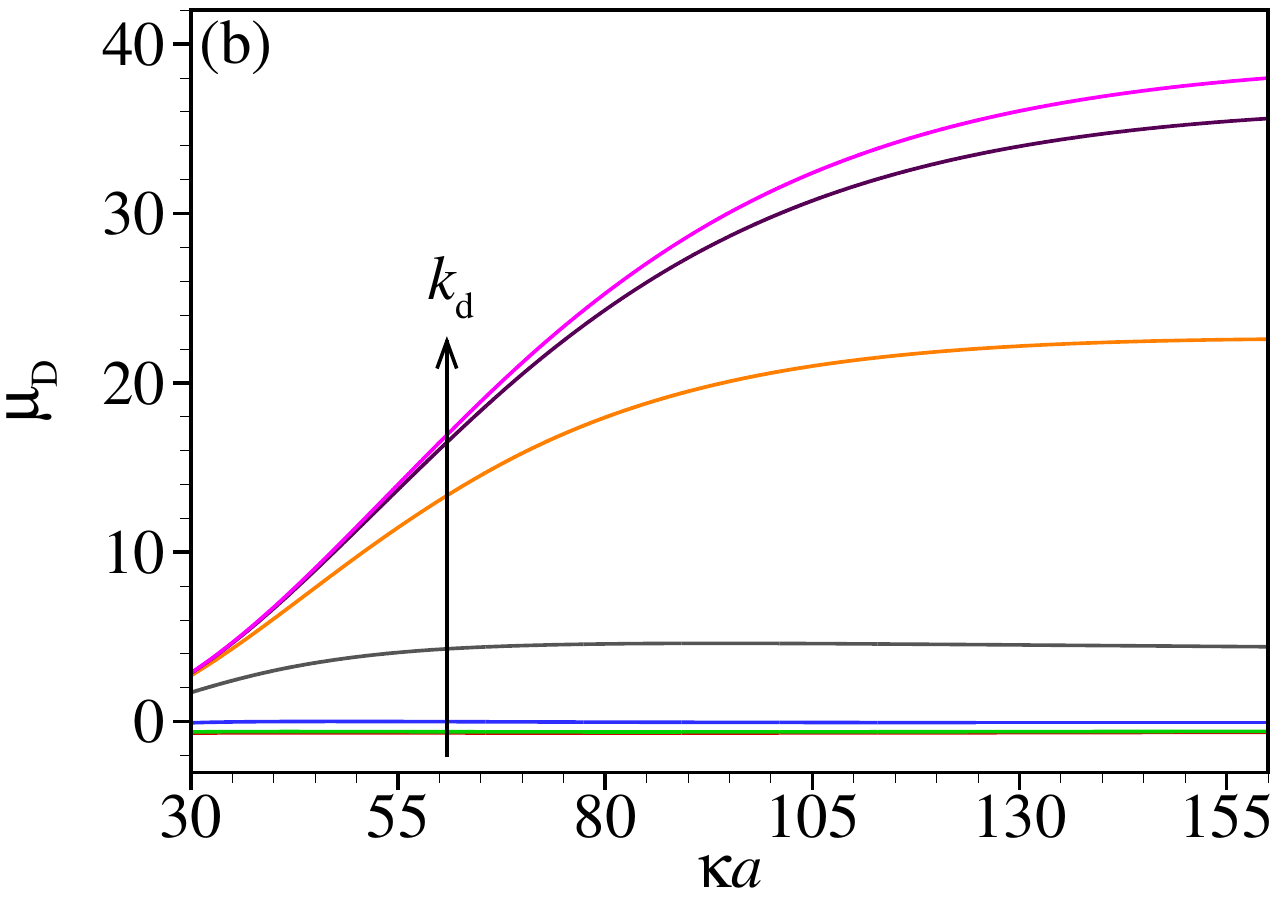}
	\includegraphics[width=1.72in]{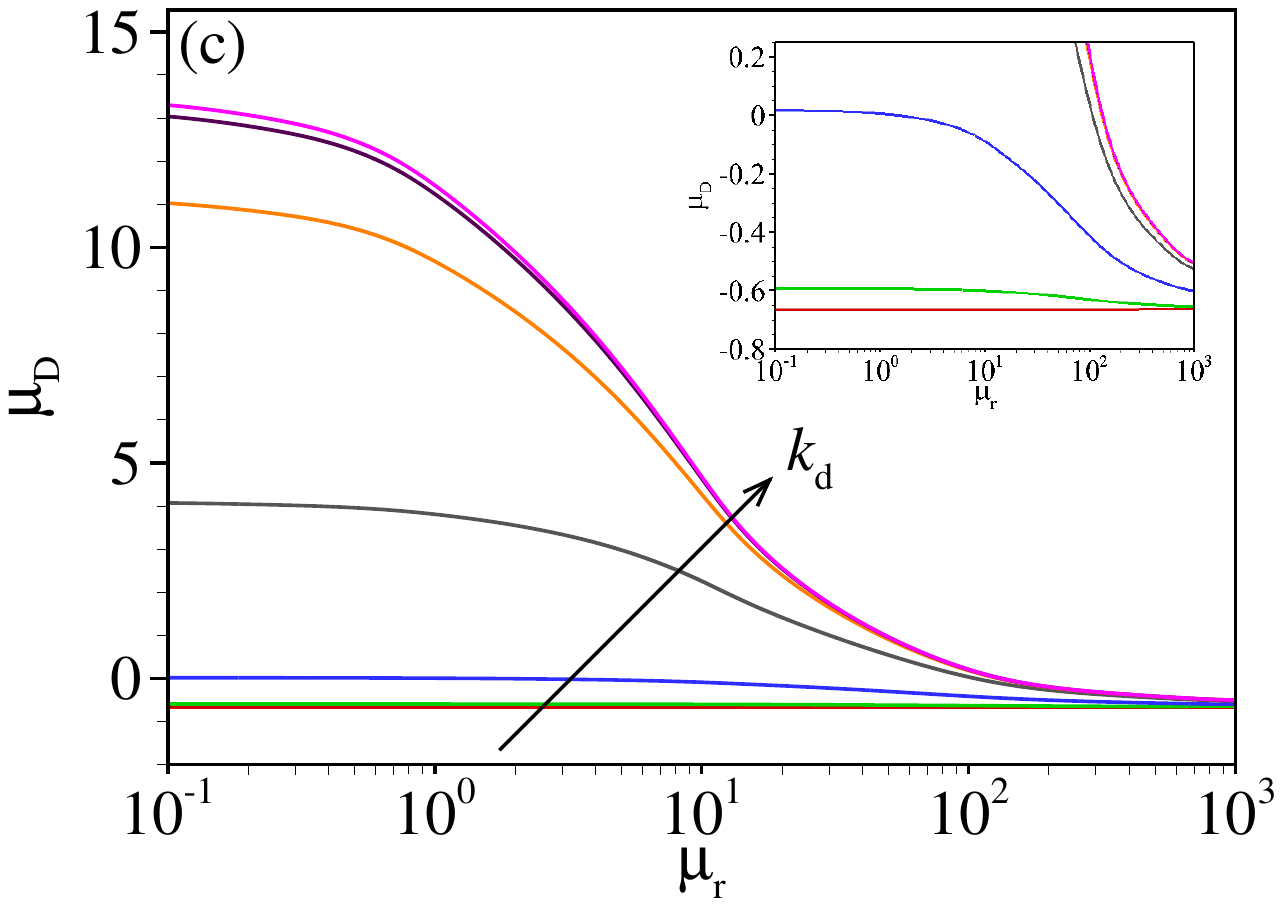}
	\caption{Variation of the mobility $\mu_{D}$ with (a) lower values of $\kappa a$ ($\leq$ 30) and (b) higher values of $\kappa a$ ($\geq$ 30) for different desorption rate constants $k_{d}=0,10^{4},10^{5},10^{6},10^{7},10^{8}, 10^{10}~\rm s^{-1}$ with $\mu_{r}=1$ and $\Gamma^{0}=0.4$. (c) Variation of $\mu_{D}$ as a function of $\mu_{r}$ for different $k_{d}=0,10^{4},10^{5},10^{6},10^{7},10^{8}, 10^{10}~\rm s^{-1}$ at $\kappa a=50$ and $\Gamma^{0}=0.4$. Here, $\Gamma^{\infty}=1~\rm nm^{-2}$, the cation is Na$^{+}$, anion is DS$^{-}$, and consequently $\beta=0.543$. The inset in panel (c) presents an enlarged view near $\mu_{D}=0$.}
	\label{fig_EE}
\end{figure}
Fig.{\ref{fig_EE}}a-c illustrates the influence of interfacial kinetic exchange of soluble surfactant on the diffusiophoretic mobility of a surfactant-laden droplet for different values of the bulk ionic concentration and viscosity ratio $\mu_r$. Fig.{\ref{fig_EE}}a,b show that the effect of the desorption rate constant $k_d$ on the mobility depends strongly on the value of $\kappa a$. In the lower range of $\kappa a$ (Fig.{\ref{fig_EE}}a), the mobility exhibits only a weak dependence on $k_d$, with all curves collapsing for the considered range of desorption rate coefficient. In contrast, for moderate to larger values of $\kappa a$ (Fig.{\ref{fig_EE}}b), the mobility increases markedly with increasing $k_d$ before eventually approaching a plateau at a sufficiently high $k_d$. Notably, in this regime the mobility undergoes a sign change at large $k_d$ as $\kappa a$ is varied, indicating a reversal in the direction the of droplet motion. The weak influence of $k_d$ at low $\kappa a$ can be directly understood from the interfacial surfactant distributions. At lower $\kappa a$, corresponding to a thick electric double layer, the perturbed surfactant concentration $\delta\Gamma$ exhibits negligible variation with $k_d$. For example, at $\kappa a=3$, the maximum magnitude of $\delta\Gamma/\alpha$ is 0.0046 and it does not vary with $k_{d}$. As a result, the surfactant redistribution creates the surface charge variation and surface tension gradients intrinsically weak, leading to small changes in Maxwell and Marangoni stresses. Consequently, change in $k_{d}$ is unable to alter significantly the interfacial stress balance, and the mobility remains largely insensitive to $k_d$.
\par 
As $\kappa a$ increases, however, the behaviour changes qualitatively. For a smaller $k_{d}$ or zero exchange of surfactant ($k_{d}=0$), the surfactant distribution remains relatively less perturbed and the diffusiophoresis process is dominated by the electrophoresis. In this case, the positive diffusion potential of NaDS i.e., $\beta>0$ produces a negative mobility of the negatively charged droplet.  As $\kappa a$ increases, the screening of surface charge as well as the concentration of soluble surfactant increases, creating a higher kinetic exchange of surfactant at the interface. This lead to the declination of $-\mu_{D}$ and eventually a change of sign from negative to positive $\mu_{D}$. At $\kappa a = 50$, the magnitude of the surfactant perturbation ($\delta\Gamma$) grows with increasing $k_d$, and a clear enhancement of surfactant accumulation near the upstream pole and depletion near the downstream pole is observed. This trend becomes even more pronounced at $\kappa a = 110$ (Fig.{\ref{fig_FF}}a), where both the magnitude of $\delta\Gamma$ and its sensitivity to $k_d$ are substantially amplified. In these thinner double layer, sharp tangential gradients of electrostatic potential and ion concentration develop along the interface, allowing substantial non-uniformities in the interfacial surfactant concentration. This enhanced non-uniformity in surfactant concentration amplify Marangoni stresses (Fig.{\ref{fig_FF}}d) and increase the interfacial slip velocity, as shown in Fig.{\ref{fig_FF}}b, which directly leads to the growth of the diffusiophoretic mobility as observed in Fig.{\ref{fig_EE}}b.
\par 
The mobility reversal observed at sufficiently large $k_d$, arises from a competition between electrophoretic and chemiphoretic contributions of opposite sign. In the present study, the equilibrium surface charge of the droplet is negative due to the adsorption of anionic surfactant, while the ambipolar diffusivity parameter $\beta$ is positive for the considered electrolyte. Consequently, the electrophoretic contribution to the mobility is negative ($\beta\sigma<0$), whereas, the chemiphoretic contribution is positive. For smaller values of $k_d$ interfacial kinetic exchange is weak, which attenuates the chemiphoresis part arises due to osmotic pressure drop. Consequently, the electrophoresis part dominates, which creates a negative mobility of the droplet. The Maxwell traction creates a negative interfacial velocity and the Marangoni stress is relatively weak. 
\begin{figure}
	\center
	\includegraphics[width=2.3in]{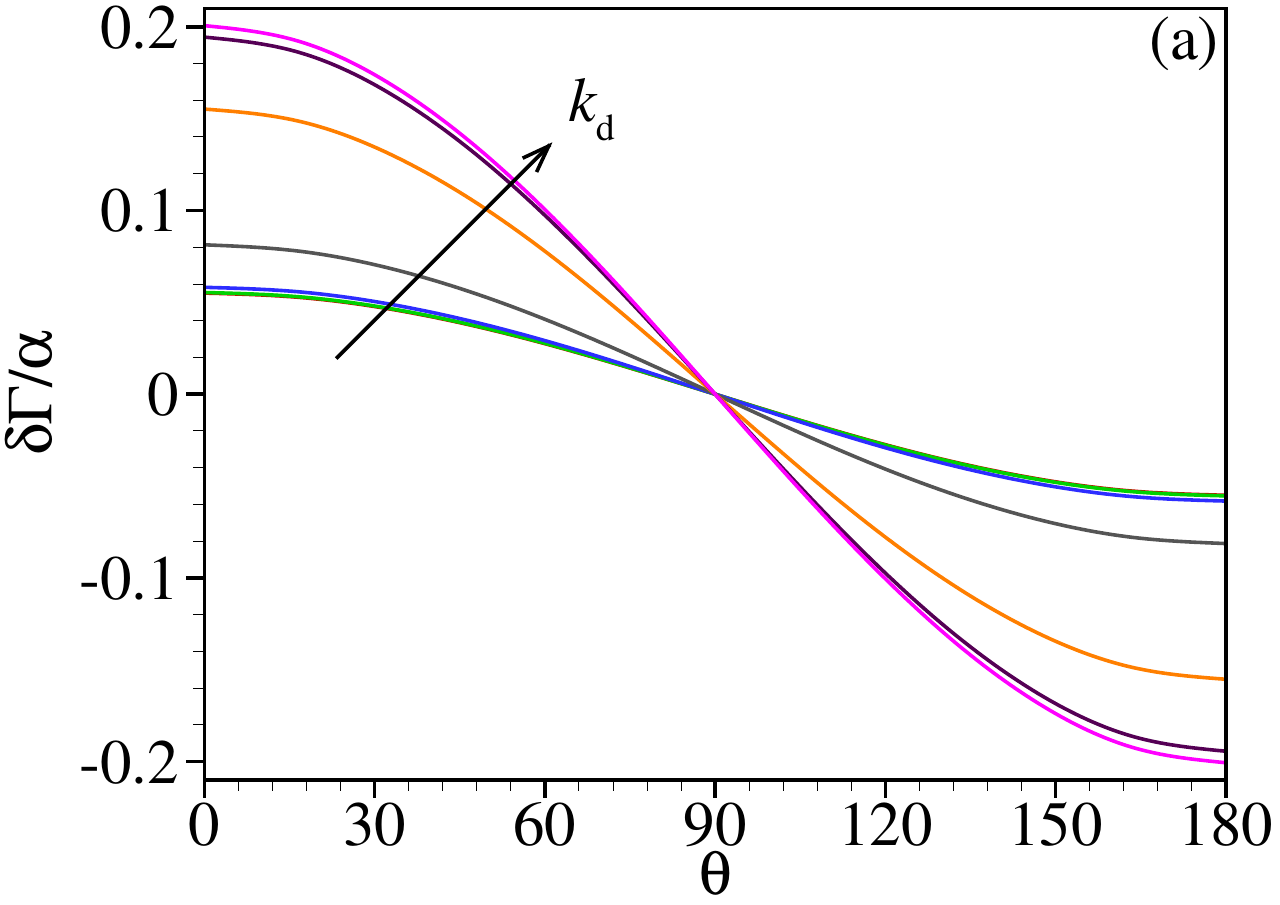}
	\includegraphics[width=2.3in]{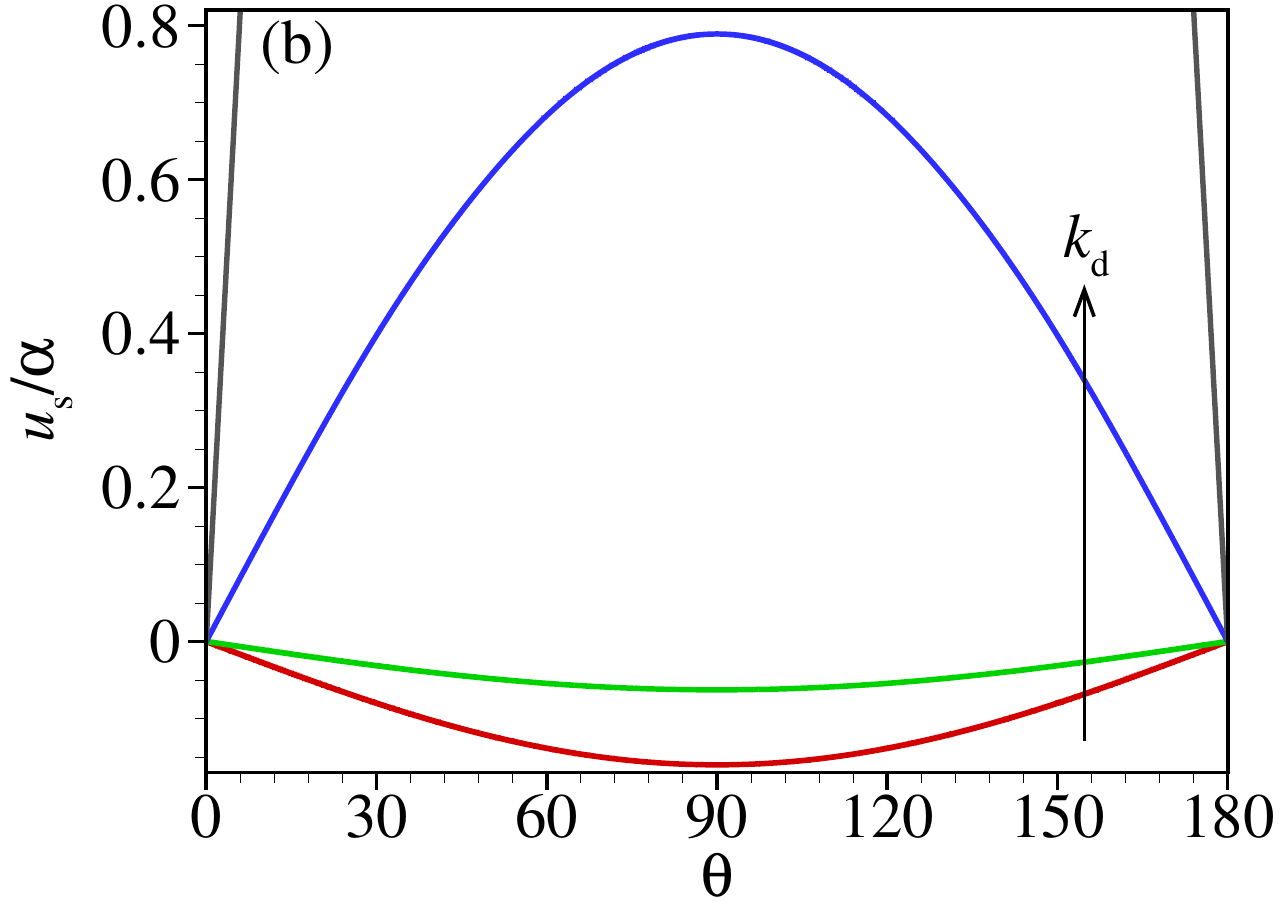}
    \includegraphics[width=2.3in]{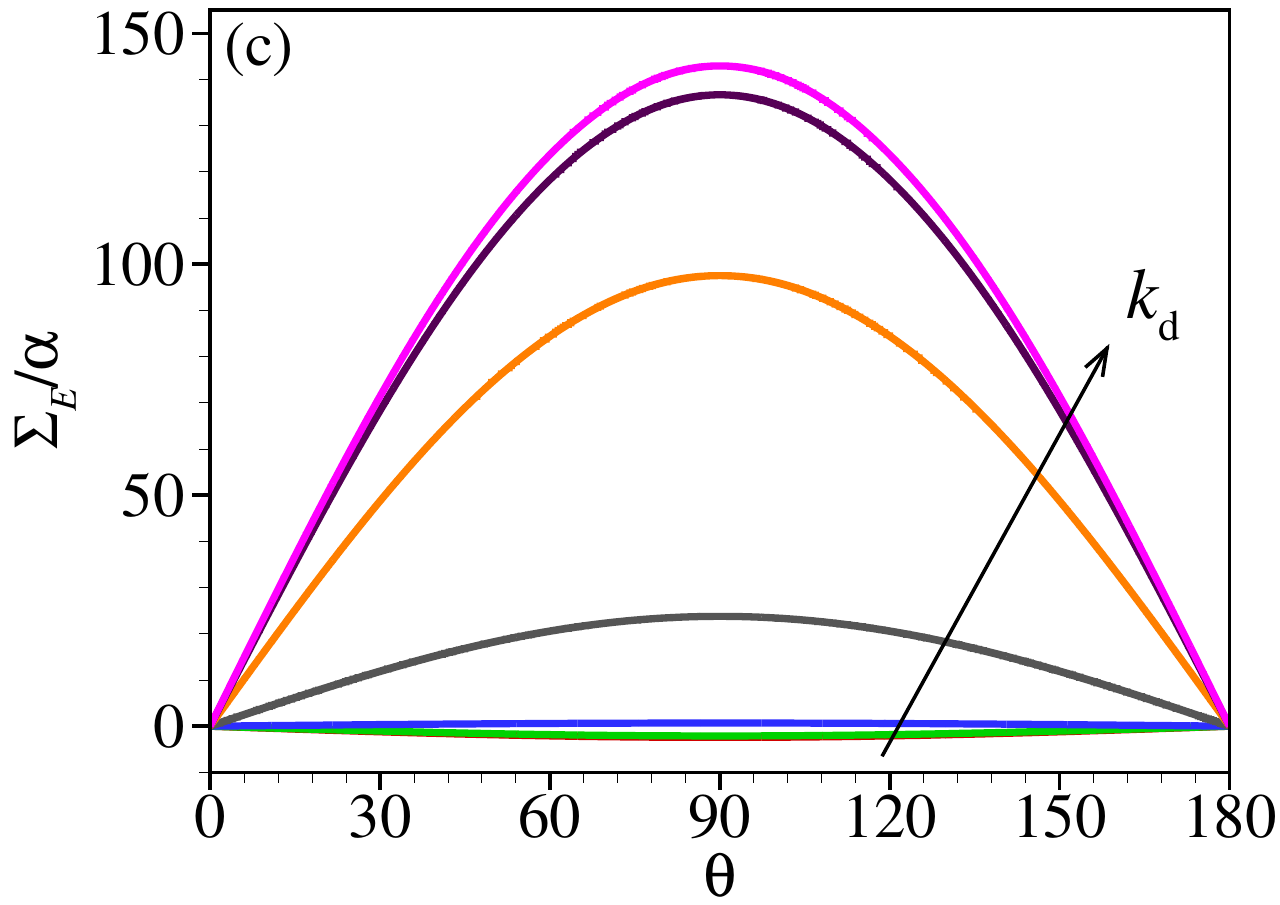}
    \includegraphics[width=2.3in]{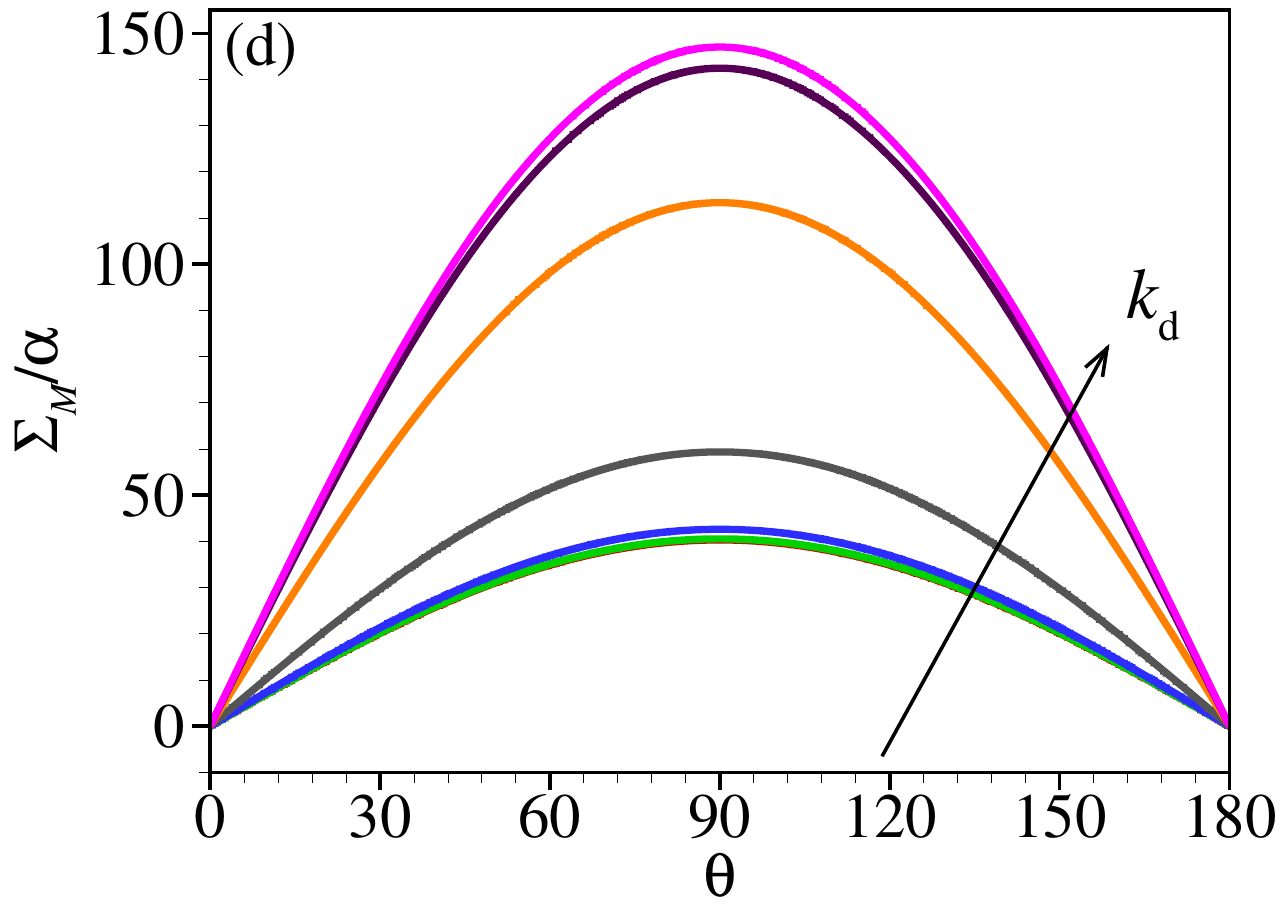}
	\caption{Distribution of (a) surfactant concentration ($\delta\Gamma$) over the interface, (b) interfacial velocity, (c) tangential Maxwell stress and, (d) Marangoni stress at $\kappa a=90$ with $\mu_{r}=1$ and $\Gamma^{0}=0.4$. In (b) $k_d = 0, 10^{4}, 10^{5},$ and $10^{6}~\mathrm{s}^{-1}$ whereas in (a,c,d) $k_{d}=0,10^{4},10^{5},10^{6},10^{7},10^{8}, 10^{10}~\rm s^{-1}$. Here, $\Gamma^{\infty}=1~\rm nm^{-2}$,  the cation is Na$^{+}$, anion is DS$^{-}$, and consequently $\beta=0.543$. The expressions for the interfacial tangential Maxwell stress, Marangoni stress and slip velocity are given by $\Sigma_{E}=-z_{s}Ma\Gamma^{0}\Psi(1)\alpha\sin\theta$, $\Sigma_{M}=Ma(1-\Gamma^{0})^{-1}\Delta_{s}\alpha\sin\theta$, and $u_{s}=h_{r}\alpha\sin\theta$, respectively, (see Section \ref{Perturbed}).}
	\label{fig_FF}
\end{figure}
As $k_d$ increases, surfactant exchange between the interface and the bulk becomes increasingly rapid, enabling strong and sustained non-uniformities in surface concentration. This enhancement of surface gradients substantially amplifies Marangoni stresses as well as the chemiphoresis part. Increase in $k_{d}$ implies the surfactant is highly soluble in the adjacent fluid. Consequently, increase in $k_{d}$ creates a reduction in the electrophoresis part. Beyond a critical value of $k_d$, the chemiphoretic contribution exceeds the negative electrophoretic contribution, causing the mobility to cross zero and reverse sign from negative to positive value. This mobility reversal thus reflects a kinetic-control-induced shift in the balance between electrophoresis and chemiphoresis rather than the emergence of the DLP-II effect creating a negative chemiphoresis, which occurs only in models assuming uniform surface charge density, as discussed before.
\par 
We find from Fig.\ref{fig_EE}a-c that at sufficiently large values of $k_d$, the mobility saturates, indicating a crossover from a kinetically controlled regime to a transport-limited regime. This transition can be understood by comparing the characteristic adsorption--desorption time scale $\tau_k \sim k_d^{-1}$ with the surface diffusion time scale $\tau_s \sim a^2/D_s$ and the electro-diffusive relaxation time across the Debye layer $\tau_b \sim \kappa^{-2}/D_s$. In the large $k_d$ limit, $\tau_k \ll \tau_s, \tau_b$, so that interfacial exchange occurs much more rapidly than either surface or bulk transport. The interfacial surfactant concentration therefore remains in quasi-local equilibrium with the surrounding electrolyte, and further increases in $k_d$ do not alter surface gradients. Consequently, the Maxwell and Marangoni stresses, as well as the interfacial velocity (Fig.{\ref{fig_FF}}b-d) saturate and attain finite limiting values, leading to saturation of the diffusiophoretic mobility $\mu_{D}$.
\par 
 Due to the symmetry of the flow the tangential velocity (Fig.{\ref{fig_FF}}b) as well as Maxwell stress and Marangoni stress (Fig.{\ref{fig_FF}}c,d) are zero at the poles. Tangential gradient of the surfactant concentration creates a viscous tangential shear stress at the interface, the Marangoni stress, which induces fluid motion from area of low surface tension to area of high surface tension. Accumulation of surfactant by adsorption locally reduces the surface tension and rises near the region of desorption. Note that a positive interfacial stress acts along the positive $\theta$-direction, while a negative value indicates the opposite direction. Fig.{\ref{fig_FF}}b shows that as $k_{d}$ is increased the interfacial velocity becomes positive and augments with $k_{d}$. At a higher $k_{d}$ the Marangoni stress enhances, which acts opposite to the Maxwell spinning force, leading to a positive interfacial velocity. 
\par
\begin{figure}
	\center
	\includegraphics[width=1.72in]{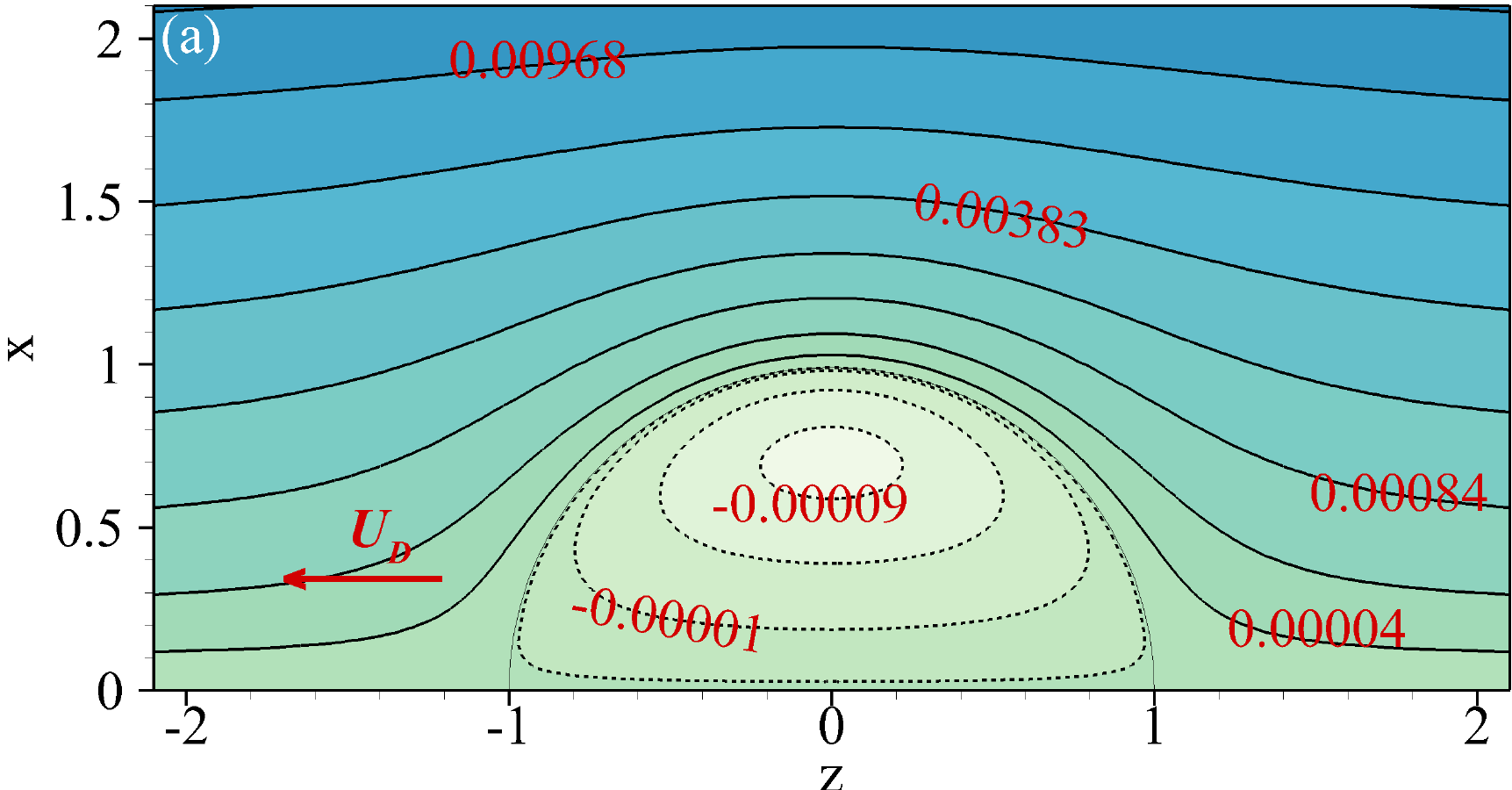}
	\includegraphics[width=1.72in]{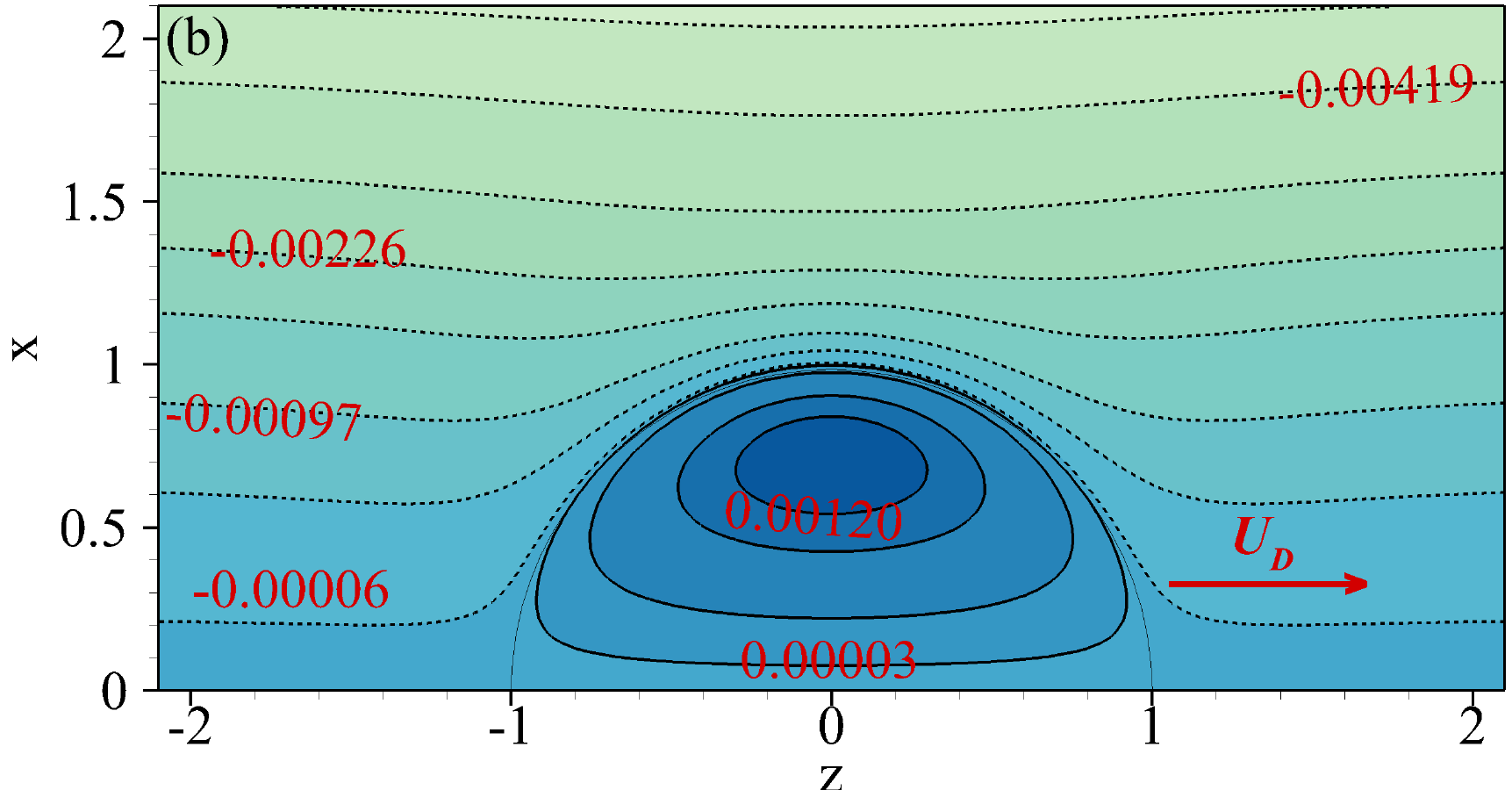}
	\includegraphics[width=1.72in]{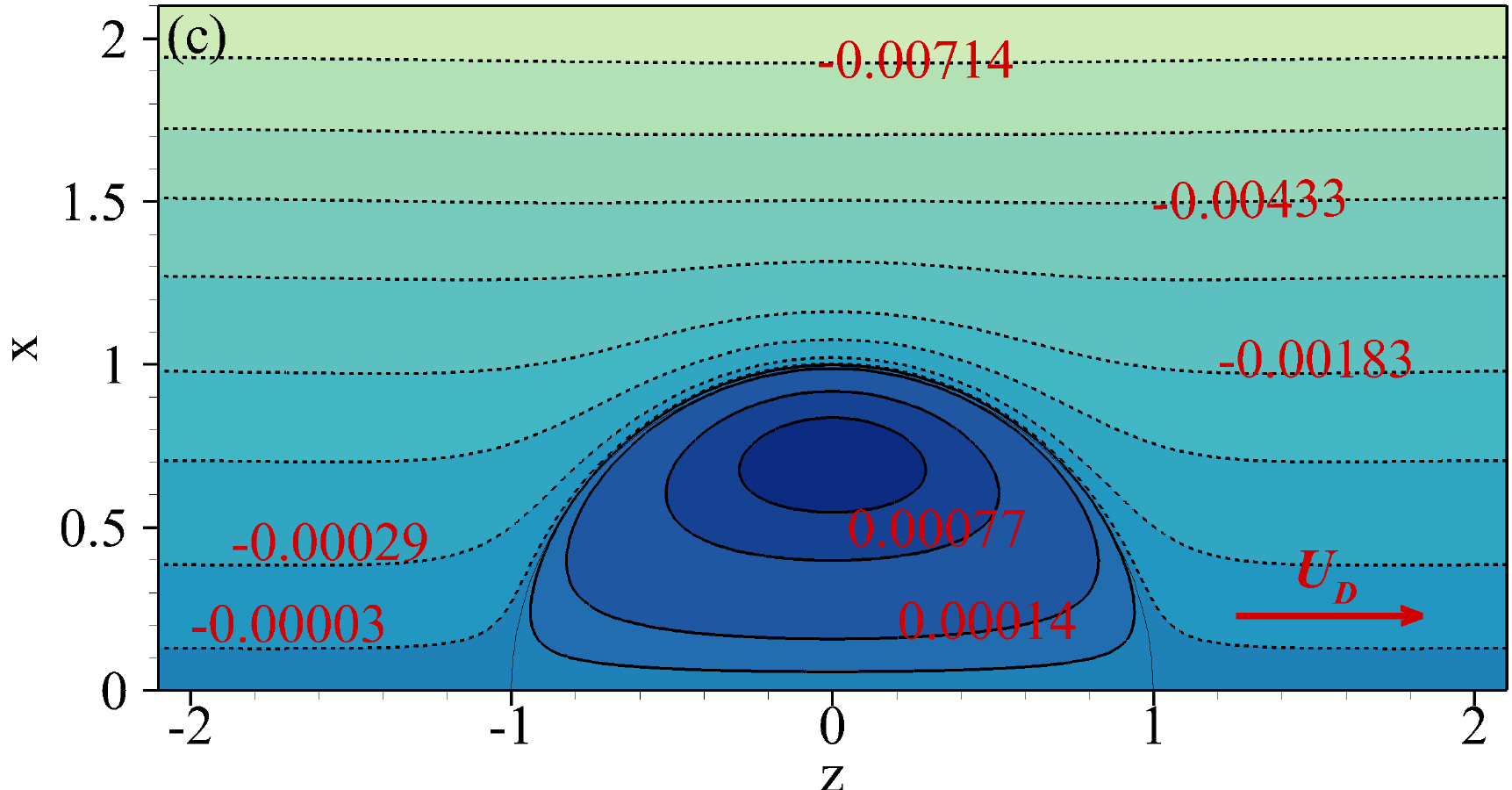}
	\caption{Streamlines pattern inside and around the droplet for $\kappa a=5$, $\mu_{r}=0.1$ at (a) $k_{d}=10^{4}~\rm s^{-1}$, (b) $k_{d}=10^{6}~\rm s^{-1}$, and (c) $k_{d}=10^{8}~\rm s^{-1}$. Here, the electrolyte is NaDS, $\Gamma^{\infty}=1~\rm nm^{-2}$, and $\Gamma^{0}=0.03$.}
	\label{fig_HH}
\end{figure}
The streamline pattern presented in Fig.\ref{fig_HH}a-c illustrate the formation of internal circulation and the flow field outside the droplet as $k_d$ is varied. At a smaller $k_d=10^{4}$, the tangential Maxwell stress acts opposite to the Marangoni stress, resulting in a weak interfacial slip that is partially opposed by the Maxwell contribution. The corresponding streamline pattern in Fig.~\ref{fig_HH}a shows a clockwise internal circulation, with the interfacial velocity directed along the negative $\theta$-direction and the background diffusioosmotic flow acting in the same direction i.e., $z>0$ direction. As $k_d$ is increased, surfactant exchange between the interface and the adjacent electrolyte becomes progressively stronger, enhancing surface-tension gradients at the interface. As a result, the Marangoni stress increases in the positive $\theta$ direction, while the magnitude of the tangential Maxwell stress decreases and eventually tends to align in the positive $\theta$-direction as well. This creates the interfacial slip along the positive direction. At a higher $k_{d}$, electrophoresis part is low and chemiphoresis dominates to create $\mu_{D}>0$ i.e, the background flow along the positive z-direction. For this, no toroidal vortex develops for these cases. 
\par 
A further insight into the role of interfacial kinetics is illustrated by examining the non-monotonic dependence of the mobility on the viscosity ratio $\mu_r$ in Fig.{\ref{fig_GG}}a for $\kappa a=5$. Dependence of $\mu_{D}$ on $\mu_{r}$ for different choice $k_{d}$ for a thinner EDL $\kappa a=50$ is already shown in Fig.{\ref{fig_EE}}c. These results reveal a qualitative change in the $\mu_E$–$\mu_r$ relationship as the desorption rate constant $k_d$ is increased. For smaller values of $k_d$, the mobility decreases monotonically with increasing $\mu_r$. At this smaller $k_{d}$ the electrophoresis part dominates, as well as the non-uniformity in interfacial distribution attenuates. This results in low Marangoni stress and the Maxwell traction together with viscous shear stress spins the droplet along the negative direction. As $k_d$ is increased, however, an intermediate regime emerges in which the magnitude of mobility increases with $\mu_r$. In this case, the interfacial velocity becomes positive whereas, the mobility remains negative. This implies that the chemiphoresis part is not strong enough to overcompensate the electrophoresis contribution so that $\mu_{D}<0$. However, the balance of Marangoni, viscous and Maxwell stresses create a positive spin of the droplet, which is translating along $z<0$ direction. An increased droplet viscosity $\mu_{r}$ attenuates the retarding droplet spin to enhance $-\mu_{D}$. In this cases, as seen in Fig.\ref{fig_II}a-c, a toroidal vortex appears at the outer face of the droplet. At a sufficiently large $k_d$, once the mobility becomes positive, it again decreases monotonically with increasing $\mu_r$. For smaller values of $k_d$, interfacial kinetic exchange is weak and the redistribution of surfactant along the interface remains limited, as evidenced by the small magnitude of surfactant flux at the interface shown in Fig.~\ref{fig_GG}b. Note that this mass flux $J_\textbf{n}$ is evaluated under the perturbation scheme as $J_\textbf{n}=\left[C_{a}(1-\Gamma^{0})n^{0}_{N}z_{4}(\Psi(1)-\Phi_{N}(1))-(C_{a}n^{0}_{N}+C_{d})\Delta_{s}\right]\alpha\cos\theta$. As $k_{d}$ is increased, surfactant exchange between the interface and the surrounding medium becomes sufficiently rapid to sustain stronger surface-charge and surface-tension gradients.
\par 
\begin{figure}
	\center
	\includegraphics[width=1.72in]{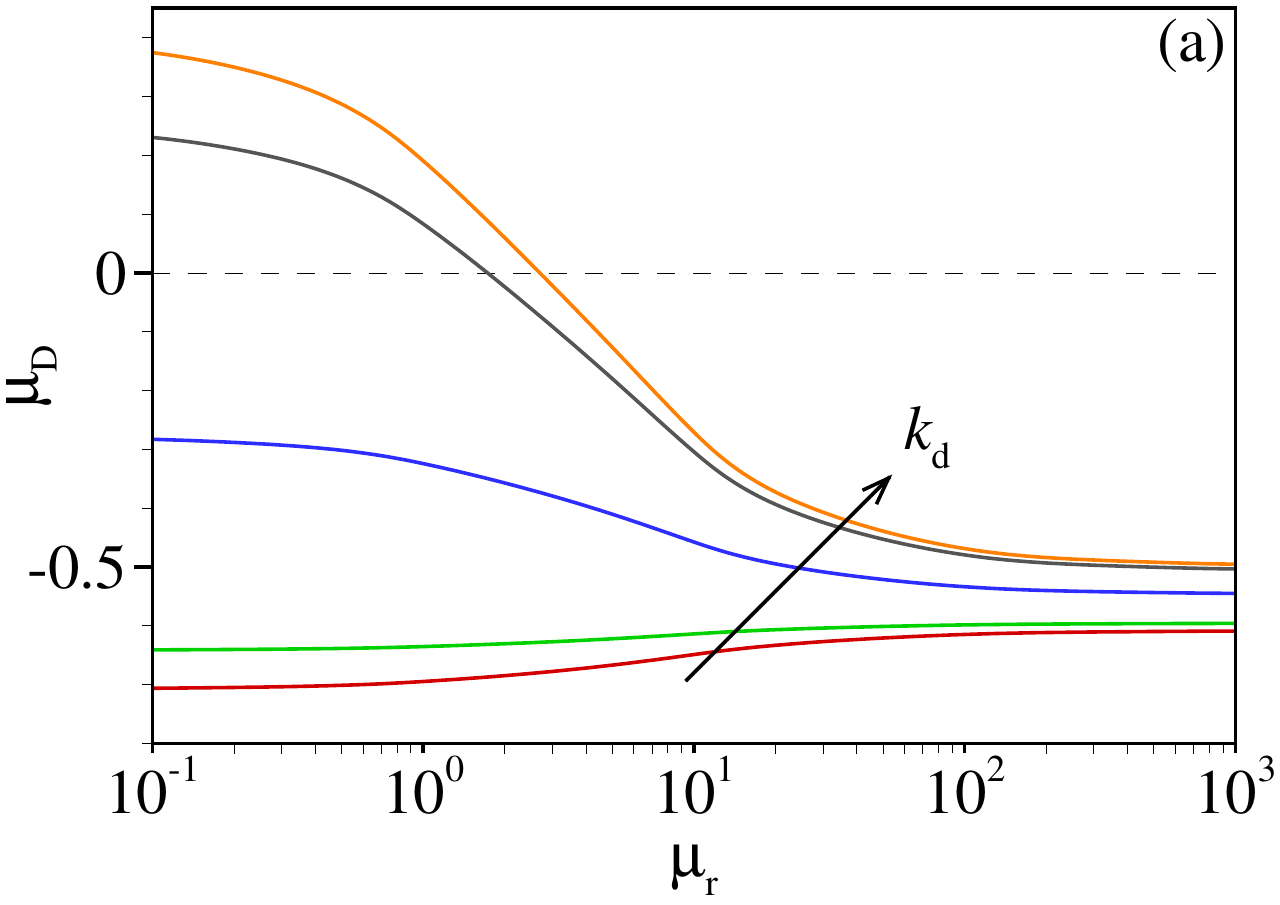}
	\includegraphics[width=1.72in]{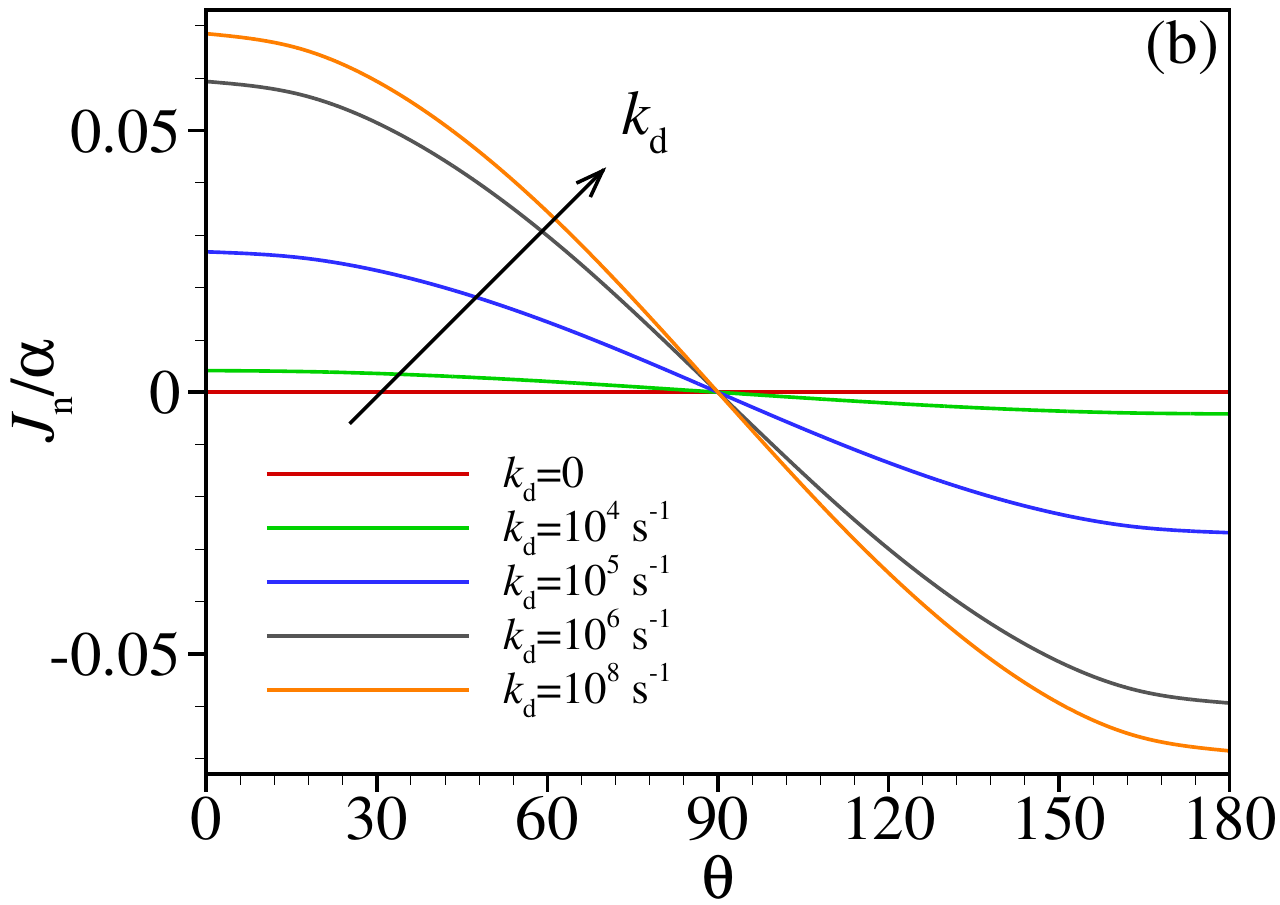}
	\includegraphics[width=1.72in]{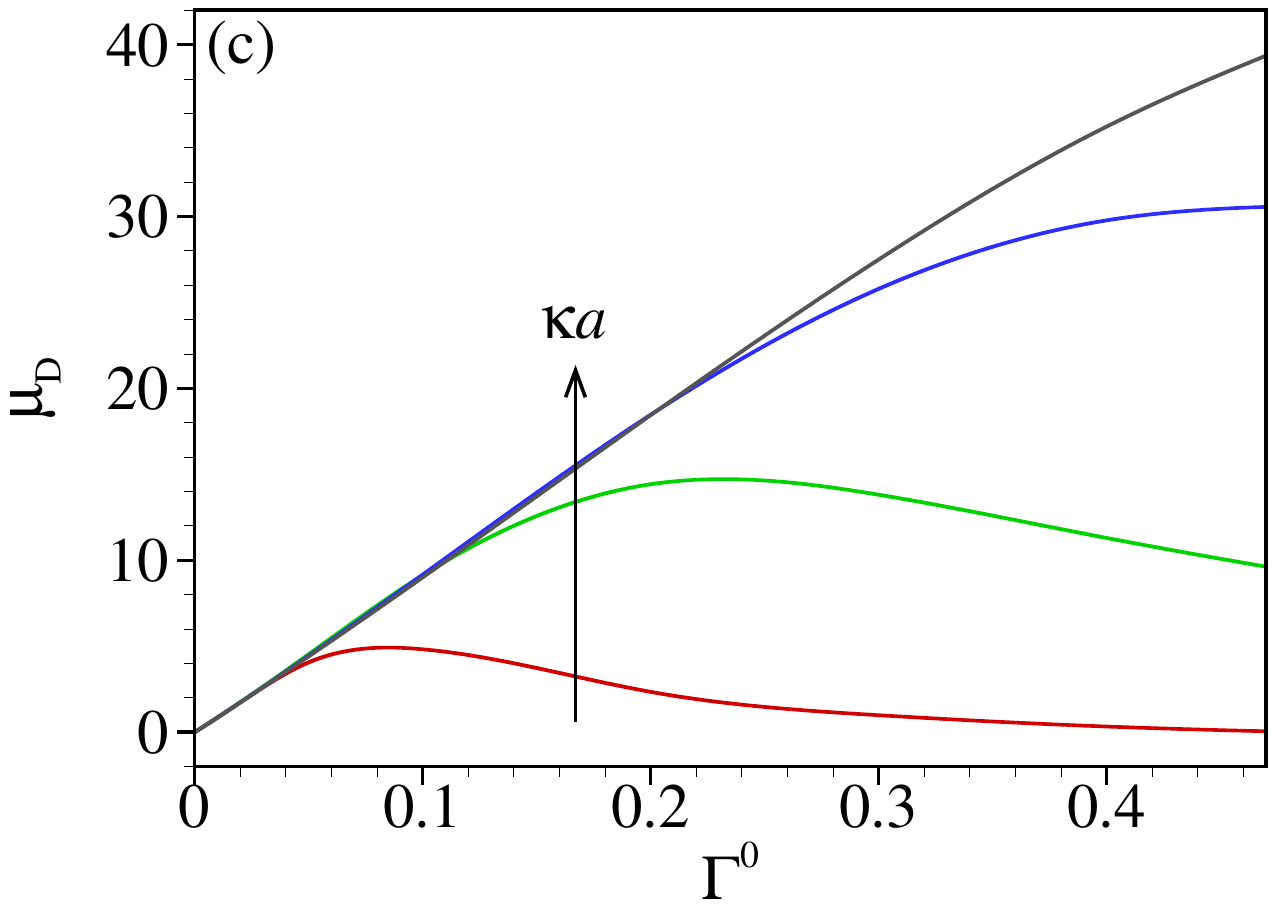}
	\caption{Variation of (a) mobility $\mu_{E}$ as a function of viscosity ratio $\mu_{r}$ and (b) normal mass flux of surfactant ($J_\textbf{n}$) at the interface for $\mu_{r}=0.1$ for different $k_{d}=0,10^{4},10^{5},10^{6},10^{8}~\rm s^{-1}$ when $\kappa a=5$ and $\Gamma^{0}=0.03$; (c) Variation of $\mu_{E}$ as a function of $\Gamma^{0}$ for different $\kappa a=20,50,100,150$ when $\mu_{r}=1$ and $k_{d}=10^{8}~\rm s^{-1}$. Here, in all figures, $\Gamma^{\infty}=1~\rm nm^{-2}$,  the cation is Na$^{+}$, anion is DS$^{-}$, and consequently $\beta=0.543$.}
	\label{fig_GG}
\end{figure}
We now illustrate the impact of equilibrium surfactant concentration $\Gamma^{0}$ on $\mu_{D}$ in Fig.\ref{fig_GG}c for different values of $\kappa a$, at a fixed viscosity ratio $\mu_r = 1$ and a large desorption rate $k_d = 10^8~\rm s^{-1}$. For lower range of $\kappa a$, the mobility exhibits a non-monotonic dependence on $\Gamma^0$. As $\Gamma^0$ is increased from small values, the mobility initially increases and reaches a maximum at an intermediate surfactant concentration, before decreasing at larger $\Gamma^0$. As $\kappa a$ increases, this non-monotonicity becomes weaker. For sufficiently large $\kappa a$, the mobility becomes a monotonic increasing function of $\Gamma^0$ for the considered range of $\Gamma^{0}$. This transition reflects the impact of retarding force arising from the surface conduction. For the lower values of $\kappa a$, corresponding to relatively thick electric double layers, an increase in $\Gamma^{0}$ enhances the surface charge density and hence, the interfacial electric potential. The increased surface potential strengthens surface conduction effects, creates a retardation of the droplet motion. Consequently, the mobility initially increases with $\Gamma^{0}$ due to the enhanced interface potential, begins to decrease once surface conduction becomes sufficiently strong, leading to a reduction in mobility after reaching a maximum. In contrast, at larger values of $\kappa a$, for which EDL is thinner, the surface conduction effects are weak due to the stronger surface charge screening. In this case a higher $\Gamma^{0}$ enhances the surface concentration gradients of surfactant give rise to strong interfacial stresses that enhance interfacial slip. As a result, the mobility increases monotonically with $\Gamma^{0}$ for $\kappa a \geq 100$. 
\par 
The streamline patterns in Fig.\ref{fig_II}a-c clearly depict the formation of a surface-adjacent recirculating vortex, indicating that the interfacial slip opposes the background diffusiophoretic flow. In this intermediate $k_d$ regime, increasing the viscosity ratio suppresses the opposing interfacial circulation more effectively than the background flow, thereby reducing the net resistance to translation of the drop and resulting in an anomalous increase in the mobility with $\mu_r$, as seen in Fig.\ref{fig_GG}a. This outer vortex disappear for higher $k_{d}$ (Fig.\ref{fig_HH}b,c), in which chemiphoresis creates a positive mobility. At a large value of $k_{d}$, both the Maxwell and Marangoni stresses act cooperatively along the interface. The interfacial slip is then dominated by Marangoni driven motion, with the Maxwell stress providing an additional reinforcing contribution.
\begin{figure}
	\center
	\includegraphics[width=1.72in]{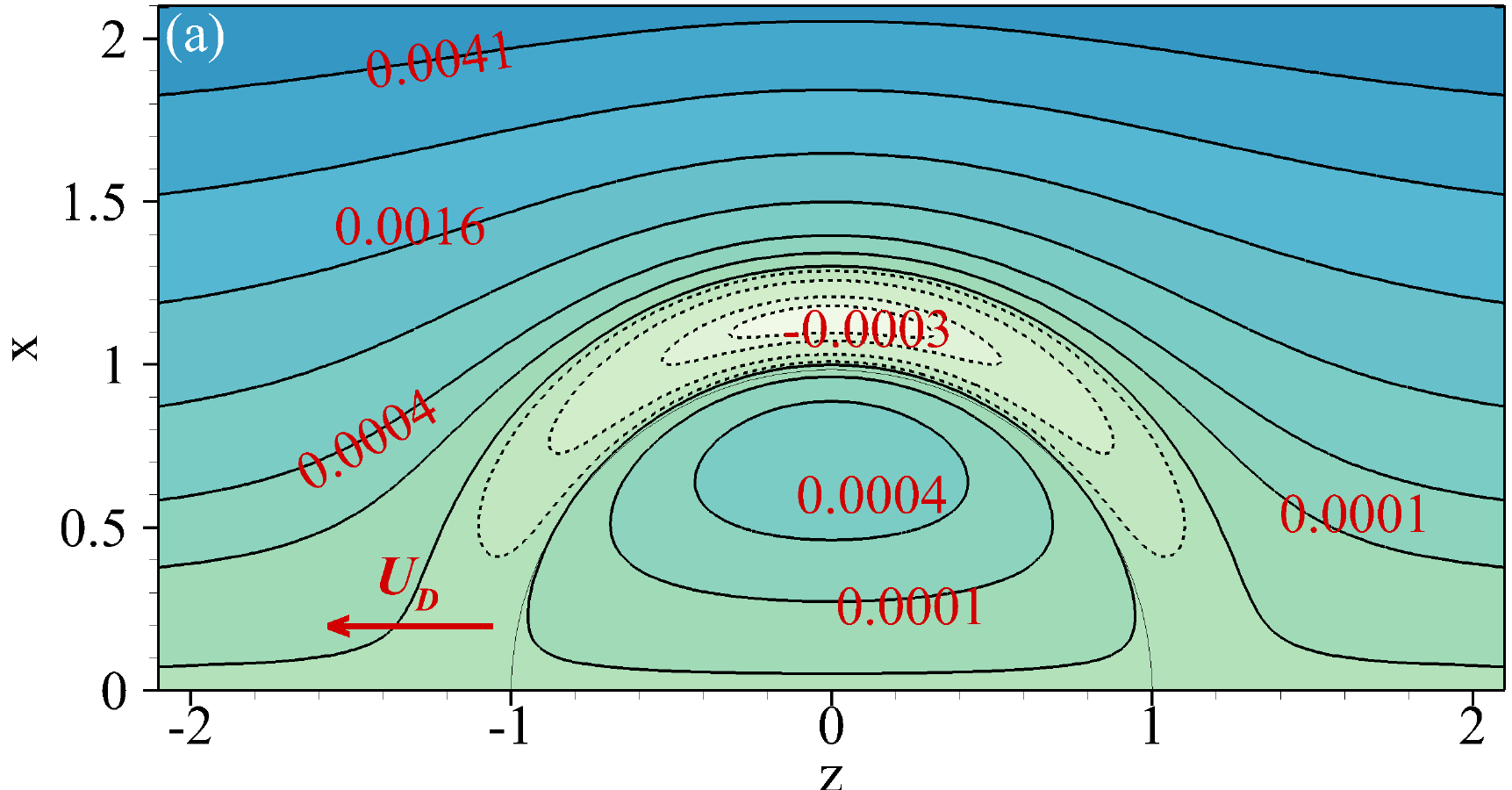}
	\includegraphics[width=1.72in]{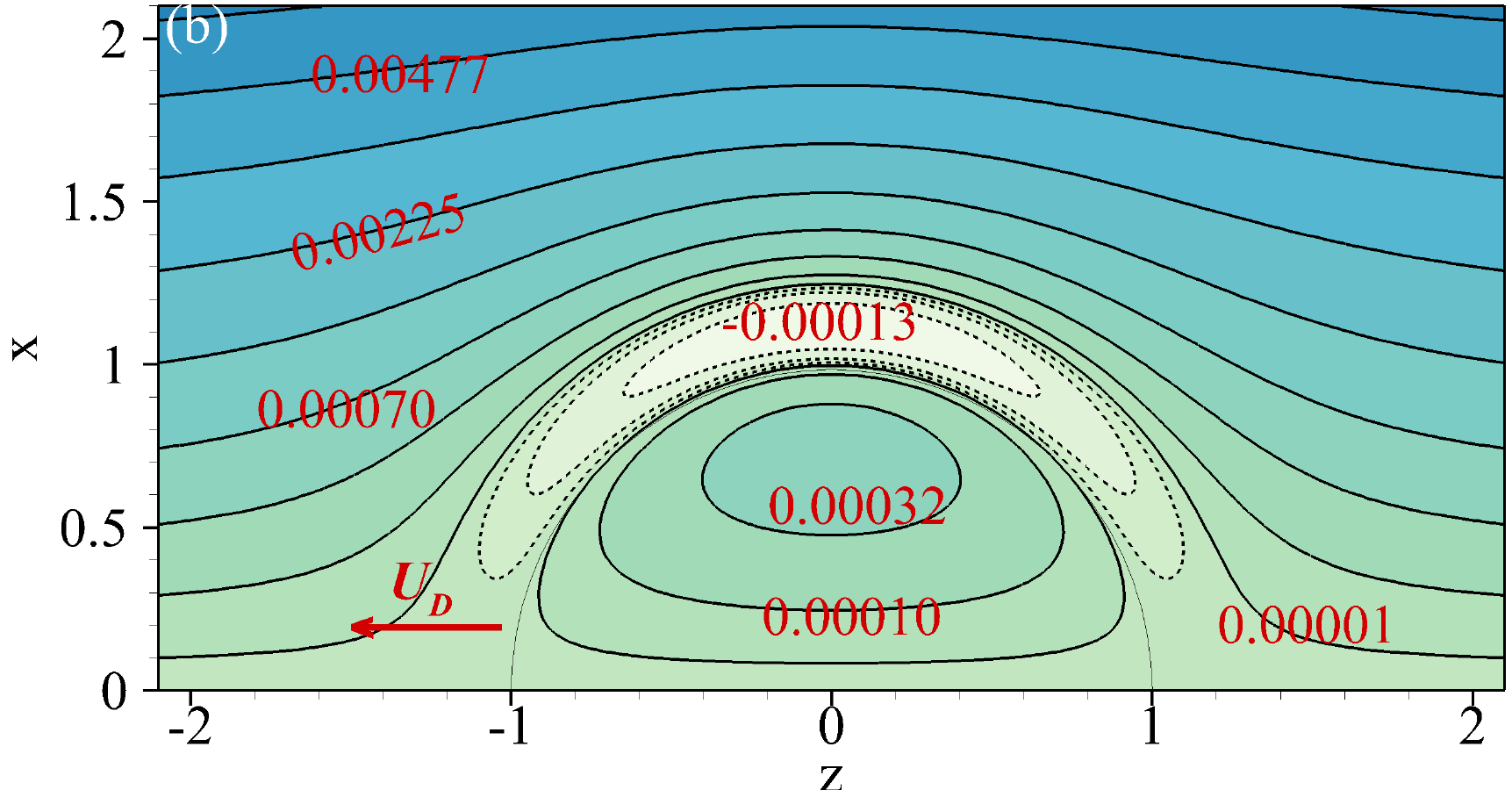}
	\includegraphics[width=1.72in]{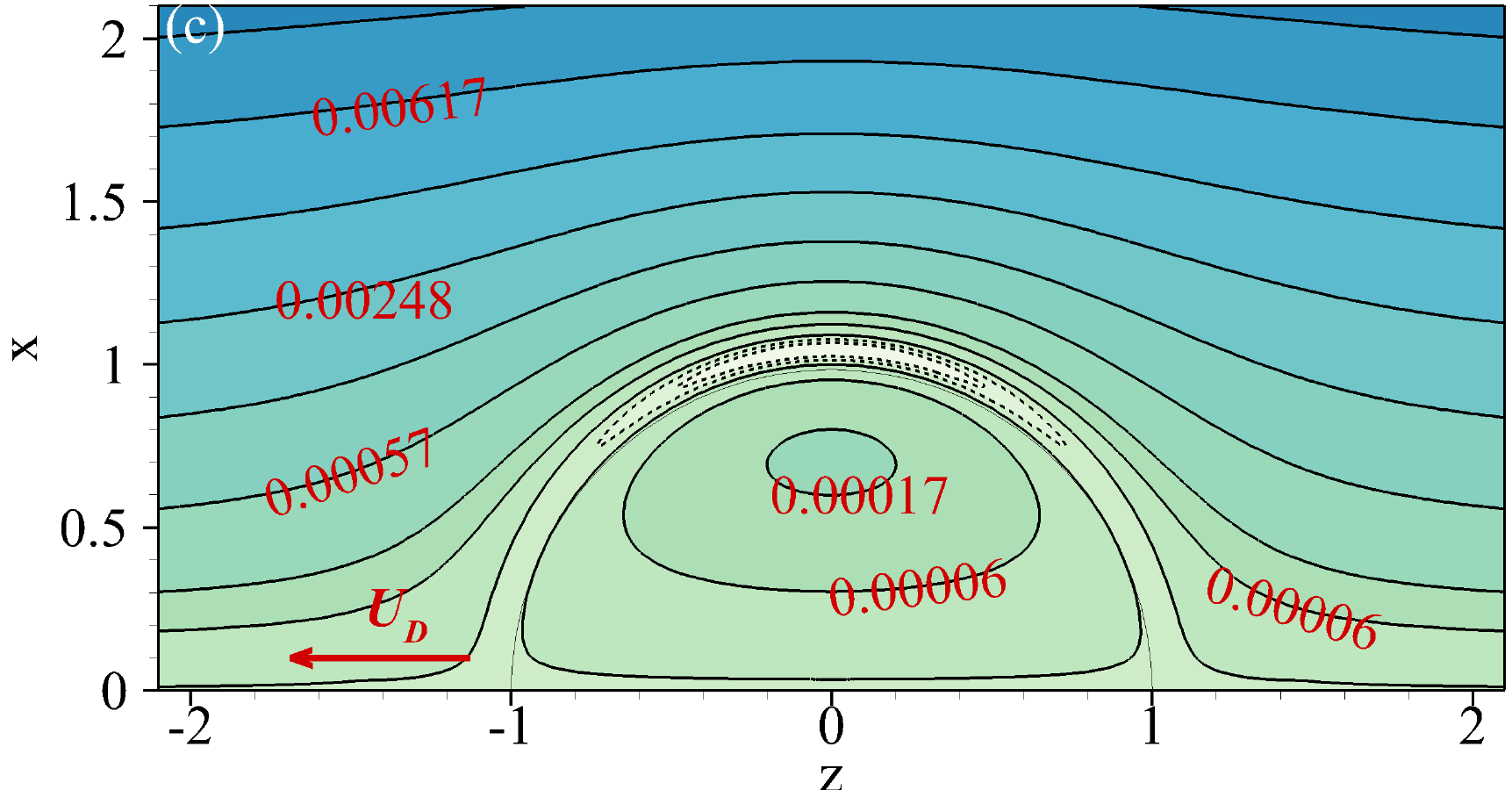}
	\caption{Streamlines pattern inside and around the droplet for $\kappa a=5$ when $k_{d}=10^{5}~\rm s^{-1}$ at (a) $\mu_{r}=0.1$, (b) $\mu_{r}=1$, and (c) $\mu_{r}=10$. Here, the electrolyte is NaDS, $\Gamma^{\infty}=1~\rm nm^{-2}$, and $\Gamma^{0}=0.03$.}
	\label{fig_II}
\end{figure}
At a large value of $k_d$, both the Maxwell and Marangoni stresses act cooperatively along the interface. The interfacial slip is then dominated by Marangoni driven motion, with the Maxwell stress providing an additional reinforcing contribution. In this high $k_d$ regime, increasing the viscosity ratio enhances viscous resistance to the dominant interfacial slip, causing the mobility to decrease monotonically with $\mu_r$, in agreement with the high $k_d$ trends observed in Fig.\ref{fig_EE}(c) and Fig.\ref{fig_GG}a.
\subsection{Interfacial electrokinetic exchange of ionic surfactants in mixed electrolytes}\label{sub:5.4}
\begin{figure}
	\center
	\includegraphics[width=1.72in]{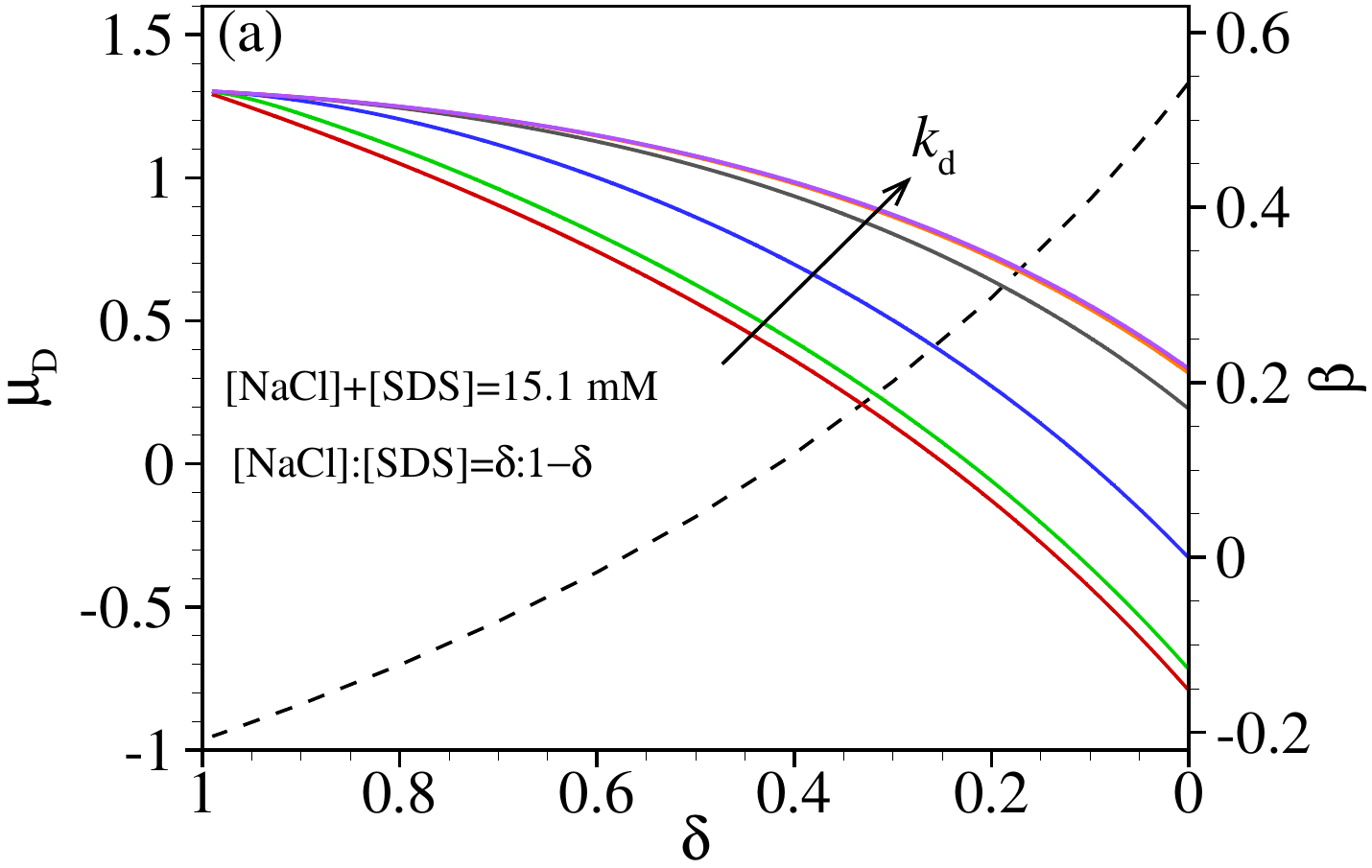}
	\includegraphics[width=1.72in]{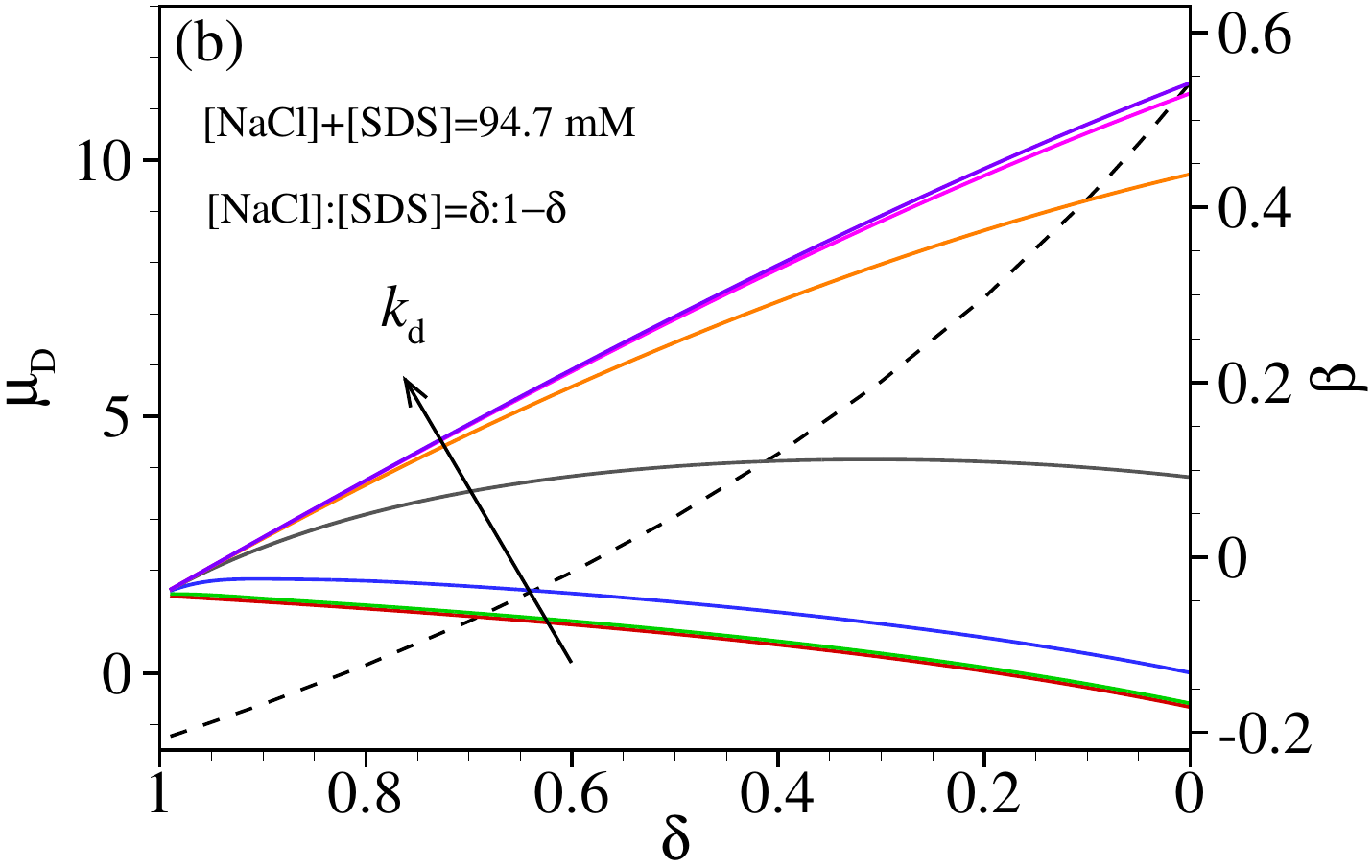}
	\includegraphics[width=1.72in]{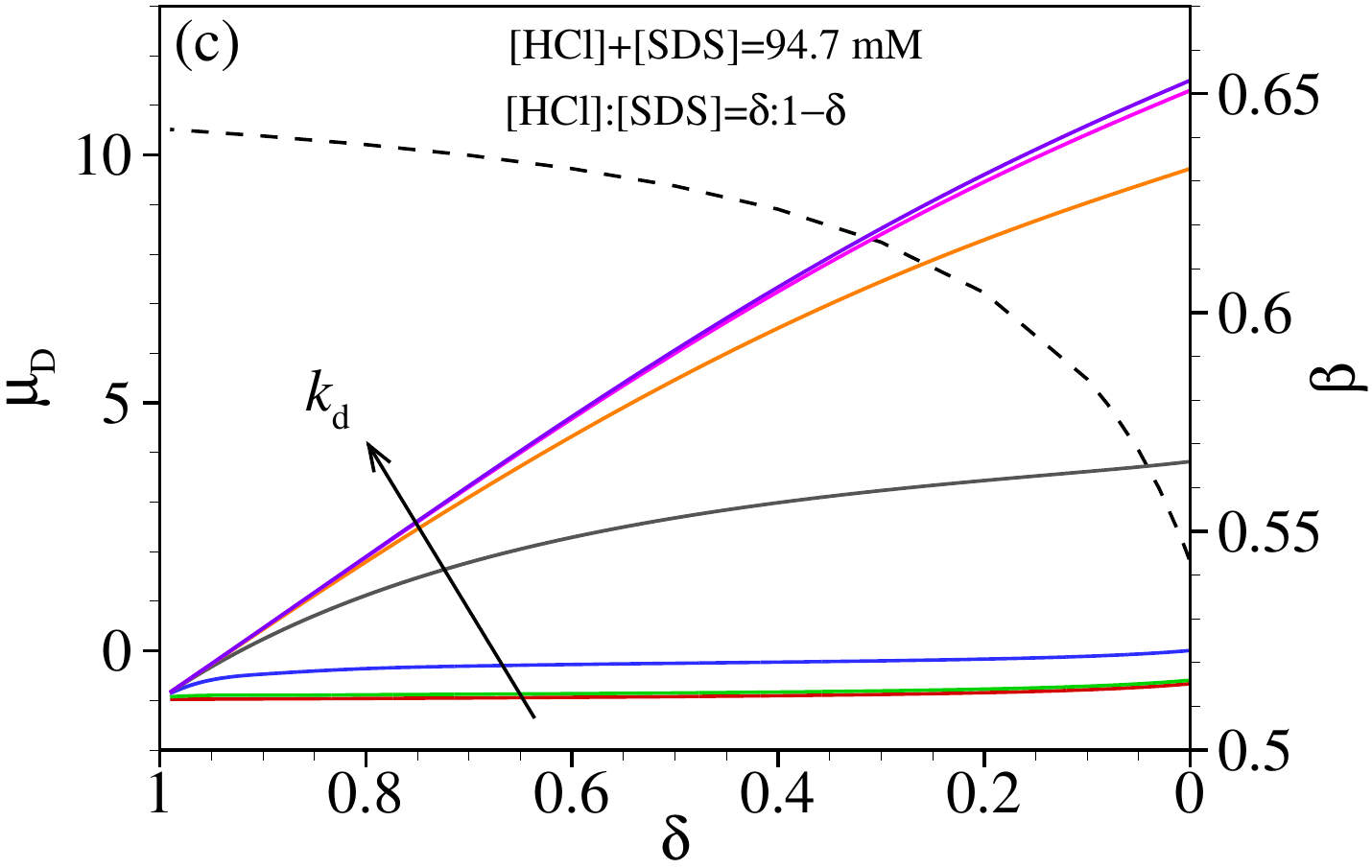}
	\caption{Variation of the diffusiophoretic mobility $\mu_{E}$ at (a) $\kappa a=20$ ($C_{\infty}=15.1~\rm mM$) and (b) $\kappa a=50$ ($C_{\infty}=94.7~\rm mM$) as a function of $\delta$ for a droplet immersed in the mixture of NaCl and SDS electrolytes. (c) Variation of $\mu_{E}$ with $\delta$ at $\kappa a=50$ when the droplet immersed in the mixture of HCl and SDS electrolytes. Here, $k_d=0,10^{4},10^{5},10^{6},10^{7},10^{8}, 10^{10}~\rm s^{-1}$, $\mu_{r}=1$, $\Gamma^{0}=0.4$, $\Gamma^{\infty}=1~\rm nm^{-2}$ and DS$^{-}$ is the adsorbing surfactant ion. Here, $\delta$ measures the fraction of (a,b) NaCl and (c) HCl in the mixture. Solid lines presents $\mu_{E}$ and black dashed lines are the diffusion potential $\beta$, which varies as $\delta$ vary.}
	\label{fig_JJ}
\end{figure}
 Fig.\ref{fig_JJ}a,b show the variation of the mobility $\mu_E$ with the variation of proportion of NaCl in a mixed NaCl-SDS electrolyte at $\kappa a = 20$ and $\kappa a = 50$, respectively, while Fig.\ref{fig_JJ}c presents the corresponding results for a mixture of HCl and SDS at $\kappa a = 50$. Here the proportion parameter $\delta$ is the ratio of concentration of $C_\text{NaCl}$ (or $C_\text{HCl}$) in the bulk concentration $C_\infty~(=n^{\infty}/N_{A})$ of the mixed electrolyte. For the NaCl-SDS mixture at $\kappa a = 20$ (Fig.\ref{fig_JJ}a), the mobility is positive when NaCl dominates the solution ($\delta > 0.5$). In this case, the diffusion potential ($\beta$) is negative, and since the equilibrium surface charge is negative, the electrophoretic contribution is positive. As $\delta$ decreases and the concentration of SDS increases, the diffusion potential varies from negative to positive, and at $\delta=0$ it becomes positive. Consequently, the electrophoretic contribution changes from positive to negative as $\delta$ decreases. For small values of $k_d$, the chemiphoretic contribution remains weak, and the mobility is therefore dominated by electrophoresis. As a result, the mobility decreases with decreasing $\delta$ and undergoes a reversal in sign from positive to negative as $\delta$ reduces from $1$. At larger values of $k_d$, although the mobility decreases with decreasing $\delta$, it remains positive throughout the entire range of $\delta$. In this case, the enhanced interfacial kinetic exchange allows stronger surfactant redistribution and larger Marangoni stress to develop, leading to a significant chemiphoretic contribution, which drags the particle along the direction of the imposed concentration gradient i.e., along the positive z-direction. For $\kappa a = 20$ the electrophoresis part has a non-negligible contribution in mobility. As $\delta$ reduces the diffusion field becomes positive creating the electrophoresis part negative leading to a reduction in the positive $\mu_{D}$ with the decrease of $\delta$. The mobility does not reverse sign as the chemiphoresis part remains dominating. This trends become more pronounced at $\kappa a = 50$, as shown in Fig.\ref{fig_JJ}b. In this thin-double-layer regime, the chemiphoretic contribution is substantially amplified and electrophoresis attenuates due to stronger screening of surface charge. For this, at a higher $k_d$ the positive mobility increases in magnitude and remains positive as $\delta$ decreases from 1. This trend of $\mu_{D}$ reflects the dominance of chemiphoresis process at large a $\kappa a$, where strong surface concentration gradients and Marangoni stress develop.
\par 
Fig.\ref{fig_JJ}c shows the corresponding behavior for a mixture of HCl and SDS at $\kappa a = 50$. In this case, the diffusion potential $\beta$ remains positive over the entire range of $\delta$, varying from $0.64$ to $0.53$. Since the surface charge is negative, the electrophoretic contribution is negative for all values of $\delta$. For small values of $k_d$, the mobility is therefore negative and decreases further as $\delta$ decreases, reflecting the reduction in electrophoresis and the weak chemiphoretic contribution at slow interfacial exchange. However, for higher $k_d$ when chemiphoresis dominates the mobility becomes positive and increases monotonically as $\delta$ decreases, which is similar to the NaCl-SDS case.
\par 
It is evident from Fig.{\ref{fig_JJ}}a-c that the impact of $k_d$ is weak when SDS concentration is low ($\delta \cong 1$), but becomes substantially stronger as the solution becomes SDS-rich ($\delta \to 0$), highlighting the crucial role of interfacial kinetic exchange in surfactant-dominated electrolytes. Variation of the tangential Maxwell stress, Marangoni stress, and velocity at the interface with the ratio of NaCl to SDS concentration in the mixed electrolyte is depicted in Fig.\ref{fig_ZZ}a-c. The Maxwell stress is negative for the lower concentration of SDS and decreases as $\delta$ decreases i.e., SDS concentration increases. At lower $\delta$ and sufficiently large $k_{d}$, Maxwell stress becomes positive (Fig.\ref{fig_ZZ}a). Fig.\ref{fig_ZZ}a-c show that increasing SDS concentration (decreasing $\delta$) in the mixed electrolytes solution of NaCl-SDS leads to a strong amplification of Marangoni stresses and, at sufficiently large $k_d$, a realignment with the positive Maxwell shear stress occurs. This cooperative action of Maxwell and Marangoni forces produces a large positive interfacial slip velocity, as shown in Fig.\ref{fig_ZZ}c. This also corroborates with the occurrence of a larger positive mobility for a lower $\delta$ (Fig. \ref{fig_JJ}b,c).
\begin{figure}
	\center
	\includegraphics[width=1.72in]{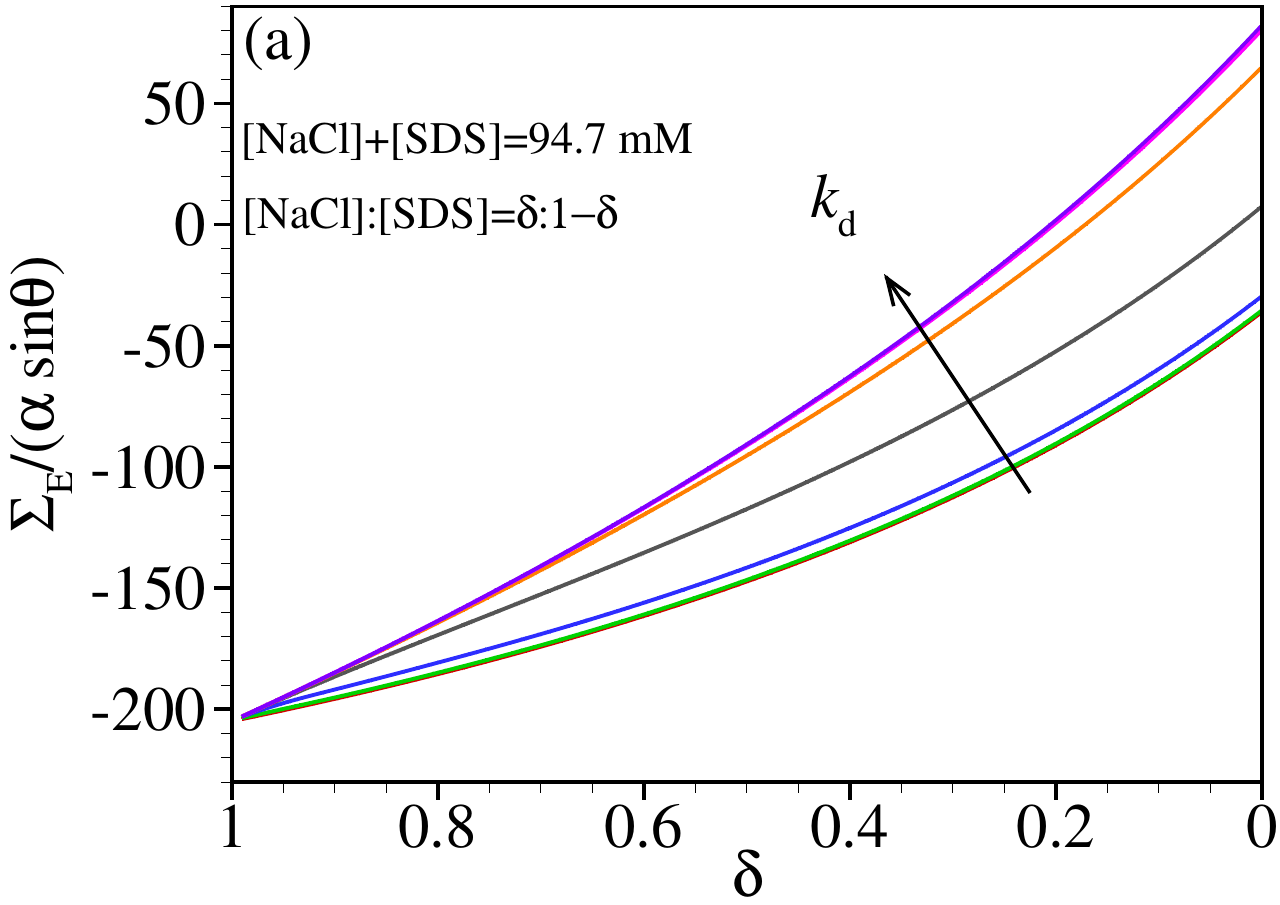}
	\includegraphics[width=1.72in]{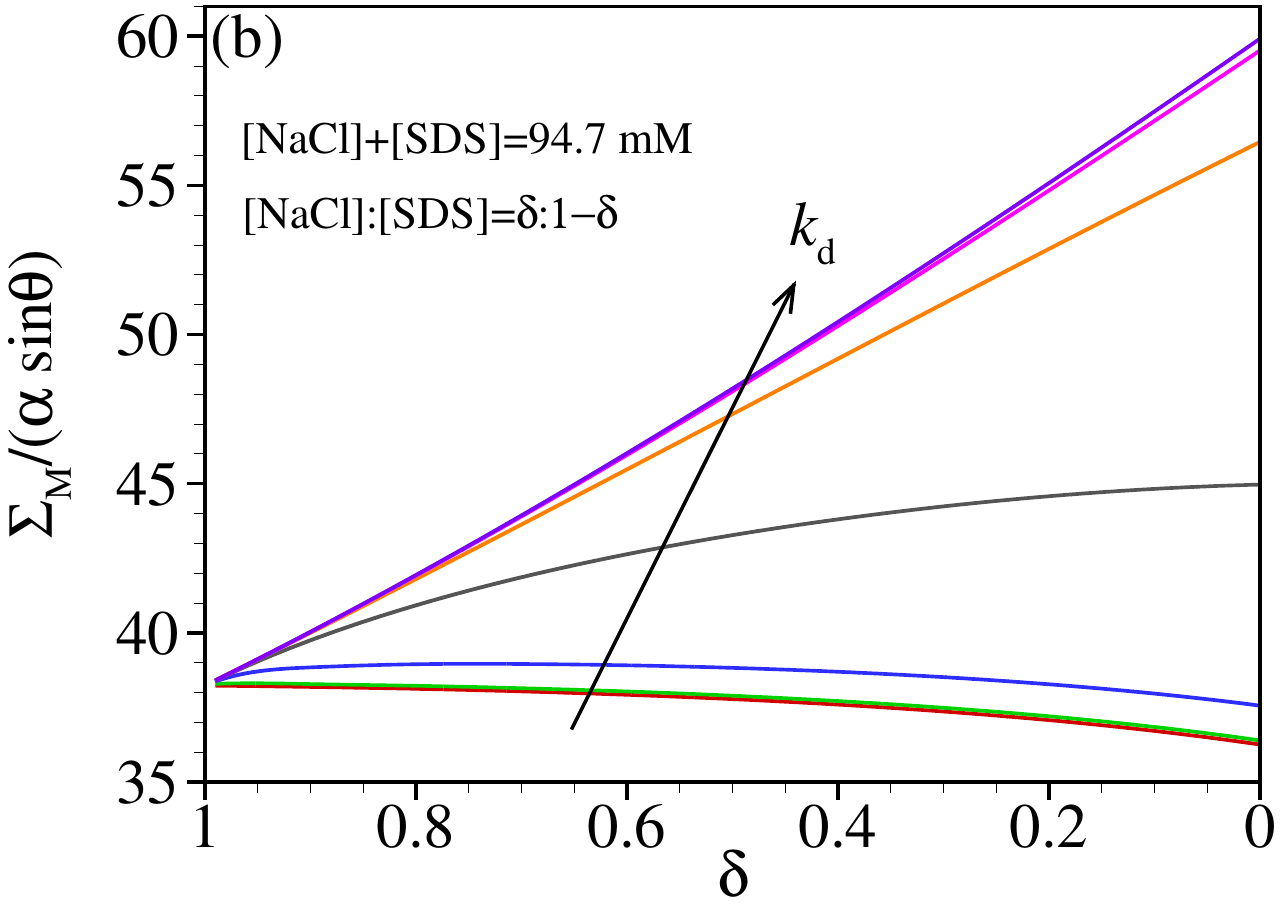}
	\includegraphics[width=1.72in]{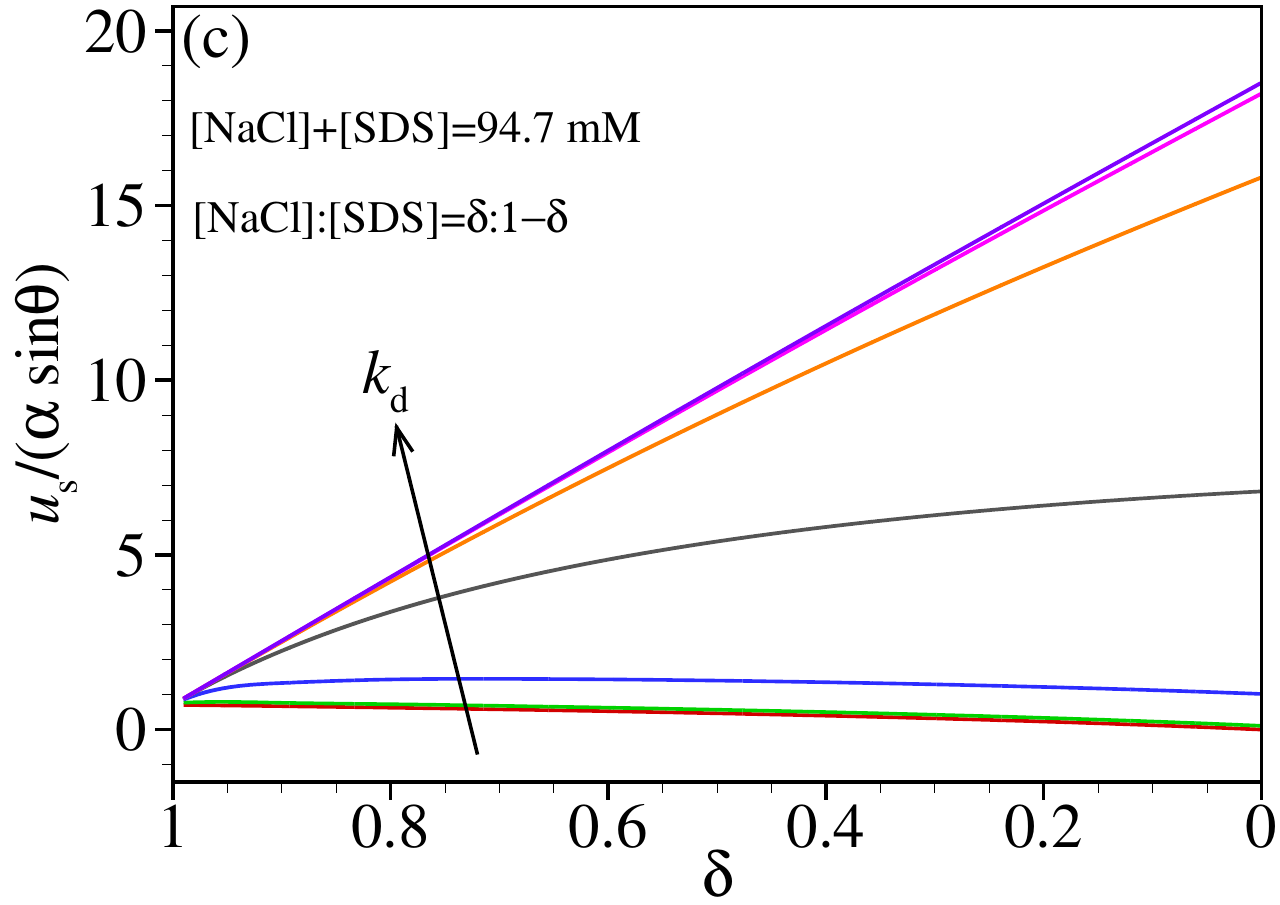}
	\caption{Variation interfacial (a) Maxwell stress, (b) Marangoni stress and (c) slip velocity as a function of $\delta$ at $\kappa a=50$ in a mixed electrolyte solution of NaCl and SDS. The parameters are $k_d=0,10^{4},10^{5},10^{6},10^{7},10^{8}$, $\mu_{r}=1$, $\Gamma^{0}=0.4$, $\Gamma^{\infty}=1~\rm nm^{-2}$ and DS$^{-}$ is the adsorbing surfactant ion. Note that a positive interfacial stress acts along the positive $\theta$-direction, while a negative value indicates the opposite direction.}
	\label{fig_ZZ}
\end{figure}
 It is evident that the variation in $\delta$ tune the relative importance of electrophoresis and chemiphoresis, while the desorption rate constant controls the ability of the interface to sustain surfactant-induced stresses. 
\section{Conclusions}
In this paper, we have presented a comprehensive theoretical and numerical investigation of the diffusiophoresis of a fluid droplet laden with soluble ionic surfactant. By coupling the electrokinetic transport of the surrounding medium with surfactant transport, adsorption--desorption kinetics, and interfacial stress balance, we have developed a framework that captures several key physical effects absent in classical models assuming immobile surface charge. The electrochemical potential of the surfactant is modeled using a Langmuir adsorption isotherm, while the dependence of surface tension on surfactant concentration is incorporated through the Gibbs isotherm. The governing equation for the interfacial surfactant concentration is derived from mass conservation, in which the divergence of the interfacial flux is balanced by the difference between adsorption and desorption fluxes. The bulk fluid flow, ionic transport, and electric potential are described through the coupled Stokes, Poisson, and Nernst--Planck equations. To numerically solve the resulting nonlinear coupled system, we employed a perturbation method with the imposed scaled concentration gradient $\alpha$ as the perturbation parameter. In general, the  experimentally adopted range of $\alpha$ is below $0.0065$. Analytical expressions for the diffusiophoretic mobility and interfacial velocity are also derived for insoluble surfactant under the Debye-H\"uckel approximation as well as in the thin-double-layer limit. Excellent agreement between the analytical and numerical results is obtained in the corresponding parameter regimes, providing strong validation of the formulation and numerical implementation.
\par 
We find that the assumption of a uniform immobile surface charge density can lead to qualitatively incorrect modeling of droplet electrokinetics. In particular, the DLP-II governed negative mobility reported in models with uniform surfactant distribution is shown to arise as an artifact of enforcing an immobile constant surface charge. When the surface charge is allowed to vary self-consistently through surfactant redistribution, the droplet mobility follows the conventional diffusiophoretic theory, which is governed by the sign of the product $\beta\sigma$ and the positive contribution of chemiphoresis. Unphysical features such as the development of negative chemiphoresis and the singular growth of mobility are thereby eliminated. The combined action of Marangoni stresses and tangential Maxwell stresses regularizes the interfacial slip and suppresses type-II double-layer polarization effects, yielding continuous, bounded and physically consistent mobility.
\par 
We further demonstrated that interfacial kinetic exchange of soluble surfactant species plays a decisive role in determining both the magnitude and the direction of droplet diffusiophoresis. At a small desorption rate, surfactant redistribution along the interface is kinetically constrained and the mobility is dominated by electrophoresis. As the desorption rate increases, enhanced adsorption--desorption kinetics of ionic surfactant species amplify the osmotic pressure drop and Marangoni stress, leading to a stronger chemiphoresis and, in certain regimes, a reversal in the direction of motion from negative to positive due to a shift in the balance between electrophoretic and chemiphoretic contributions. At a sufficiently large desorption rate, chemiphoresis becomes dominant and the mobility attains a large positive value irrespective of the sign of the equilibrium surface charge and diffusion field, before eventually saturating in the transport-limited regime.
\par 
The dependence of mobility on the viscosity ratio reveals additional complexity arising from the interplay between Maxwell, Marangoni and viscous shear stresses at the interface. Distinct regimes of the desorption rate constant are identified, in which the mobility decreases monotonically with viscosity ratio, exhibits a non-monotonic behavior, or reverts to monotonic decay. An increase in the interfacial kinetic exchange rate alters the tangential Maxwell and Marangoni stresses, which together determine whether interfacial slip reinforces or opposes the background diffusiophoretic flow and, consequently, whether increasing viscosity suppresses the droplet motion or enhances it over an intermediate range.
\par
We showed that electrolyte composition provides an effective means of controlling droplet motion. For mixed electrolytes, variation in the diffusion potential tune the electrophoretic contribution, while the presence of soluble surfactant and their kinetic exchange with the surrounding medium regulate the strength of chemiphoresis. We find that, by varying the proportion of an added salt in the SDS solution, the droplet mobility can be tuned, leading to a mobility reversal from positive to negative and trapping of the droplet (zero mobility) can be created. These trends in a mixed electrolyte arises due to the occurrence of competitive or cooperative Maxwell and Marangoni stresses at the interface.
\par 
The present study provides a unified and physically consistent framework for understanding the diffusiophoresis of surfactant-laden droplets, highlighting the inseparable coupling between surface chemistry, electrostatics, and hydrodynamics. The model developed here can serve as a foundation for future studies of droplet diffusiophoresis in non-dilute ionic solutions, and multivalent electrolytes, where ion-steric interactions and correlations among ions may become important. Extensions to droplets in polyelectrolyte gel media and other complex environments also represent promising directions for connecting diffusiophoretic transport to practical applications.\\

\textbf{Acknowledgements.} The authors thank Tobias Baier from Technische Universit\"at Darmstadt, Fachgebiet Nano- und Mikrofluidik, for his suggestions and comments on earlier versions of the manuscript.
\bibliographystyle{jfm_v2}
\bibliography{reference}
\end{document}